%% file: FTB_v07.tex
\documentclass[onecolumn,amsmath,amssymb,nofootinbib,12pt]{article}
 
 \ifx\pdfoutput\not\undefined
 \pdfoutput=1
\fi
\usepackage{jheppub}
\usepackage{ifpdf}
\usepackage{graphicx,subcaption}
\graphicspath{ {figures/} }
 \usepackage{amsfonts}
 \usepackage{amsmath}
 \usepackage{amssymb}
 \usepackage{mathtools}
\usepackage{epsfig}
\usepackage{pdfpages}
\usepackage{bbm}
\usepackage{graphicx,epstopdf}
\usepackage[numbers]{natbib}
\usepackage[makeroom]{cancel}
\usepackage{hyperref}
\hypersetup{pdftex,colorlinks=true,allcolors=blue}
\usepackage{array}
\usepackage[export]{adjustbox}

\usepackage[normalem]{ulem}

\usepackage[utf8]{inputenc}

\definecolor{holo1}{HTML}{800080}
\definecolor{holo2}{HTML}{0000fe}
\definecolor{holo3}{HTML}{f60000}

\newcommand{\xm}{{x_\QES^-}}
\newcommand{\xp}{{x_\QES^+}}
\newcommand{\ym}{{y_\QES^-}}
\newcommand{\yp}{{y_\QES^+}}
\newcommand{\tinf}{{t_{\infty}}}
\newcommand{\Teff}{{T_{\rm{eff}}}}
\newcommand{\HP}{{\mt{HP}}}

\newcommand{\xb}{\bar{x}}

\newcommand{\eg}{{\it e.g.,}\ }
\newcommand{\ie}{{\it i.e.,}\ }
\newcommand{\reef}[1]{(\ref{#1})}

\newcommand{\mt}[1]{\textrm{\tiny #1}}

\newcommand{\AdS}{\mathrm{AdS}}
\newcommand{\CFT}{\mathrm{CFT}}
\newcommand{\Bath}{\mathrm{bath}}
\newcommand{\bath}{\mathrm{b}}
\newcommand{\QML}{QM$_{\mt{L}}$}
\newcommand{\QMR}{QM$_{\mt{R}}$}
\newcommand{\GN}{G_{\mt{N}}}
\newcommand{\twist}[1]{
	\ifthenelse{\equal{#1}{1}}%
	{\text{1pt}}%
	{\text{#1pt}}
}
\newcommand{\Bek}{\mathrm{BH}}
\newcommand{\gen}{\mathrm{gen}}

\newcommand{\vn}{\mathrm{vN}}
\newcommand{\QES}{\mt{QES}}
\newcommand{\QM}{\mathrm{QM}}
\newcommand{\Page}{\mt{Page}}
\newcommand{\IR}{\mt{IR}}
\newcommand{\halfLine}{{\frac{1}{2}\text{-line}}}
\newcommand{\rec}{\mathrm{rec}}
\newcommand{\nrec}{\text{non-rec}}
\newcommand{\bif}{{\mt{QES}''}}
\newcommand{\turn}{\mathrm{turn}}
\newcommand{\pure}{\overline{\Bath}}
\newcommand{\ypure}{\tilde{y}}
\newcommand{\sigmapure}{\tilde{\sigma}}
\newcommand{\upure}{\tilde{u}}
\newcommand{\UV}{\mathrm{UV}}
\newcommand{\shock}{\mathrm{shock}}

\newcommand{\Rone}{\uppercase\expandafter{\romannumeral1}}
\newcommand{\Rtwo}{\uppercase\expandafter{\romannumeral2}}
\newcommand{\Rthree}{\uppercase\expandafter{\romannumeral3}}
\newcommand{\Rfour}{\uppercase\expandafter{\romannumeral4}}

\newcommand{\beq}{\begin{equation}}
\newcommand{\eeq}{\end{equation}}
\newcommand{\beqs}{\begin{equation}\begin{aligned}}
\newcommand{\eeqs}{\end{aligned}\end{equation}}

\newcommand{\bea}{\begin{eqnarray}}
\newcommand{\eea}{\end{eqnarray}}

\newcommand{\beqa}{\begin{eqnarray}}
\newcommand{\eeqa}{\end{eqnarray}}

\newcommand{\aemm}{AEMM}
\newcommand{\aims}{AEM$^4$Z}

\renewcommand{\(}{\left(}
\renewcommand{\)}{\right)}
\renewcommand{\[}{\left[}
\renewcommand{\]}{\right]}

\newcommand{\Tb}{T_\mathrm{b}}

\usepackage[thinlines]{easytable}
\usepackage{textcomp}

\title{Evaporating Black Holes Coupled to a Thermal Bath}

\author[a,b]{Hong Zhe Chen,}
\author[a]{Zachary Fisher,}
\author[a,b]{Juan Hernandez,}
\author[a]{Robert C. Myers}
\author[a,b]{and Shan-Ming Ruan}

\affiliation[a]{Perimeter Institute for Theoretical Physics, Waterloo, ON N2L 2Y5, Canada}
\affiliation[b]{Dept.~of Physics $\&$ Astronomy, University of Waterloo, Waterloo, ON N2L 3G1, Canada}

\emailAdd{hchen2@perimeterinstitute.ca}
\emailAdd{me@zachfisher.com}
\emailAdd{jhernandez@perimeterinstitute.ca}
\emailAdd{rmyers@perimeterinstitute.ca}
\emailAdd{sruan@perimeterinstitute.ca}

\date{\today}

\abstract{We study the doubly holographic model of \cite{Almheiri:2019hni} in the situation where a black hole in two-dimensional JT gravity theory is coupled to an auxiliary bath system at arbitrary finite temperature. Depending on the initial temperature of the black hole relative to the bath temperature, the black hole can lose mass by emitting Hawking radiation, stay in equilibrium with the bath or gain mass by absorbing thermal radiation from the bath. In all of these scenarios, a unitary Page curve is obtained by applying the usual prescription for holographic entanglement entropy and identifying the quantum extremal surface for the generalized entropy, using both analytical and numeric calculations. As the application of the entanglement wedge reconstruction, we further investigate the reconstruction of the black hole interior from a subsystem containing the Hawking radiation. We examine the roles of the Hawking radiation and also the purification of the thermal bath in this reconstruction.}

\begin{document}

\maketitle

\section{Introduction}\label{sec:intro}
\input{sections/intro.tex}

\section{Background and setup}\label{sec:background}
\input{sections/background.tex}

\section{Thermal equilibrium}\label{sec:equilibrium}
\input{sections/equilibrium.tex}

\section{Taking Black Holes from the Fridge to the Oven}\label{sec:nonzeroTb}
\input{sections/nonzeroTb.tex}

\section{Summary and Discussion} \label{discuss}
\input{sections/discuss.tex}

\section*{Acknowledgments}
We would like to thank Ahmed Almheiri, Raphael Bousso, Prem Kumar, Hugo Marrochio, Dominik Neuenfeld, Geoff Penington, Ignacio Reyes, Joshua Sandor, Antony Speranza,  Jingxiang Wu and Beni Yoshida for useful comments and discussions. Research at Perimeter Institute is supported in part by the Government of Canada through the Department of Innovation, Science and Economic Development Canada and by the Province of Ontario through the Ministry of Colleges and Universities. HZC is supported by the Province of Ontario and the University of Waterloo through an Ontario Graduate Scholarship. RCM is also supported in part by a Discovery Grant from the Natural Sciences and Engineering Research Council of Canada. JH is also supported by the Natural Sciences and Engineering Research Council of Canada through a Postgraduate Doctoral Scholarship. RCM also received funding from the BMO Financial Group. RCM and ZF also received funding from the Simons Foundation through the ``It from Qubit'' collaboration.

\bibliographystyle{JHEP}
\bibliography{references}

\end{document}

%% file: sections/intro.tex

In the past year, new models of black hole evaporation~\cite{Penington:2019npb,Almheiri:2019hni,Almheiri:2019psf} have given fresh insight into one of the longest-standing puzzles in quantum gravity, the black hole evaporation paradox~\cite{Haw76a,AMPS,AMPSS,Mathur:2009hf,Mat10,MatPlu11,Mathur:2011uj}. The black hole information paradox is essentially the problem that in Hawking's famous calculation, black hole evaporation appears non-unitary, in conflict with the standard rules of quantum mechanics. A black hole may be formed in the collapse of a pure quantum state, however, the evaporation process appears to leave only thermal Hawking radiation in a mixed state. That is, quantum information seems to be destroyed by this process. The newly constructed models, however, have convincingly demonstrated for the first time the entropy decreases after the Page time and unitarity is maintained in quantum gravity. Although the models are semiclassical, they exhibit novel saddle points, first observed in~\cite{Penington:2019npb,Almheiri:2019psf}, which take into account large corrections from quantum fields and produce a Page curve consistent with unitary evaporation. This result represents the first major progress toward resolving the famous paradox in many years.

The model of Almheiri, Engelhardt, Marolf and Maxfield~\cite{Almheiri:2019psf} examines black holes in two-dimensional Jackiw-Teitelboim (JT) gravity theory coupled to conformal matter. Later, Almheiri, Mahajan, Maldacena and Zhao~\cite{Almheiri:2019hni} made a small but important modification: instead of only assuming conformal symmetry for the bulk matter, they also assume that the matter theory is holographic. In this paper, we will use the initials of the original paper ({\aemm}) to denote the original model, and the initials of both papers combined (\aims) to denote the model with holographic matter.

We now give a brief description of the setup for both models. One begins with a two-sided equilibrium black hole, which is a solution of JT gravity coupled to a CFT in two-dimensional anti-de Sitter (AdS$_2$) spacetime. The black hole is allowed to evaporate by changing the asymptotic boundary conditions with a `joining quench' to a nongravitational region containing the same CFT. That is, at time zero, the asymptotic boundary on one side is joined to a semi-infinite interval $[0, \infty)$. The conformal matter in the latter space acts as an auxiliary bath system, which absorbs the Hawking radiation emitted from the evaporating black hole. The dynamics of this model can be solved analytically, including the gravitational backreaction and the von~Neumann entropy of the Hawking radiation. One can study the entropy of the black hole or its complementary subsystem (containing the Hawking radiation) as a function of time, using the Engelhardt-Wall prescription~\cite{EngWal14} (see also \cite{FLM13}) for calculating von~Neumann entropy -- a generalization the Hubeny-Ryu-Rangamani-Takayanagi (HRT) prescription~\cite{RyuTak06,HubRan07} to incorporate quantum corrections. The important distinction between the HRT prescription and the Engelhardt-Wall prescription is that the former computes entropy using codimension-two surfaces with stationary areas, whereas the latter asks us to instead find minimal values of the generalized entropy,\footnote{Counterterms are required to render this quantity finite. For a thorough discussion of how the renormalization of entropy works, see the appendix of~\cite{BouFis15a} for example.} defined by 
\begin{equation}\label{eq:Sgen}
  S_{\rm{gen}} = \frac{A}{4\GN \hbar} + S_{\rm{out}}~.
\end{equation}
That is, to leading order in $\GN \hbar$, this quantity is simply the area $A$,\footnote{Note that in the following we examine two-dimensional JT gravity where the Bekenstein-Hawking contribution is replaced by ${\phi}/(4 G_N\hbar)$, where $\phi$ denotes the value of the dilaton evaluated on the QES.} but the functional receives a quantum correction $S_{\rm{out}}$ given by the entropy of quantum fields of the spatial region outside the surface. The surface which extremizes $ S_{\rm{gen}}$ is referred to as the quantum extremal surface (QES).
In the {\aims} model, the calculation of the generalized entropy is purely geometric using holography. That is, assuming the bath system is described by a holographic CFT$_2$, $S_{\rm{out}}$  can be found using the HRT prescription in the AdS$_3$ dual, while the Bekenstein-Hawking term becomes an additional boundary contribution (from the JT gravity) for HRT surfaces ending in the gravitational region, \ie on the Planck brane -- see \cite{Chen:2020uac} for further discussion.

Using this approach, at early times, the entropy of the Hawking radiation grows in a manner consistent with Hawking's original calculation of information loss. However, later in the evolution, a new class of extremal surfaces appears and the QES computing the entropy switches to this new class. These new surfaces, which lie close to the black hole's horizon, exist and dominate the values of the generalized entropy $S_{\rm{gen}} $ due to large entropy gradients that come from the contribution of Hawking radiation at late times. With the QES approaching the horizon at late times, the result is a phase transition in the entropy, producing a downward-sloping Page curve consistent with unitary evolution towards a pure state. Recovering a unitary Page curve for old black holes is a major step towards resolving the information paradox. It indicates that the semiclassical gravity path integral knows more about unitarity than previously believed.

This result is surprising from the perspective of the two-dimensional theory. In particular, the above phase transition indicates that at late times, the standard calculation of the von~Neumann entropy of the Hawking radiation is incorrect because of gravitational effects. Instead, one must modify the usual prescription for computing the entropy with the so-called `island formula' \cite{Almheiri:2019hni}, which accounts for the contributions of quantum extremal islands (QEIs). The QEIs are gravitational regions that may contribute to reducing the (entanglement) entropy of a non-gravitational region by creating new stationary points for the generalized entropy, \ie the sum of the gravitational and matter entropies. In particular, for a QEI, a change in area from perturbing the boundary of a QEI is exactly compensated for with an equal and opposite change in the entropy of the quantum fields inside the island. The HRT prescription in the three-dimensional bulk theory implies that the correct generalized entropy in the two-dimensional theory should be computed by including these islands, whenever they exist, to the entangling region, if doing so results in a smaller entropy. In the present context, the phase transition where the QEIs appear corresponds to the time when the thermal bath encodes (part of) the black hole interior, a manifestation of the ER = EPR principle~\cite{MalSus13}. See \cite{Almheiri:2019yqk,Chen:2019uhq,Chen:2020uac,Almheiri:2019psy,Chen:2019iro,Gautason:2020tmk,Anegawa:2020ezn,Balasubramanian:2020hfs,Hartman:2020swn,Hashimoto:2020cas,Hollowood:2020cou,Alishahiha:2020qza,Geng:2020qvw,Li:2020ceg,Chandrasekaran:2020qtn,Almheiri:2020cfm,Bak:2020enw,Hollowood:2020kvk,QEItwo} for recent explorations on the island formula in different black hole geometries and \cite{Rozali:2019day,Almheiri:2019qdq,Bousso:2019ykv,Penington:2019kki,Akers:2019nfi,Sully:2020pza,Chen:2020wiq,Giddings:2020yes,Kim:2020cds,Verlinde:2020upt,Liu:2020gnp,Marolf:2020xie,Almheiri:2020cfm,Bousso:2020kmy} for more associated studies on information paradox and Page curve from various aspects. 

These models are clearly rich with new physics, and with fascinating implications for quantum gravity. The present work furthers the direction of our earlier work~\cite{Chen:2019uhq} exploring these models. There, we initiated a study of the flow of quantum information during black hole evaporation. In this earlier work, and indeed in most of the literature on these models, the AdS black hole evaporates completely due to the coupling to a bath system prepared at zero temperature. In this paper, we study the dynamics of coupling the initial equilibrium black hole to a bath BCFT which is initially in a finite non-zero temperature state instead. Similar situations were studied in \cite{Almheiri:2019yqk,Hollowood:2020cou}, but we do not make the assumption that the black hole and bath are initially at the same temperature. We study the resulting dynamics numerically and analytically.

As in~\cite{Almheiri:2019yqk}, we find that the quantum extremal surfaces can lie outside the black hole horizon, and correspondingly the QEIs can include part of the exterior of the black hole. For thermal baths with a temperature around the same temperature as the initial black hole, the late time QES is already outside the horizon around the Page time. On the other hand, with arbitrary bath temperatures, the late time QES are initially inside of the event horizon and eventually cross the event horizon, remaining outside for the rest of the equilibration process.

Similar to our analysis in our previous work~\cite{Chen:2019uhq}, we compute the Page curve for the dynamic black hole coupled with a thermal bath at arbitrary temperatures or equivalently, that of the complementary subsystem to the black hole \ie the {\QML} together with (parts of) the bath and its purification. Taking the bath to be at finite temperature changes the flow of quantum information in important ways. The bath has its own purification and thus must be accounted for in the computation of the generalized entropy. We study the role of the purification in altering the flow of quantum information as the black hole and bath exchange radiation. 

In section~\ref{sec:background}, we review the \aims\ model and set up the model for a black hole in contact with an auxiliary bath at finite temperature, finding the generalized entanglement entropy of various different intervals during the equilibration process. The equilibrium case, \ie the black hole temperature immediately after the quench matches that of the bath, is analyzed in section~\ref{sec:equilibrium}, and we find the constraints for finite bath intervals, together with {\QML}, to recover the black hole interior. Interestingly, the purification of the bath is essential for the reconstruction of the black hole interior. The case of general temperatures is studied in section~\ref{sec:nonzeroTb}, and the results smoothly interpolate between the evaporating case of~\cite{Almheiri:2019hni,Almheiri:2019psf,Chen:2019uhq} and the equilibrium case in section~\ref{sec:equilibrium}.

%% file: sections/background.tex

\begin{figure}[t]
	\centering\includegraphics[width=3.5in]{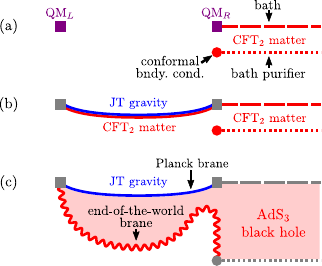}
	\caption{In the {\aims} model, the holographic principle is invoked twice, resulting in three different pictures of the same physical system. In our present analysis of this model, we include  a thermal bath at finite temperature. In the top picture (a), there are two quantum mechanical systems (\textcolor{holo1}{\QML} and \textcolor{holo1}{\QMR}), as well as a two copies of the field theory (\textcolor{holo3}{CFT$_2$}) on a half-line (\textcolor{holo3}{dashed} and \textcolor{holo3}{dotted}). Both of the quantum mechanical and field theory systems are prepared in independent thermofield double states. The middle picture (b) introduces the two-dimensional holographic geometry (\textcolor{holo2}{JT gravity}) dual to the entangled state of \textcolor{holo1}{\QML} and \textcolor{holo1}{\QMR}. This gravitating region also supports the same \textcolor{holo3}{CFT$_2$} that appears in the bath region. The last picture (c) contains the doubly-holographic description, where the holographic CFT is replaced by an \textcolor{holo3}{ AdS$_3$ bulk}, and in particular, the thermofield double is replaced by a bulk region with the geometry of an \textcolor{holo3}{ AdS$_3$ black hole}.}\label{trio}
\end{figure}

The \aims\ model~\cite{Almheiri:2019hni} has three holographic descriptions -- see figure \ref{trio}. The boundary perspective describes the system as two quantum mechanical systems  {\QML} + {\QMR} in a thermofield double (TFD) state which is connected to a bath via a quantum quench. In the present analysis, the bath consists of two copies of a two-dimensional holographic CFT on a half-line, which is initially prepared in an independent TFD state, with a temperature $T_\bath$. After the quench, the system evolves towards a new equilibrium between the quantum mechanical and bath systems,  during which three different phases are distinguished by the position of the quantum extremal surface. The TFD in {\QML} + {\QMR} is dual to a two-dimensional black hole in JT gravity, and this gravitational region also supports the same holographic CFT matter as appears in the bath. The third description replaces the holographic CFT with a three-dimensional AdS bulk and in particular, the TFD is replaced by a AdS$_3$ black hole geometry. From this bulk perspective, the joining quench~\cite{Shimaji:2018czt} connecting the systems has a holographic description as an end-of-the-world brane pinching off the AdS$_2$/bath boundary and falling into AdS$_3$ spacetime. 

\begin{figure}
	\centering
	\includegraphics[width=0.6\textwidth]{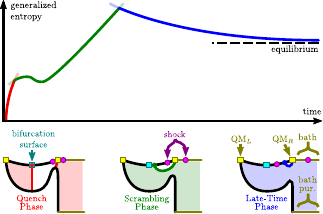}
	\caption{A cartoon illustration of the three phases for the entanglement entropy of $\QM_{\mt{R}}$ or of $\QM_{\mt{L}}$, (a semi-infinite interval in) the thermal bath, and the (entire) bath purifier, after the quench where $\QM_{\mt{R}}$ is connected to the bath. The darker colors indicate the true generalized entropy, while the lighter colors indicate the general shape of each of the branches slightly beyond the regime where it provides the minimal value for the generalized entropy. Below the plot is a sketch of the shape of the extremal HRT surfaces in AdS$_3$ which contribute to the generalized entropy in each phase.}
	\label{fig:sertraline}
\end{figure}

The three phases of the equilibration process are illustrated in figure~\ref{fig:sertraline}. The QES remains at the bifurcation surface during the quench phase. At the transition to the scrambling phase, the QES shifts outwards by a very small distance. The generalized entropy in these two phases increases, consistent with the original information loss calculations. However, at the Page transition, the QES is instead located at a new minimum outside of the infalling shock. The generalized entropy at the Page transition then begins to asymptote towards the expected entropy of a black hole in equilibrium with the bath, completing a correct Page curve of the equilibration process. In the example shown in figure~\ref{fig:sertraline}, the temperature of the bath is less than that of the black hole so the entropy decreases in the late time phase, similarly to the evaporating black hole. Note that a bath with temperature greater than that of the black hole instead heats up the black hole, giving a Page curve as in figure~\ref{fig:Page_curve}.

The central quantity necessary for studying the Page curve and the behaviour of the extremal surface throughout the equilibration process is the generalized entanglement entropy $S_\gen$. Similar to previous work in the evaporating AdS$_2$ black hole in JT gravity~\cite{Almheiri:2019psf}, we break the process of calculating the generalized entropy into three steps:\footnote{Note that we have adapted the notation in eq.~\eqref{eq:Sgen} to our specific system, in which the area of the HRT surface is given by the value of the dilaton. Further, we specify that the quantum corrections $S_{\rm out}$ are given by the von-Neumann entropy $S_{\rm vN}$ of the CFT matter on either side of the bipartition.}
\begin{itemize}
	\item Calculating the von Neumman entropy of the CFT matter $S_\vn$
	\item Calculating the backreaction of the quench onto the dilaton $\phi$
	\item Extremizing the resulting $S_\gen = \frac{\phi}{4 \GN} + S_\vn$ 
\end{itemize}
Conveniently, these steps are very similar to the evaporating models in~\cite{Almheiri:2019psf,Almheiri:2019hni,Chen:2019uhq}, the only change coming from the details of the time reparametrization function in eq.~\eqref{eq:fmap} and the extra conformal transformation in eq.~\eqref{eq:icing} required to map the vacuum on upper half plane to our quenched system. We now proceed to carry out each one of these steps in the rest of this section.

\input{sections/background/bcft}
\input{sections/background/jt}
\input{sections/background/thermal}


%% file: sections/background/bcft.tex
\subsection{Entropy of holographic CFT$_2$}
To calculate the von Neumann entropy of the CFT matter, we proceed in a similar way to previous work on evaporating models~\cite{Almheiri:2019psf,Almheiri:2019hni,Chen:2019uhq} and map the corresponding quantum state to the vacuum of the CFT on the upper half plane by a local Weyl rescaling and a coordinate transformation. The details of the required coordinate transformation will be explained in section~\ref{sec:quenching}, but for now, we simply specify that we will be working in Poincar\'e coordinates for the AdS$_2$ spacetime
\beq
ds^2_\AdS = - \frac{4 L_\AdS^2}{(x^+-x^-)^2}dx^+dx^-\,, \quad (x^\pm = t\pm s)\,,
\label{eq:citalopram}
\eeq
and in flat coordinates for the bath
\beq
ds^2_{\rm bath} = -\frac{L_\AdS^2 dy^+ dy^-}{\epsilon^2} \,, \quad (y^\pm = u \mp \sigma)\,.
\label{eq:fluoxetine}
\eeq
The two spaces are glued together at one-dimensional boundary with $\sigma = -\epsilon$, $s = \epsilon f' $, $g_{uu}=\frac{L_\AdS^2}{\epsilon^2}$ where $\epsilon$ corresponds to the UV cutoff in the dual boundary theory, and $f$ is the coordinate reparametrization function $x = f(y)$, given below in eq.~\eqref{eq:fmap}.\footnote{In section \ref{sec:smoke}, we also introduce analogous coordinates $\ypure^\pm = \upure \pm \sigmapure$ for the purification of the bath. These are related to $x^\pm$ in eq.~\reef{wacked}, which is then the analog of eq.~\reef{wacked0}.} In the rest of the paper, we simply set $L_\AdS=1$.

The CFT matter state can then be mapped to the CFT vacuum via the local Weyl rescaling
\begin{equation}
\label{eq:Weyl}
\begin{split}
ds^2_\AdS &\to \Omega(x^+,x^-)^2ds^2_\AdS= dz d\bar{z}\,,\\
ds^2_{\rm bath} &\to \Omega'(y^+,y^-)^2ds^2_{\rm bath}= dz d\bar{z}\,,
\end{split}
\end{equation}
where
\begin{equation}
\Omega = \frac{x^+-x^-}{2}\sqrt{z'(x)\bar{z}'(\bar{x})}\,, \quad \Omega' = \epsilon \sqrt{z'(y)\bar{z}'(\bar{y})}\,,
\end{equation}
where we have introduced the Euclidean coordinates $x=-x^-$, $\bar{x} = x^+$ and similarly for $y$ and $\bar{y}$. The coordinate transformations relating the $x$, $y$ and $z$ coordinates in eqs.~\eqref{eq:icing}, \eqref{eq:icecream} and \eqref{eq:fmap} are all derived in section~\ref{sec:quenching}. In the rest of this subsection, we focus on deriving the von Neumann entropy of the CFT matter in the $z$ coordinates. 

To begin, one can consider the von Neumann entropy of a finite interval with one end-point being the boundary of the BCFT and the other $(z,\bar{z})$ residing in the interior. Equivalently, this is the entropy for the semi-infinite interval beginning at $(z,\bar{z})$ and extending to infinity. This can be calculated using twist operator one-point functions in the upper half plane, but by the method of images, the latter resembles a two-point function of a CFT on the entire plane. Correspondingly, the von Neumann entropy resembles that of an interval with length $-i(z-\bar{z})$:
\begin{align}
	S_{\twist{1}} = \frac{c}{6} \log[-i(z-\bar{z})]+ \log g
	\label{eq:beyondBurger}
\end{align}
where $\log g$ is the Affleck-Ludwig boundary entropy~\cite{Affleck_1994}.

The entanglement entropy of an interval in a two-dimensional  CFT in the presence of a conformal boundary at $z-\bar{z}=0$ is~\cite{Cardy:1984bb,Calabrese:2007rg,DiFrancesco:1997nk,Coser_2014} 
\beq\label{eq:crepe}
S_{\twist{2}} = \frac{c}{6} \log \(|z_1-z_2|^2 \eta\) + \log G(\eta)\,,
\eeq
where $\eta = \frac{(z_1-\bar{z}_1)(z_2-\bar{z}_2)}{(z_1-\bar{z}_2)(z_2-\bar{z}_1)}$ is the conformally invariant cross ratio and $G(\eta)$ is an undetermined function that depends on the theory and boundary conditions. The $G(\eta)$ function has two limits that can be determined by either a bulk OPE or an operator-boundary expansion: $G(\eta\to 1) = 1$ from the OPE limit, and $G(\eta\to 0) = g^2$ from the operator-boundary expansion. 

In the following, we adopt the holographic framework describing boundary conformal field theory (BCFT) \cite{Takayanagi:2011zk,Fujita:2011fp}. In this setup, the JT gravity plus bath system lives on the boundary of an $\text{AdS}_3$ geometry. From this bulk perspective, the boundary defect at the moment of quenching anchors an end-of-the-world (ETW) brane hanging into the holographic direction. After the quench, the ETW brane detaches from the asymptotic boundary falls off into the bulk. For this system, the entanglement entropy is determined using the Ryu-Takayanagi prescription \cite{Rangamani:2016dms}, \ie for a two-dimensional CFT on the asymptotic AdS boundary, the entanglement entropy is simply given by evaluating the bulk length of the corresponding geodesics connecting the end-points on the boundary, with the added possibility of having geodesics ending at the ETW brane. In the $z$ coordinates, this corresponds to evaluating the length of the geodesics connecting the end-points in a flat asymptotic boundary of $\text{AdS}_3$ with the possibility of having geodesics ending at a flat ETW brane intersecting the asymptotic boundary at $z-\bar{z}=0$ at an angle determined by the boundary entropy $\log g$. In this case, eq.~\reef{eq:crepe} reduces to the following simple form
\begin{equation}
\label{eq:2twists}
S_{\twist{2}} = \begin{cases}
\frac{c}{3} \log\(  |z_1-z_2|  \)&\text{if $\eta >\eta_*$}\\
\frac{c}{6} \log \(|z_1-\bar{z}_1||z_2-\bar{z}_2|\) +2\,\log g &\text{if $\eta<\eta_*$}
\end{cases},
\end{equation}
where $\eta_* = \frac{1}{1+g^{12/c}}$ is the value of the conformal cross ratio at which the transition between HRT surfaces occur. Let us note that with the simple choice $g=1$ (\ie $\log g=0$ and a tensionless ETW brane), the latter simplifies to $\eta_*=1/2$. Equivalently, the $G(\eta)$ function for a holographic BCFT is given by
\beq
G(\eta) = \theta(\eta -\eta_*)\, \eta^{-c/6} + \theta(\eta_*-\eta)\,\frac{g^2}{(1-\eta)^{c/6}}\,.
\label{eq:bagel}
\eeq
It is straightforward to verify that $G(\eta \to 1) = 1$ and $G(\eta\to 0) = g^2$. For simplicity, we take the case of zero boundary entropy $g=1$ (and $\eta^*=1/2$) in the following. As was noted in~\cite{Chen:2019uhq}, for a general $g$, the quench to scrambling phase transition gets shifted, while the Page transition remains unaffected.

The von Neumann entropies in eqs.~\eqref{eq:beyondBurger} and \eqref{eq:2twists} correspond to intervals of the vacuum of the BCFT. To find the von Neumann entropies of the CFT matter in our black hole thermalization model, we simply have to include the effect of the local Weyl transformation in eq.~\eqref{eq:Weyl}. Under a Weyl transformation $g_{\mu\nu}\to \Omega^{-2} g_{\mu\nu}$, the transformation of twist operators induces the following transformation on entropy:
\begin{align}
	S_{\Omega^{-2}g}
	=& S_g - \frac{c}{6}\sum_{\text{endpoints}} \log \Omega(\text{endpoint}).
	\label{eq:waffle}
\end{align} The above transformation may be interpreted as resulting from the rescaling of UV cutoffs with respect to which the entropy is defined.


%% file: sections/background/jt.tex
\subsection{Jackiw-Teitelboim gravity}\label{sec:escitalopram}

The brane perspective of the \aims\ model -- see figure \ref{trio}b -- describes the system as a black hole in two-dimensional JT theory coupled to holographic conformal matter which is connected to a bath with a joining quench, and allowed to evaporate. We refer the reader to~\cite{Almheiri:2019psf} for a more detailed discussion of this description. In this subsection, we summarize the essential parts of our analysis. 

The dynamics of the black hole and CFT matter are governed by the action
\begin{equation}\label{eq:action}
I = \frac{1}{16 \pi \GN} \[\int_{\cal M} d^2x  \sqrt{-g}\,\phi \(R+\frac{2}{L_\AdS^2}\) +2\int_{\partial {\cal M}} \phi_b K \]+ I_{\rm top}+ I_{\rm CFT}\,,
\end{equation}
where
\begin{equation}
I_{\rm top} = \frac{\phi_0}{16 \pi \GN} \[\int_{\cal M} d^2x  \sqrt{-g} \,R +2\int_{\partial {\cal M}} K \]
\end{equation}
is a topological term which provides a large constant contribution $S_0 = \frac{\phi_0}{4\GN}$ to the entropy of the black hole. The last term in eq.~\eqref{eq:action} is the action of the holographic CFT matter to which JT gravity is coupled.

The dilaton equation of motion imposes the geometry to be locally $\text{AdS}_2$ with radius $L_\AdS$, as described by the metric in eq.~\eqref{eq:citalopram}. The metric equations of motion give the coupling of the dilaton to the CFT stress tensor
\beqs
2\partial_{x^+} \partial_{x^-} \phi +  \frac{4 \phi}{(x^+-x^-)^2}  &= 16 \pi \GN \langle T_{x^+ x^-} \rangle\,,\\
-\frac{\partial_{x^+}\((x^+-x^-)^2\partial_{x^+}\phi\)}{(x^+-x^-)^2} & = 8 \pi \GN \langle T_{x^+ x^+} \rangle\,,\\
-\frac{\partial_{x^-}\((x^+-x^-)^2\partial_{x^-}\phi\)}{(x^+-x^-)^2} & = 8 \pi \GN \langle T_{x^- x^-} \rangle\,.
\eeqs
Before the quench, the CFT matter is in the vacuum of the generator of $t$ translations (see eq.~\eqref{eq:citalopram}) \ie $\langle T_{x^+x^+} \rangle = \langle T_{x^-x^-} \rangle = \langle T_{x^+x^-} \rangle = 0$,\footnote{In principle, we should have one non-zero component $\langle T_{x^+x^-} \rangle =\frac{c}{12\pi (x^+-x^-)^2}$ due to the trace anomaly. But this extra term can be absorbed by shifting the value of the dilaton field as $ \tilde{\phi}=\phi -\frac{c\GN}{3}$ -- see discussion in \cite{Chen:2020uac}. So we simply ignore the trace anomaly in the following.} however, this can also be seen as a TFD state for the generator of $u$ translations (see eq.~\eqref{eq:fluoxetine}). Here we have continued the $y$ coordinates into a Rindler patch of $\text{AdS}_2$ with
\beq\label{wacked0}
x^\pm = \frac{1}{\pi T_0}\, \tanh \(\pi T_0 y^\pm\)\,.
\eeq
The dilaton profile is given by
\beq
\label{eq:dilatonbefore}
\phi = 2\bar{\phi}_r \frac{1-\(\pi T_0\)^2 x^+ x^-}{x^+-x^-} = 2\bar{\phi}_r \pi T_0 \coth \(\pi T_0\(y^+-y^-\)\)\,.
\eeq
After the quench, the dilaton receives a contribution from the back-reaction of the matter stress tensor
\beq
\label{eq:dilatonafter}
\phi =  \bar{\phi}_r \,\frac{2-2\(\pi T_1\)^2 x^+ x^-+ k I_0}{x^+-x^-} \,,
\eeq
where
\begin{equation}
I_0 = -\frac{24\pi }{c} \int_0^{x^-} dt\, (x^+-t)(x^--t)\, \langle T_{x^- x^-} (t) \rangle\,,
\end{equation}
accounts for the matter back-reaction and $k = \frac{c \GN}{3\bar{\phi}_r}$ controls the strength of the back-reaction, which we take to be very small. The dilaton profile in eqs.~\eqref{eq:dilatonbefore} and \eqref{eq:dilatonafter} give the leading contribution to the generalized entanglement entropy. The details of the dilaton profile after the quench in eq.~\eqref{eq:dilatonafter} and the resulting generalized entropy are calculated in section~\ref{sec:quenching}.


%% file: sections/background/thermal.tex
\subsection{Coupling to a thermal bath}
\label{sec:quenching}

\begin{figure}[htbp]
	\centering\includegraphics[width=5.7in]{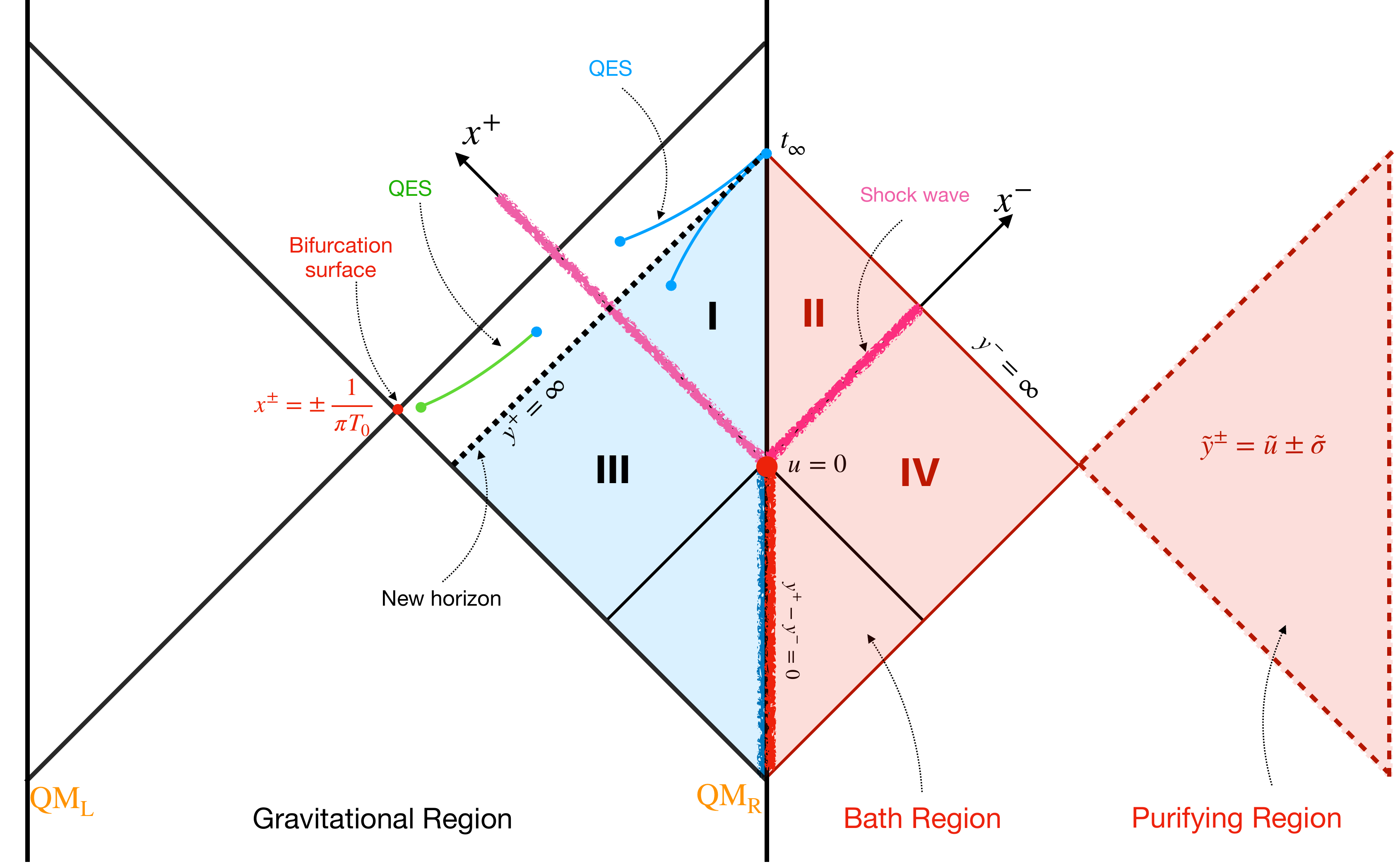}
	\caption{The Penrose diagram for the AdS$_2$ black hole  coupled with a thermal bath and its purification in flat spacetime at time $u=0$. The (thick) pink lines are the shock waves propagating into the gravitating and bath regions, which are generated by this joining quench. The bifurcation surface of the initial equilibrium black hole is indicated by the red dot. The new horizon is indicated by the black dashed line, $\ie y^+=\infty$. Note that only the blue and red shaded regions are covered by the $y^\pm, \tilde{y}^\pm$ coordinates, respectively. The evolution of quantum extremal surface in three phases is presented by the corresponding colored curves, as indicated in figure \ref{fig:sertraline}.}\label{fig:Penrose}
\end{figure}

The setup which we wish to consider is very similar to the one constructed in \cite{Almheiri:2019psf}: a two-sided $\AdS_2$ black hole prepared at some temperature $T_0$ coupled by a joining quench to a bath consisting of a CFT on a half-line. Again, the key difference will be that our bath will be at some finite temperature $T_\bath$, rather than zero temperature as in \cite{Almheiri:2019psf}. The corresponding Penrose diagram is shown in figure \ref{fig:Penrose}. Up until an initial time, we imagine two decoupled systems. Firstly, we have the $\AdS_2$ black hole solution\footnote{Note that an appropriate choice of coordinates, \eg{} those spanning the trajectory of the JT boundary particle, furnishes a pure $\AdS_2$ with Rindler horizons --- we are treating the $\AdS_2$ spacetime as a black hole in the usual sense for JT gravity \cite{Almheiri:2014cka,Engelsoy:2016xyb,Maldacena:2016upp}.} of JT gravity with the metric and dilaton profile in eqs.~\eqref{eq:citalopram} and \eqref{eq:dilatonbefore}, respectively. This gravitating region also supports the same two-dimensional CFT as appears in the bath region. The right side of this black hole will have a boundary given by an IR cutoff introduced by the JT boundary particle. Additionally, we have a separate bath system supporting an identical $\CFT_2$ (but in a different state), prepared on a half-line $\sigma = \frac{y^--y^+}{2}>-\epsilon$ on the flat spacetime \eqref{eq:fluoxetine}. The boundaries in the two systems initially impose reflecting boundary conditions. But, at some initial time $u=t=0$, we perform a joining quench. This is done by identifying $\sigma=-\epsilon$ in the bath with the $\AdS_2$ IR cutoff surface, allowing CFT matter to flow freely across the now transparent division between the $\AdS_2$ and bath systems. The details of this gluing are specified by the trajectory $t=f(u)$ -- that is, we identify the time parameter of the $\AdS_2$ boundary with the time coordinate of the bath. Further demanding that the induced metrics along the $\AdS_2$ and bath boundaries match to leading order in $\epsilon$, we have
\begin{align}
 x^\pm = f(y^\pm)
\end{align}
along the gluing. For convenience, we shall further extend the above equality to hold everywhere, so that we may alternatively describe patches of $\AdS_2$ and the bath using either $x^\pm$ or $y^\pm$ coordinates. Later in this section, we will determine the trajectory $t=f(u)$ of the JT boundary particle by tracking the exchange of energy between the $\AdS_2$ and bath systems.

While we have described the physical evolution of the system above, it is practically useful also to consider a Euclidean preparation of the CFT state at $u=t=0$. Thus, we imagine preparing the CFT in a Hartle-Hawking state on the JT black hole with a path integral over Euclidean $\AdS_2$ (with an appropriate dilaton profile). Similarly, we prepare the CFT in the bath (and the purifying copy) in a thermofield state with a path integral on Euclidean half-spaces. (The details will be elaborated below.) Both systems have reflecting boundary conditions, except in an infinitesimal neighborhood of $iu=it=0$, where the two spacetimes are joined. The size of this neighborhood provides a regulator for the shock energy $E_S$ produced by the joining quench --- recall that stripping the vacuum entanglement along an entangling surface (in this case, the point at the $\AdS$-bath boundary) produces an infinite amount of energy. This construction produces the CFT state at $u=t=0$, from which analytic continuation provides the correct Lorentzian evolution according to the joined Hamiltonian. We note that this joined evolution, obtained by analytic continuation, does not match the physical decoupled evolution of the $\AdS$ and bath systems to the past of the point of the joining quench. In particular, we expect the time-reversal symmetry of the Euclidean path integral to carry over to Lorentzian time upon analytic continuation; in contrast, the physical Lorentzian evolution is manifestly not time-reversal symmetric due to the change in boundary conditions at the quench. However, results obtained by analytic continuation will be adequate for our purposes as we are primarily interested in the Lorentzian physics beyond the past light-cone of the quench point where the $\AdS_2$ and bath boundaries are joined.

Our point of departure from \cite{Almheiri:2019psf,Chen:2019uhq} lies with the generalization to thermal baths prepared at finite temperature. To put the bath at a finite temperature $T_\bath$, we take the Euclidean $y$ coordinates for the bath and\footnote{Note that this identification makes $T_\bath$ the temperature associated with the unit time-like vector in the geometry $dy d\bar{y}$, as opposed to the physical geometry $\frac{L_\AdS^2 dy d\bar{y}}{\epsilon^2}$. In the doubly holographic language of Figure \ref{trio}c, the former is the CFT metric of the asymptotic boundary of $\AdS_3$ while the latter is the induced metric on a cutoff surface which becomes the asymptotic boundary in the $\epsilon\to 0$ limit. Similarly, in \eqref{wacked0}, $T_0$ describes a temperature with respect to the parametric time $u = \frac{y^++y^-}{2}$ of the boundary particle, which does not correspond to a unit vector in the $\AdS_2$ geometry \eqref{eq:citalopram}.} identify $y \sim y + \frac{i}{T_\bath}$. We still take the bath to be the half-space $\frac{y+\bar{y}}{2} \le \epsilon$. As expected for a thermal state, this results in a non-zero stress tensor expectation value in $y$ coordinates. Although the $x$ coordinates of AdS$_2$ are stitched to the $y$ coordinates of the bath (\ie $x=f(y)$), it will nonetheless be convenient in the following to introduce a conformal transformation after which the stress tensor becomes trivial. This can be achieved by transforming the thermal half-cylinder, with coordinates $y$, to the left half-plane,\footnote{
  Strictly speaking, we should take the bath to be the half-space $\frac{y+\bar{y}}{2}\le \epsilon$ and \eqref{eq:icing} would map this region to the plane minus a large disk in the right half-plane. Similarly, Euclidean preparation of the $\AdS_2$ system, in the $x,\bar{x}$ coordinates analytically continued from \eqref{wacked0}, does not occur on a full Euclidean Poincar\'e $\AdS_2$, but rather on a large disk-like subregion. Note that the stress tensor still vanishes in these subregions of the $dY d\bar{Y}$ and $\frac{2 dx d\bar{x}}{x+\bar{x}}$ geometries, since a flat disk (or its complement) is related to a flat half-plane by a Mobius transformation. (The rescaling of $dx d\bar{x}$ by the Poincar\'e Weyl factor $\frac{2}{x+\bar{x}}$ does not introduce an extra anomalous contribution to the stress tensor.)
}
 with coordinates $Y$, via
\begin{equation}
	Y = \frac{1}{\pi T_\bath} \tanh(\pi T_\bath y).
	\label{eq:icing}
\end{equation}
This is simply the composition of an exponential map $y'=e^{2\pi T_\bath y}$ taking the thermal half-cylinder to a unit disk, and a Mobius map $Y=\frac{1}{\pi T_\bath} \frac{y'-1}{y'+1}$ pushing a point on the boundary of the disk to $\infty$.

It will be useful, \eg to make use of the entropy formula \eqref{eq:waffle}, to write down another map which maps the joint system of AdS, with Poincare coordinates $x$, and the bath, with Euclidean coordinates $y$ or equivalently the coordinates $Y$ found above, to the upper half plane, with coordinates $z, \bar{z}$. Just prior to coupling the AdS and bath systems, the AdS system is in the Hartle-Hawking state with vanishing stress tensor in $x$ coordinates. Meanwhile, by construction, the stress tensor of the bath vanishes upon conformal transformation to $Y$ coordinates. Finally, the stress tensor in the half-plane with coordinates $z$ must also vanish. By demanding that the conformal anomalies of the map from $x$ and $Y$ to $z$ vanish respectively in AdS and the bath, together with boundary conditions, fixes this map. Following \cite{Almheiri:2019psf}, we choose boundary conditions such that the AdS$_2$ space is mapped to the region $(0, i z_0)$ and the bath to $(i z_0, i \infty)$. The map is piecewise-Mobius: 
\begin{equation}
	z = \begin{cases}
		\frac{-i z_0^2}{x-i z_0} & x > 0\,,\\
		z_0 -i Y & x < 0\,.
              \end{cases}
              \label{eq:fluvoxamine}
\end{equation}
The discontinuity at $z=z_0$ produces the shock wave  contributions to the stress tensor components $\langle T_{xx}\rangle= E_S\,\delta(x)$ and $\langle T_{\xb\xb}\rangle= E_S\,\delta(\xb)$, with
\begin{align}
  E_S \simeq& \frac{c}{12\pi (-iz_0)}.
\end{align}
In the limit $E_S \rightarrow \infty$ (\ie $-iz_0 \rightarrow 0$), the map \eqref{eq:fluvoxamine} becomes 
\begin{equation}
	z = \begin{cases}
		(\frac{12\pi}{c} E_S)^{-2} \frac{i}{x} & x > 0\,,\\
	-i Y & x < 0\,.
	\end{cases}
	\label{eq:icecream}
\end{equation}

The next step is to determine $f$ by demanding the conservation of energy between the AdS and bath systems \cite{Engelsoy:2016xyb,Almheiri:2019psf}:
\begin{align}
	\partial_u E(u)
	= f'(u)^2 (T_{x^- x^-} - T_{x^+ x^+}).
	\label{eq:pancake}
\end{align}
From the conformal anomaly associated with the Weyl transformation \eqref{eq:icecream}, \ie 
\begin{equation}
 \langle T_{xx} \rangle  =   \( \frac{d z}{ d x} \)^2  \langle T_{zz}  \rangle-\frac{c}{24 \pi} \{ z, x\}  \,,
 \end{equation}
 we can find that the stress tensor in AdS region satisfies\footnote{This result does not apply in the causal past of the junction point. Further, note that the Schwarzian is defined by $\{f(y),y\} \equiv \frac{f'''}{f'}-\frac{3}{2}\left(\frac{f''}{f'}\right)^2$.}
\begin{equation}\label{eq:stresstensor}
\begin{split}
	\langle T_{x^\pm x^\pm}(x^\pm)\rangle_\AdS
=& E_S \,\delta(x^\pm) - \frac{c}{24\pi}\, \{Y^\pm,x^\pm\}\,\Theta\!\(\mp x^\pm\) \\
=& E_S\, \delta(x^\pm) - \frac{c}{24\pi}\, \Theta\!\(\mp x^\pm\)\left[
\{y^\pm,x^\pm\} - 2\left(\frac{\pi T_\bath}{f'(y^\pm)}\right)^2
\right] \,,
\end{split}
\end{equation}
where we have used the Schwarzian composition rule
\begin{align}
	\{Y,x\}
	=& \{y,x\} + \left(\frac{dy}{dx}\right)^2\{Y,y\}.
\end{align}
For completeness, from eq.~\eqref{eq:icecream}, we also write the stress tensor in the bath region:
\begin{align}
  \langle T_{y^\pm y^\pm}(y^\pm)\rangle_{\rm bath}
  =& E_S\, \delta(y^\pm) - \frac{c}{24\pi}\, \left[
     \Theta\!\(\pm y^\pm\) \{x^\pm,y^\pm\}
     -\Theta\!\(\mp y^\pm\)\, 2(\pi T_\bath)^2 
     \right]
     \label{eq:vilazodone}
\end{align}
As mentioned below eq.~\eqref{eq:fluvoxamine}, the $\delta$-function contributions in eqs.~\eqref{eq:stresstensor} and \eqref{eq:vilazodone} may be interpreted as the positive-energy shockwaves produced by the quench. The Schwarzian terms have a similar simple interpretation: $T_{x^-x^-}\sim -\frac{c}{24\pi}\,\{y^-,x^-\}<0$ describes a negative energy flux from the bath experienced by the black hole, while $T_{y^+y^+}\sim -\frac{c}{24\pi}\,\{x^+,y^+\}>0$ describes a positive energy flux from the black hole experienced by the bath. Considering for simplicity the $T_\bath=0$ case, note that the quanta described by these fluxes are the result of vacuum fluctuations in their native geometries. In particular, on the initial time slice, these quanta register as vanishing stress-energy, which is to be expected in the Hartle-Hawking vacuum of $\AdS_2$ and the flat half-space vacuum. It is only when these quanta cross over the $\AdS_2$-bath interface that they register as non-vanishing stress energy. Finally, in the case of nonvanishing bath temperature $T_\bath>0$, the last terms in eqs.~\eqref{eq:stresstensor} and \eqref{eq:vilazodone} can be interpreted as the contribution to the stress-energy of the bath's thermal radiation.

To determine the $f$ function, we next note that the ADM energy of the AdS$_2$ JT system
\begin{align}
\label{eq:energy}
	E(u)
	=& - \frac{\bar{\phi}_r}{8\pi \GN}\{f(u),u\}.
\end{align}
 can also be expressed in terms of the Schwarzian of $f$, we have, from solving eq.~\eqref{eq:pancake}, the Schwarzian equation
\begin{equation}
	\left\{f(u), u\right\} = - 2\pi^2 \left[T_\bath^2 + (T_1^2 - T_\bath^2)e^{- k u}\right]\,, \quad \text{with} \quad k \equiv \frac{c\GN}{3\bar{\phi}_r} \ll 1\,.
	\label{eq:chocolate}
\end{equation}
From initial conditions $f(0)=0, f'(0)=1, f''(0)=0$, we can solve this differential equation to obtain the map between $y$ and $x$:\footnote{We note that the same differential equation appears in the analysis of \cite{Hollowood:2020cou}, although differences arise since their work involves different boundary conditions for $f(u)$.}
\begin{equation}
\begin{aligned}
\label{eq:fmap}
f(u, T_{b}) 
	&= \frac{2}{ka}\,
	\frac{I_\nu(a)\, K_\nu(a e^{-ku/2}) - K_\nu(a)\,I_\nu(a e^{-ku/2})}{I_{\nu}'(a)\, K_\nu(a e^{-ku/2}) - K_{\nu}'(a)\,  I_\nu(a e^{-ku/2})}
	\end{aligned}
\end{equation}where 
\begin{equation}
	a = \frac{2 \pi}{k} \sqrt{T_1^2 - T_\bath^2}\qquad{\rm and} \qquad \nu = \frac{2\pi T_\bath}{k}\,.
\end{equation}
The above function is also well-defined and always real for complex $a$, \ie $T_1 < T_{\bath}$. 

Given the map \eqref{eq:icecream} and the function $f$, we may compute the von Neumann entropy of various intervals in the AdS-bath system by applying the transformation rule \eqref{eq:waffle} to the formulas \eqref{eq:beyondBurger} or \eqref{eq:crepe} for entropy of intervals in a half-plane. 


First, we divide the spacetime of interest into four regions according to 
\begin{equation}
\begin{split}
x^\pm\in
\begin{cases}
\rm{\Rone}
&\text{\,post-shock in AdS\,,} \,\quad  x^+ \ge  x^-\ge 0\,,
\\
\rm{\Rtwo}
&\text{post-shock in bath\,,} \,\quad  x^- \ge  x^+\ge 0\,,
\\
\rm{\Rthree} &\text{pre-shock in AdS\,,}\qquad  x^+ \ge 0 \ge  x^-\,,
\\
\rm{\Rfour}
& \text{pre-shock in bath\,,}\qquad  x^- \ge 0 \ge x^+\,.
\end{cases}
\end{split}
\end{equation}
Applying the entropy transformation rule \eqref{eq:waffle} to eq.~\eqref{eq:beyondBurger}  with these Weyl factors and the form \eqref{eq:icecream} of the map, we obtain the following formulas for the von Neumann entropy computed with a single twist operator at $x^\pm$:
\begin{equation}\label{eq:absolut}
	S_{\twist{1}}(x^\pm) =
\log g
+\frac{c}{6}\log
\begin{cases}
\frac{24 E_S}{c T_\bath}
\frac{
	x^+ \sinh(\pi T_\bath y^-) \sqrt{f'(y^-)}
}{
	x^+ - x^-
}\,,
& x^\pm \in \rm{\Rone}\,,
\\
\frac{12 E_S}{c \epsilon T_\bath}
\frac{x^+ \sinh(\pi T_\bath y^-)}{\sqrt{f'(y^+)}}\,,
&x^\pm \in \rm{\Rtwo}\,,
\\
2\,, &x^\pm \in \rm{\Rthree}\,,
\\
\frac{
	\sinh[\pi T_\bath(y^- - y^+)]
}{
	\pi \epsilon T_\bath
}\,,& x^\pm \in \rm{\Rfour}\,.
\end{cases}
\end{equation}
Note that in the pre-shock cases, we recover the expected entropy formulas in AdS and a thermal half-line. In particular, if we take $\frac{y^- - y^+}{2} \to \sigma_\IR$ for some IR cutoff $\sigma_\IR$, we get the entropy of the whole thermal half-line:
\begin{align}
	S_\halfLine
	=& \log g
	+ \frac{c}{6}\left[
	2\pi T_\bath \sigma_\IR
	+ \log\left(\frac{1}{2\pi \epsilon T_\bath}\right)
	\right].
	\label{eq:rum}
\end{align}
Of course, these three terms are interpreted as: the boundary entropy, the thermal entropy of the CFT at temperature $T_\bath$, and the log divergent contribution associated with the endpoint of the interval.

We obtain entropy formulas derived from two-point function by transforming eq.~\eqref{eq:crepe}:
\begin{equation}\label{eq:scone}
	\begin{split}
S_{\twist{2}}\(x_1^\pm \in {\rm{\Rtwo}}, x_2^\pm\) 
		= \log G(\eta) 
		+\frac{c}{6}\log\begin{cases}
			\frac{2}{\pi \epsilon T_\bath}
		\frac{\sinh[\pi T_\bath(y_2^- - y_1^-)] (x_1^+ - x_2^+)}{x_2^+ - x_2^-}
		\sqrt{\frac{f'(y_2^-)}{f'(y_1^+)}}\,,
		&x^\pm_2 \in \rm{\Rone}\,,
		\\
		\frac{1}{\pi \epsilon^2 T_\bath}
		\frac{
		-\sinh[\pi T_\bath(y_1^- - y_2^-)](x_1^+ - x_2^+)
		}{
		\sqrt{f'(y_1^+)f'(y_2^+)}
		}\,,
		& x^\pm_2 \in \rm{\Rtwo}\,,
		\\
		\frac{24 E_S}{c\epsilon T_\bath}
		\frac{\sinh(\pi T_\bath y_1^-) x_1^+ x_2^- (x_1^+-x_2^+)}{x_2^+(x_1^+-x_2^-)\sqrt{f'(y_1^+)}}\,,
		& x^\pm_2 \in \rm{\Rthree}\,,
		\\
			\!\begin{aligned}
	&\textstyle\frac{12 \pi E_S}{c \epsilon^2}
	\frac{
		-x_1^+ Y_2^+ (Y_2^- - Y_1^-)
		\eta
	}{\sqrt{f'(y_1^+)}}
	\\
	&\times
	\textstyle{
		\cosh(\pi T_\bath y_1^-)
		\cosh(\pi T_\bath y_2^+)
		\cosh(\pi T_\bath y_2^-)
	}\,,
	\end{aligned}
	&x^\pm_2 \in \rm{\Rfour}\,,
	\end{cases}
  \end{split}
\end{equation}
where the cross-ratio is determined by 
\begin{align}
	\eta(x^\pm_1 \in {\rm{\Rtwo}}, x_2^\pm)
	=& \begin{cases}
	\frac{Y_1^-(Y_2^- - Y_2^+)}{Y_2^-(Y_1^- - Y_2^+)}\,,
	&x^\pm_2 \in \rm{\Rfour}\,,
	\\
	\frac{x_1^+(x_2^+-x_2^-)}{x_2^+(x_1^+ - x_2^-)}\,,
	&x^\pm_2 \in \rm{\Rthree}\,,
	\\
	1 \,,&x^\pm_2 \in \rm{\Rone},\rm{\Rtwo}\,.
	\end{cases}
	\\
\end{align}
Note that this agrees with eq.~(3.30) of \cite{Almheiri:2019psf} in the limit when $T_\bath \to 0$ and the $x_1^\pm$ endpoint is taken to the AdS-bath boundary. With the holographic formula \eqref{eq:bagel} for $G$, the pre-shock cases of \eqref{eq:scone} with $x_1^\pm \in {\rm{\Rtwo}}$ become\footnote{The $x_2$ dependence of the bulk entropy is identical to that found for a bath with vanishing temperature, \eg see eqs.~(3.2) and (3.3) in \cite{Chen:2019uhq}. This immediately implies that the position of the quantum extremal surfaces in the quench and scrambling phases are the same  as for the $T_\bath =0$ case.}
\begin{equation}
	S_{\twist{2}}=\begin{cases}
	S_{\twist{1}}(x_1^\pm) + S_{\twist{1}}(x_2^\pm)\,,
	&\text{if $\eta\le \eta^*$}
	\\
	\frac{c}{6} \log\left(
	\frac{24 E_S}{c\epsilon T_\bath}
	\frac{(x_2^+-x_1^+)x_2^-\sinh(\pi T_\bath y_1^-)}{(x^+_2-x_2^-)\sqrt{f'(y_1^+)}}
  \right)\,,
	& \text{if $\eta > \eta^*$, $x^\pm_2 \in {\rm{\Rthree}}$}\,,
		\\
	\frac{c}{6} \log\left(
	\frac{12 E_S}{c \pi \epsilon^2 T_\bath^2}
	\frac{
		x_1^+
		\sinh(\pi T_\bath y_2^+)
		\sinh[\pi T_\bath (y_1^- - y_2^-)]
	}{
		\sqrt{f'(y_1^+)}
	}
	\right)\,,
	&\text{if $\eta > \eta^*$, $ x^\pm_2 \in {\rm{\Rfour}}$}\,.
	\end{cases}
\label{eq:vanilla}
\end{equation}


%% file: sections/equilibrium.tex

From eq.~\eqref{eq:chocolate}, we see that the main effect of a finite temperature $T_\bath>0$ for the bath is that the black hole does not evaporate completely, but rather equilibriates with the bath. That is, it tends towards a stationary black hole with temperature $T_\bath$, for which
\begin{align}
	\{f(y),y\}
	=& -2\pi^2 T_\bath^2\,.
\end{align}
Indeed, when $T_1=T_\bath$, the black hole does not change at all and instead, after consuming the shock, the black hole remains a stationary black hole at temperature $T_1=T_\bath$. In this case, $f$ takes same form as for the eternal black hole solution
\begin{align}
	f(y)
	= \frac{1}{\pi T_1}\, \tanh(\pi T_1 y)\,.
\end{align}
Note that, from eq.~\eqref{eq:icing}, we then have $x=Y$ which agrees with the intuition that the radiation emitted by the bath mimics the radiation that would have been reflected from the AdS boundary in the Hartle-Hawking state had the bath not been attached. 

The entropy formulas \eqref{eq:absolut}, \eqref{eq:scone}, and \eqref{eq:vanilla} also become simple. More explicitly, the one-point function \reef{eq:absolut} reduces to 
\begin{equation}	\label{eq:jackDaniels}
	S_{\twist{1}}(x^\pm) =
	\log g
	+ \frac{c}{6} \log \begin{cases}
		\frac{24 \pi E_S}{c}
	\frac{x^+ x^-}{x^+ - x^-}\,,
	&x^\pm \in \rm{\Rone}\,,
	\\
	\frac{12 \pi E_S}{c \epsilon}
	\frac{x^+ x^-}{\sqrt{[1-(\pi T_1 x^+)^2][1-(\pi T_1 x^-)^2]}}\,,
	&x^\pm \in \rm{\Rtwo}\,,
	\\
	2\,, &x^\pm \in \rm{\Rthree}\,,
	\\
		\frac{x^- - x^+}{
		\epsilon
		\sqrt{
			[1-(\pi T_1 x^+)^2]
			[1-(\pi T_1 x^-)^2]
		}
	},
	&x^\pm \in \rm{\Rfour}\,.
\end{cases}
\end{equation}
According to the position of endpoints $x_1,x_2$, the entanglement entropy based on two-point function reads 
\begin{equation}\label{eq:equil2}
\begin{split}
	S_{\twist{2}}\(x_1,x_2\)=
	\begin{cases}
	S_{\twist{1}}(x_1^\pm) + S_{\twist{1}}(x_2^\pm)\,,
	& \text{if $\eta\le \eta^*$}
	\\
	\frac{c}{6}\log\left\{
	\frac{24 \pi E_S}{c\epsilon}
	\frac{x_1^- x_2^- (x_1^+ - x_2^+)}{
	(x_2^+ - x_2^-)
	\sqrt{[1-(\pi T_1 x_1^+)^2][1-(\pi T_1 x_1^-)^2]}
	}
	\right\}\,,
	&\text{if $\eta>\eta^*$, $x_2^\pm\in \rm{\Rthree}$},
	\\
		\frac{c}{6}\log\left\{
	\frac{12\pi E_S}{c \epsilon^2}
	\frac{x_1^+ x_2^+ (x_1^- - x_2^-)}{
		\sqrt{
			[1-(\pi T_1 x_1^+)^2]
			[1-(\pi T_1 x_1^-)^2]
			[1-(\pi T_1 x_2^+)^2]
			[1-(\pi T_1 x_2^-)^2]
		}
	}
	\right\}\,,
	&\text{if $\eta>\eta^*$, $x_2^\pm \in\rm{\Rfour}$},
\end{cases}
\end{split}
\end{equation}
for $x_1^\pm \in {\rm{\Rone}}$ and also 
\begin{equation}
S_{\twist{2}}
=\log G(\eta) 
+\frac{c}{6}\log
\begin{cases}
\frac{2}{\epsilon}
\frac{(x_1^+ - x_2^+)(x_2^- - x_1^-)}
{(x_2^+ - x_2^-)\sqrt{[1-(\pi T_1 x_1^+)^2][1-(\pi T_1 x_1^-)^2]}}\,,
&x^\pm_2 \in \rm{\Rone}\,,
\\
\frac{1}{\epsilon^2}
\frac{
	-(x_1^+ - x_2^+)(x_1^- - x_2^-)
}{
	\sqrt{
		[1-(\pi T_1 x_1^+)^2]
		[1-(\pi T_1 x_1^-)^2]
		[1-(\pi T_1 x_2^+)^2]
		[1-(\pi T_1 x_2^-)^2]
	}
}\,,
& x^\pm_2 \in \rm{\Rtwo}\,,
\\
\frac{24 \pi E_S}{c\epsilon}
\frac{x_1^+ x_1^- x_2^- (x_1^+-x_2^+)}
{x_2^+(x_1^+-x_2^-)\sqrt{[1-(\pi T_1 x_1^+)^2][1-(\pi T_1 x_1^-)^2]}}\,,
& x^\pm_2 \in \rm{\Rthree}\,,
\\
\frac{12 \pi E_S}{c \epsilon^2}
\frac{
	-x_1^+ x_2^+ (x_2^- - x_1^-) \eta
}{
	\sqrt{
		[1-(\pi T_1 x_1^+)^2]
		[1-(\pi T_1 x_1^-)^2]
		[1-(\pi T_1 x_2^+)^2]
		[1-(\pi T_1 x_2^-)^2]
	}
}\,,
&x^\pm_2 \in \rm{\Rfour}\,,
\end{cases}
\label{eq:equil1}
\end{equation}
when $x_1 \in \rm{\Rtwo}$. As noted below eq.~\eqref{eq:absolut}, the before-shock single-twist entropy formulas are the standard ones in AdS and the thermal half-line, which are invariant under translations in time $u=\frac{y^+ + y^-}{2}$. For the thermal case at hand, the two-twist formulas, with both twists inserted to the future of the shock, are also time-translation invariant. This can be made manifest by writing those cases of \eqref{eq:equil1} in $y^\pm$ coordinates:
  \begin{equation}\label{eq:smoke}
	S_{\twist{2}}\(x_1^\pm \in {\rm{\Rtwo}}\)
	=
	\frac{c}{6}\log\begin{cases}
	\frac{
	2
	\sinh[\pi T_1(y_2^+ - y_1^+)]
	\sinh[\pi T_1(y_1^- - y_2^-)]
	}{
	\pi T_1 \epsilon
	\sinh[\pi T_1(y_2^+ - y_2^-)]
	}\,,
	&
	x_2^\pm \in \rm{\Rone}\,,
	\\
		\frac{
		\sinh[\pi T_1(y_2^+ - y_1^+)]
		\sinh[\pi T_1(y_1^- - y_2^-)]
	}{
		(\pi T_1\epsilon)^2
	}\,,
	&
	x_2^\pm \in \rm{\Rtwo}\,.
\end{cases}
\end{equation}
Moreover, the above is also invariant under `time-reversal' $u_1-u_2 \leftrightarrow -(u_1-u_2)$. These properties will be helpful in finding the late-time QES. Indeed, eq.~\eqref{eq:smoke} is the same entropy formula as for an eternally-coupled black hole and bath system, as studied in \cite{Almheiri:2019yqk}. For simplicity, we shall take $g=1$ and $\eta^*=1/2$ in the following sections.  

\begin{figure}
  \centering
  \includegraphics[width=0.9\textwidth]{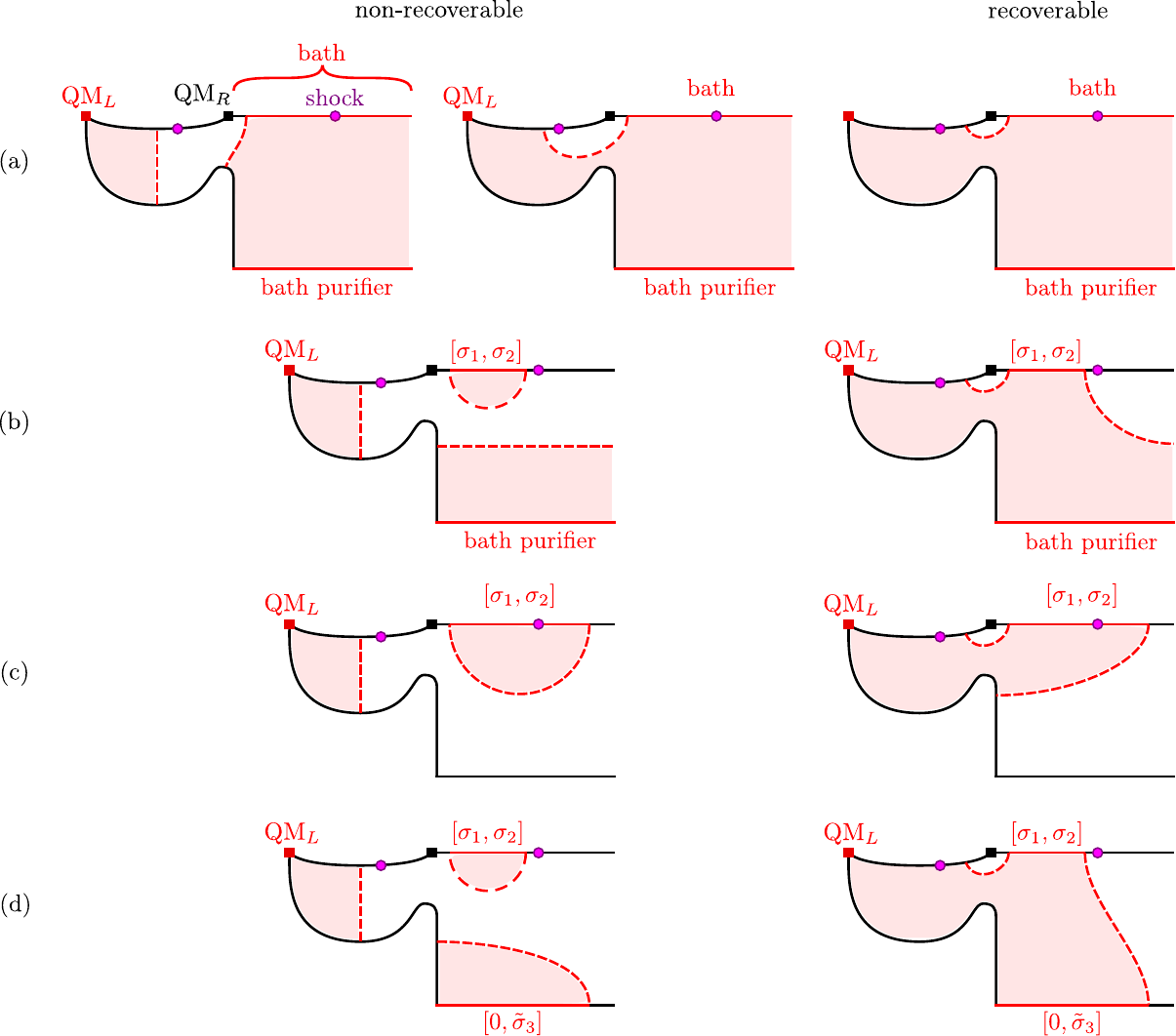}
  \caption{Competing channels computing the generalized entropy of various subsystems (solid red) and the corresponding bulk RT surfaces (dashed red) and entanglement wedges (light red). In each case, the R-channel where the black hole interior is recoverable or reconstructible is shown on the left. On the right, we show the N-channel where the interior is non-recoverable or non-reconstructible. The corresponding generalized entropies for these channels are denoted $S_{\mt{R}}$ and $S_{\mt{N}}$, respectively. In the top row (a), we consider the generalized entropy of $\QM_{\mt{L}}$, the thermal bath, and the bath's purifier. In row (b), we keep only a finite interval $[\sigma_1,\sigma_2]$ of the bath. In row (c), we further trace out the purifier. Finally, in row (d), we include a finite interval $[0,\sigmapure_3]$ of the purifier. Note that in this last case, we can also vary $\upure_3$, the time slice of the purifier interval, and we find the minimal $\sigmapure_3$ depends on $\upure_3$ --- see section \ref{sec:smoke}.}
  \label{fig:desvenlafaxine}
\end{figure}

In the following sections, we apply the RT formula to the calculation of entropy for various subregions in the full system consisting of $\QM_{\mt{L}}$, $\QM_{\mt{R}}$, the thermal bath, and an auxiliary system purifying the bath. We begin in Section \ref{sec:depression} by considering the entropy of $\QM_{\mt{L}}$, the bath system, and the purifier, recovering the Page curve, discussed previously in Section \ref{sec:background} and illustrated in figure \ref{fig:sertraline} --- the corresponding bulk RT surfaces are also shown in figure \ref{fig:desvenlafaxine}a. Next, in Section \ref{sec:lonely}, we trace out the majority of the bath, as shown in figure \ref{fig:desvenlafaxine}b, finding that only a finite bath interval of some minimal length is required to recover the black hole interior. Finally, in Section \ref{sec:smoke}, we evaluate the importance of the bath's purifier. In particular, we find that if the purifier is completely traced out, as shown figure \ref{fig:desvenlafaxine}c, the black hole interior can no longer be recovered, regardless of the size of the bath interval that one can access; at the very least, a finite interval of the purifier is required, as shown in figures \ref{fig:desvenlafaxine}d and \ref{fig:duloxetine}.

\subsection{Semi-infinite interval of the bath}\label{sec:depression}
First, we consider the evolution of (generalized) entropy for the subsystem consisting of \QML, a semi-infinite interval of the bath with endpoint $x_1^\pm$ after the shock, and the purifier of the bath. The corresponding HRT surfaces and the time evolution are illustrated in figure \ref{fig:sertraline}, and we shall find three phases for the generalized entropy, corresponding to different portions of a Page curve. Note that, tracing out the purifier would also produce an infinite thermal entropy for the semi-infinite bath interval, \ie the infinite entanglement entropy between the semi-infinite interval and the purifier. 

Initially, in the quench phase, the QES in the gravitating region simply sits at the bifurcation surface of the original eternal black hole geometry,
\begin{align}
x_\QES^\pm = \pm \frac{1}{\pi T_0}\,.
\label{eq:bacardi}
\end{align}
The corresponding generalized entropy is obtained by sum the Bekenstein-Hawking entropy
\begin{align}
	S_\Bek(T_0)
	=& \frac{c(\phi_0 + 2\pi T_0 \phi_r)}{12 k\phi_r}
     \label{eq:bekenstein}
\end{align}
and the von Neumann entropy \eqref{eq:equil2} evaluated holographically with endpoints $x_1^\pm$ and $x_\QES^\pm$, which picks out the $\eta< \eta^*=1/2$ channel:
\begin{align}
	S_\gen
	=& S_\Bek(T_0) + S_{\twist{1}}(\AdS) + S_{\twist{1}}(x_1^\pm) \,,  	\label{eq:rockBottom}
\end{align}
with $S_{\twist{1}}(\AdS)$ and $S_{\twist{1}}(x_1^\pm)$ given by eq.~\eqref{eq:absolut}. Note that, in the $\eta<\eta^*$ channel, the von Neumann entropy \eqref{eq:equil2} (and more generally eq.~\eqref{eq:vanilla}) has no dependence on $x_2^\pm$; this justifies a posteriori choosing the QES to simply be the classical one at the bifurcation point. This was also the case in the zero-temperature bath case \cite{Chen:2019uhq}.

Transitioning to the scrambling phase, the QES jumps from the bifurcation point to another saddle of generalized entropy, which is still located before the shock but now with $\eta > \eta^*=1/2$. Since the $x_2^\pm$ dependence of eq.~\eqref{eq:equil2} (and more generally eq.~\eqref{eq:vanilla}) in this channel is also identical to the zero temperature case \cite{Chen:2019uhq}, we obtain the same QES:
\begin{align}
	x_\QES^+
	=& \frac{1}{\pi T_0}\left[
	1-\frac{k}{\pi T_0}+ O\left(\frac{k^2}{T_0^2}\right)
	\right]
	 \\
	x_\QES^-
	=& \frac{1}{\pi T_0}\left[
	-1
	+ \frac{k}{\pi T_0}
	\frac{1+\pi T_0 x_1^+}{1-\pi T_0 x_1^+}
	+ O\left(\frac{k^2}{T_0^2}\right)
	\right].
\end{align}
Evaluating eq.~\eqref{eq:equil2} for this QES, one finds
\begin{align}
	S_\gen-S_\Bek(T_0)
	\sim \frac{c}{6}\log\left\{
	\frac{12 E_S}{c\epsilon T_0}
	\frac{x_1^-(1-\pi T_0 x_1^+)}
	{\sqrt{[1-(\pi T_1 x_1^+)^2][1-(\pi T_1 x_1^-)^2]}}
	\right\}.
	\label{eq:zirkova}
\end{align}
Comparing eq.~\eqref{eq:zirkova} with the $\eta<\eta^*$ channel of eq.~\eqref{eq:equil2}, we find that the quench-to-scrambling transition occurs at the same point as in the zero-temperature bath case:
\begin{align}
	x_1^+
	\sim& \frac{1}{3\pi T_0}\,.
\end{align}
Note that this is essentially the instant at which the bifurcation point \eqref{eq:bacardi} reaches $\eta=\eta^*=1/2$. At later times, eq.~\eqref{eq:zirkova} exhibits a growth linear in the physical time $u=\frac{y_1^+ + y_1^-}{2}$:
\begin{align}
	S_\gen-S_\Bek(T_0)
	\sim \frac{c}{6}\left\{
	\log\left[
	\frac{3 E_S (T_1-T_0)}{c \epsilon \pi T_0 T_1^2}
	\right]
	+ 2\pi T_1 u
	\right\}
	\label{eq:pizza}
\end{align}
with
\beq
	x_1^\pm
	= \frac{1}{\pi T_1}\left[1+O\left(\frac{k}{T_1}\right)\right]\,.
\label{pizza2}	
\eeq
We note that this growth has a rate double that of the zero-temperature bath case. Physically, this can be explained by the fact that, in addition to absorbing radiation from the AdS black hole, the semi-infinite interval of the bath is also sending radiation into the black hole which itself (and the purifier of the bath) purifies.

Finally, there is a transition to the late-time phase, with the QES jumping to a saddle point after the shock in AdS. As noted around eq.~\eqref{eq:smoke}, the relevant post-shock two-point entropy formula is the same as if the black hole and bath were eternally coupled. Since the after-shock AdS geometry is also the same as for an eternal black hole, the late-time generalized entropy is identical to the eternally coupled case studied in \cite{Almheiri:2019yqk}. This matching with the eternally-coupled case suggests that, by the Page time, the black hole and bath have reached equilibrium. 

As in the eternally-coupled case, time translation invariance (in $u$) simplifies the determination of the QES. In particular, the QES must be on the same time-slice as $\frac{y_1^+ + y_1^-}{2}=u_1$:
\begin{align}
	u_\QES
	=& u_1.
	\label{eq:homeless}
\end{align}
Hence, it remains only to determine the spatial location of the QES. Substituting eq.~\eqref{eq:homeless} into the entropy formula \eqref{eq:smoke} with $ y^{\pm}_1=u \mp \sigma_1$ gives
\begin{align}
	S_\gen
	=& \frac{c}{12k}\left[
	\frac{\phi_0}{\phi_r}
	- 2\pi T_1 \coth(2\pi T_1 \sigma_\QES)
	\right]
	+\frac{c}{6}\log\left[
	\frac{
	2\sinh^2[\pi T_1(\sigma_1-\sigma_\QES)]
	}{
	\pi T_1 \epsilon
	\sinh(-2\pi T_1 \sigma_\QES)
	}
	\right],
	\label{eq:fire}
\end{align}
in agreement with (19) in \cite{Almheiri:2019yqk}. By setting the $\sigma_\QES$-derivative of eq.~\eqref{eq:fire} to zero, we find
\begin{align}
	-\sigma_\QES
	=&\ \sigma_1 + \frac{1}{2\pi T_1}\log\left(\frac{2\pi T_1}{k}\right)
	+ \frac{1}{T_1}O\left(\frac{k}{T_1}\right),
	\label{eq:soot}
\end{align}
reproducing eq.~(21) in \cite{Almheiri:2019yqk}. Hence the QES sits outside of the black hole horizon. 

As an aside, time translation invariance permits a natural measure of proper distance between the QES and the horizon along a constant time slice. Using eq.~\eqref{eq:soot}, we have in units of $L_\AdS$:
\begin{align}
	\int_{\frac{\tanh(-\pi T_1 \sigma_\QES)}{\pi T_1}}^{\frac{1}{\pi T_1}}
	\frac{ds}{s}
	=& \frac{k e^{-2\pi T_1\sigma_1}}{\pi T_1}
	+ O\left[\left(\frac{k}{T_1}\right)^2\right],
\label{eq:paroxetine}
\end{align}
from which we see that the QES is an order $k/T_1$ proper distance outside the horizon\footnote{By measuring the distance \eqref{eq:paroxetine} between the QES and the horizon along a constant Killing time slice, we have implicitly extended the after-shock geometry to before the shock. The bifurcation surface of the final stationary black hole does not actually exist in the physical spacetime.}.
Using the location of the QES given by eqs.~\eqref{eq:homeless} and \eqref{eq:soot}, we can evaluate the generalized entropy of the late time phase. Again, by similarity to the eternally coupled case, this is a constant:
\begin{align}
	S_\gen \( T_1, \sigma_1 \) 
	\sim S_\Bek(T_1) + \frac{c}{6}\left[
	\log\left(\frac{1}{\pi T_1 \epsilon}\right)
	+2\pi T_1 \sigma_1
	\right],
	\label{eq:pasta}
\end{align}
 Interestingly, the above von Neumann part of the generalized entropy matches the entropy obtained from placing a twist operator at a large separation $\sigma_1$ from the boundary of a thermal half-line (see eq.~\eqref{eq:rum}) plus $S_{\twist{1}}(\AdS)$. Comparing with the generalized entropy given by eq.~\eqref{eq:pizza} for the scrambling phase, we see that the transition between the scrambling and late time phases occurs when $y_1^+$ hits a Page time of
\begin{align}
	y_\Page^+
	\approx& \frac{1}{2k}\left(1-\frac{T_0}{T_1}\right)
	- \frac{1}{2\pi T_1}\log\left[
	\frac{3 E_S (T_1-T_0)}{c T_0 T_1}
	\right].
	\label{eq:ImDrunk}
\end{align}


For later use, we note that more exact formulas may be obtained in the late time phase in the simple case where $x_1^\pm$ is placed on the boundary of $\AdS_2$, \ie we consider the entanglement wedge of $\QM_{\mt{L}}$ plus the entire bath and its purifier. As previous works \cite{Almheiri:2019psf,Chen:2019uhq}, we can ignore the correction from the position of the cut-off surface and set $x^+_1=x^-_1=t= f(u)= \frac{1}{\pi T_1} \tanh \( \pi T_1 u \)$. With this simplification, one can exactly solve, in $x^\pm$ coordinates, the equations
\begin{equation}\label{eq_QES_equilibrium}
\begin{split}
k \left(x^+-x^-\right)^2 \left(\frac{1}{x^+-x_1^+}-\frac{1}{x^+-x^-}\right) &=1-\left(\pi  T_1 x^-\right){}^2\,,\\
k\left(x^+-x^-\right)^2 \left(\frac{1}{x^+-x^-}-\frac{1}{x_1^--x^-}\right)  &=\left(\pi  T_1 x^+\right){}^2 -1\,,
\end{split}
\end{equation}
obtained from minimization of the late-time generalized entropy. These admit two trivial solutions $x^{\pm}_{\QES}=\pm \frac{1}{\pi T_1}$ and two non-trivial ones. Because of the constraints $x^+_{\QES}>x^-_{\QES}>0$, the only relevant solution for the position of QES reads
\begin{equation}\label{QES_solution01}
\begin{split}
x^+_{\QES}(t)&= \frac{\sqrt{k^2+\pi ^2 T_1^2} \left((\pi T_1 t)^2-1\right)+k\((\pi T_1 t)^2+1\)}{\pi ^2 T_1^2 \left(\pi ^2T_1^2 t^2+2 k t-1\right)}\,, \\
x^-_{\QES}(t)&= \frac{\sqrt{k^2+\pi ^2 T_1^2} \left((\pi  T_1 t)^2-1\right)+k \left((\pi  T_1 t)^2+1\right)}{\pi ^2 T_1^2 \left(-\pi ^2 T_1^2t^2+2 k t+1\right)}   \,. \\
\end{split}
\end{equation}
As a consistency check, we can use the time map $t= \frac{1}{\pi T_1} \tanh (\pi T_1 u)$ and find our solution \eqref{QES_solution01} for QES  satisfies eq.~\eqref{eq:homeless}:
\begin{equation}
\frac{1}{2}\(\yp + \ym\) \equiv \frac{1}{2\pi T_1}\, \text{arctanh} \( \pi T_1 \xp \) +\frac{1}{2\pi T_1}\, \text{arctanh} \( \pi T_1 \xm \) =u \,.
\end{equation}
Noting that the above solution of QES is not always physical, we need to impose the restrictions on parameters $k, t$ as
\begin{equation}\label{equi_inequalities}
\begin{split}
\pi ^2T_1^2 t^2+2 k t-1 &> 0 \,,\\
\sqrt{k^2+\pi ^2 T_1^2} \left((\pi T_1 t)^2-1\right)+k\((\pi T_1 t)^2+1\)&>0\,,
\end{split}
\end{equation}
which implies $t$ is extremely near $t_{\infty}$, \ie late time phase.\footnote{This is why the small $k$ expansion does not work for $\xm$ at late time phase because we have another much smaller value $(1- \pi T_1t)$ except for $k$. For example, $1- \tanh(\pi T_1 u) |_{\pi T_1 u=10}\approx 4\times10^{-35}$ does not depend on $k$ and is much smaller than $k$ at late time phase.} Moreover, it is direct to show $x^{\pm}_{\QES}(t_\infty) = t_{\infty}=\frac{1}{\pi T_1}$
and the simple monotonic behavior due to the fact
\begin{equation}
\frac{dx_{\QES}^+(t)}{dt}=\frac{2 k}{\pi ^2 T_1^2t \left(t\sqrt{k^2+\pi ^2 T_1^2}+2\right)+\sqrt{k^2+\pi ^2 T_1^2}+k(1-t^2\pi ^2 T_1^2)} >0 \,,
\end{equation}
from which we verify that the QES is located outside the horizon, as described around eq.~\eqref{eq:paroxetine}. (Note that any apparent spatial motion of the QES is purely an artifact of the choice of coordinates here -- due to time translation invariance in $u$, the QES is spatially stationary in $\sigma = \frac{y^- - y^+}{2}$, as indicated in eq.~\eqref{eq:soot}.)
So this is the first difference with the case under zero temperature bath where the QES is located inside the horizon.
With the exact solution, one can obtain the generalized entropy
\begin{equation}\label{SgenT1}
\begin{split}
S_{\gen,\rm{late}} (T_1)&= \frac{\bar{\phi}}{2\GN}  \( \sqrt{k^2+ \pi^2 T_1^2} - k \log\[\epsilon\!\( k+\sqrt{k^2+ \pi^2 T_1^2}\)\]   \)  \,,\\
\end{split}
\end{equation}
where the first term is the thermal entropy of a one-sided black hole with temperature $T_1$ and the second term describes the von Neumann entropy of bulk matter with the same temperature. It is obvious that the above generalized entropy is exactly a constant, indicating this is a thermal equilibrium state.

\subsection{Finite interval of the bath}
\label{sec:lonely}
In this section, we consider the question of whether the interior of the black hole can be recovered by a finite-sized interval in the bath, together with QM$_\mt{L}$ and the bath's purifier. We shall write the endpoints of the finite bath interval as $y_1^\pm = u\mp \sigma_1$ and $y_2^\pm = u \mp \sigma_2$ where $\sigma_1,\sigma_2\ge -\epsilon$.

To begin, we consider the case where we have access to the entire purifier for the thermal bath --- this is illustrated in Figure \ref{fig:desvenlafaxine}a. (In Section \ref{sec:smoke}, we shall see that the purifier is crucial for recovering the black hole interior.)
To stand a chance of recovering the black hole interior, we take $y_1^+ \ge y_\Page^+$, with $y_\Page^+$ given in eq.~\eqref{eq:ImDrunk}. We also take $y_2^\pm$ to be in the bath to the future of the shock, as we will see that this is sufficient to recover the black hole interior.

The two competing channels of generalized entropy in the holographic limit, corresponding to recoverability and non-recoverability of the black hole interior, are\footnote{There is in fact another channel where the black hole interior is recoverable, $S^\gen_{\QES-1} + S_2 + S_\halfLine$, but comparison of this with $S_{\mt{N}}$ in eq.~\eqref{eq:beer} reduces to a problem where the purifier of the bath has been traced out --- we deal with this case later in this section.}
\begin{align}
	S_{\mt{R}}
	=& S^\gen_{\QES-1} + S_{2-\IR},
	&
	S_{\mt{N}}
	=& S^\gen_\bif + S_{1-2} + S_\halfLine,
	\label{eq:beer}
\end{align}
where $S^\gen_i,\ S_i$ denote generalized and von Neumann entropies calculated with a single twist operator at $x_i^\pm$, while $S^\gen_{i-j},\ S_{i-j}$ denote generalized and von Neumann entropies of the interval with endpoints $x_i^\pm, x_j^\pm$. Further, subscripts QES, QES$''$ and IR denote the late-time QES associated to $y_1^\pm$, the (original) bifurcation point, and the IR point at $\sigma = \frac{y^- - y^+}{2}=\sigma_\IR$, respectively. Recall that the entropy $S_\halfLine$ of the thermal half-line is given in eq.~\eqref{eq:rum}. The entropy $S_{2-\IR}$, like $S_\halfLine$, is IR divergent as $\sigma_\IR\to\infty$; below, these divergences cancel in the differences of the entropies in the distinct channels.

To determine whether the black hole interior is recoverable, we ask whether $S_{\mt{R}} < S_{\mt{N}}$, or equivalently,
\begin{align}
	\begin{split}
	0 &> S_{\mt{R}}- S_{\mt{N}} \\
	&\approx\	\frac{c}{6}\Bigg\{
		\frac{\pi(T_1-T_0)}{k}
		+2\pi T_1 \sigma_1
		\\
	&\qquad+\log\left[
	\frac{3 E_S}{c \pi T_1^2}
		\frac{
		x_2^+ (1-\pi T_1 x_2^-)
		\sqrt{
		[1-(\pi T_1 x_1^+)^2]
		[1-(\pi T_1 x_1^-)^2]
		}
		}{
		(x_2^- - x_1^-)(x_1^+ - x_2^+)
		}
		\right]
		\Bigg\},
	\end{split}
		\label{eq:kaput}
\end{align}
where we have used eq.~\eqref{eq:pasta} to approximate $S^\gen_{\QES-1}$, eq.~\eqref{eq:rockBottom} for the one-point generalized entropy at the bifurcation point, and eq.~\eqref{eq:rum} for the entropy of the thermal half-line. The remaining entropies were obtained from the appropriate cases in eqs.~\eqref{eq:equil1} and \eqref{eq:equil2}. (Recall we are taking here $y_2^\pm$ to the future of the shock.) Since we have $y_1^\pm, y_2^- \gg \frac{1}{\pi T_1}$, we may use the following approximations,
\begin{align}
	x_1^\pm
	\approx&
	\frac{1}{\pi T_1}\left(
	1 - 2 e^{-2\pi T_1 y_1^\pm}
	\right),
	&
	x_2^-
	\approx&
	\frac{1}{\pi T_1}\left(
	1 - 2 e^{-2\pi T_1 y_2^-}
	\right).
	\label{eq:alcohol}
\end{align}
In this limit of large $y_1^\pm, y_2^- \gg \frac{1}{\pi T_1}$, the RHS of eq.~\eqref{eq:kaput} becomes
\begin{align}
	S_{\mt{R}}- S_{\mt{N}}
	\approx& \frac{c}{6}\left\{
	\frac{\pi(T_1-T_0)}{k}
	-4\pi T_1 \sigma_2
	+\log\left[
	\frac{6 E_S}{c T_1}
	\frac{1}{
	(e^{-2\pi T_1\sigma_1}
	-e^{-2\pi T_1\sigma_2})^2
	}
	\right]
	\right\}.
	\label{eq:sociopath}
\end{align}
Hence, the recoverability of the black hole interior is equivalent to $\sigma_2-\sigma_1> \Delta_\turn$, where
\begin{align}
	\Delta_\turn
	\approx& \frac{1}{4\pi T_1}\left[
	\frac{\pi(T_1-T_0)}{k}
	+ \log\left(
	\frac{6 E_S}{c T_1}
	\right)
	\right].
	\label{eq:areYouSure}
\end{align}
Comparing terms in eqs.~\eqref{eq:ImDrunk} and \eqref{eq:areYouSure} leading order in $k$, note that $\Delta_\turn \approx y_\Page^+/2$.

\subsection{Importance of the bath's purifier}
\label{sec:smoke}

As hinted earlier, the purifier of the bath is crucial to the reconstruction of the black hole interior. Let us briefly attempt a similar calculation to the above, now additionally tracing out this purifier. The generalized entropy for QM$_\mt{L}$ and an interval of the bath from $y_1^\pm$ to $y_2^\pm$ then has the competing channels -- see figure \ref{fig:desvenlafaxine}
\begin{align}
	S_\mt{R}
	=& S^\gen_{\QES-1} + S_2,
	&
	S_\mt{N}
	=& S^\gen_\bif + S_{1-2},
\end{align}
as illustrated in Figure \ref{fig:desvenlafaxine}c. Now, we take $y_2^\pm$ to the past of the shock, since we shall momentarily show that, even as the interval is extended by taking $\sigma_2=\frac{y_2^- - y_2^+}{2}$ arbitrarily large, the N-channel with entropy $S_{\mt{N}}$ will nonetheless remain favorable.\footnote{If one attempts a similar exercise with $y_2^\pm$ after the shock, then naively one finds with our entropy formulas that when this endpoint is placed at $O(c/E_S)$ away from the shock, it is possible for $S^\gen_\rec < S^\gen_\nrec$. However, this is an artifact of the fact that our setup is incapable of probing distances of such small scales. Since extending the interval of the bath can only increase the entanglement wedge, the argument presented in the main text precludes the possibility of the black hole interior being recovered from shorter intervals which stop to the future of the shock.} 
Using the formulas \eqref{eq:pasta} for $S^\gen_{\QES-1}$, \eqref{eq:rockBottom} for $S^\gen_\bif$, \eqref{eq:absolut} for $S_2$, and \eqref{eq:scone} for $S_{1-2}$, we have
\begin{align}
	\begin{aligned}
	S_{\mt{R}} - S_{\mt{N}}
	\approx
	\frac{c}{6}\Bigg\{&
	\frac{\pi(T_1-T_0)}{k}
	+ 2 \pi T_1 \sigma_1 \\
	&\qquad+
	\log\left[
	\frac{c}{24 \pi^2 T_1 E_S}
	\frac{
	(x_2^+ -x_2^- )
	\sqrt{[1-(\pi T_1 x_1^+)^2][1-(\pi T_1 x_1^-)^2]}
	}{
	x_1^+x_2^+ (x_2^- - x_1^-)
	}
	\right]
	\Bigg\}.
	\end{aligned}
\end{align}
Using again the $y_1^\pm, y_2^- \gg \frac{1}{\pi T_1}$ approximations \eqref{eq:alcohol}, we find
\begin{align}
	S_{\mt{R}} - S_{\mt{N}}
	\approx
	\frac{c}{6}\Bigg\{&
	\frac{\pi(T_1-T_0)}{k}
        + 2 \pi T_1 \sigma_1
	+ \log\left(\frac{c T_1}{12 \pi E_S}\right)\notag\\
	&\qquad+ \log\left[
	\frac{
	x_2^+ -x_2^-
	}{
	T_1 x_1^+ x_2^+
	(e^{-2\pi T_1 \sigma_1} - e^{-2\pi T_1 \sigma_2})
	}
	\right]
	\Bigg\}.
\end{align}
On the RHS, the first three terms sum to a large positive number (note that the third term, though negative, scales like the logarithm of the first term). Moreover, the logarithm of the last term is bounded from below by $\log \pi$. It follows that $S_{\mt{R}} > S_{\mt{N}}$. We thus conclude that when the purifier of the bath is traced out, no matter how large an interval of the bath one has access to, the N-channel is favorable and the black hole interior cannot be recovered. 

\begin{figure}
  \centering
  \includegraphics[width=\textwidth]{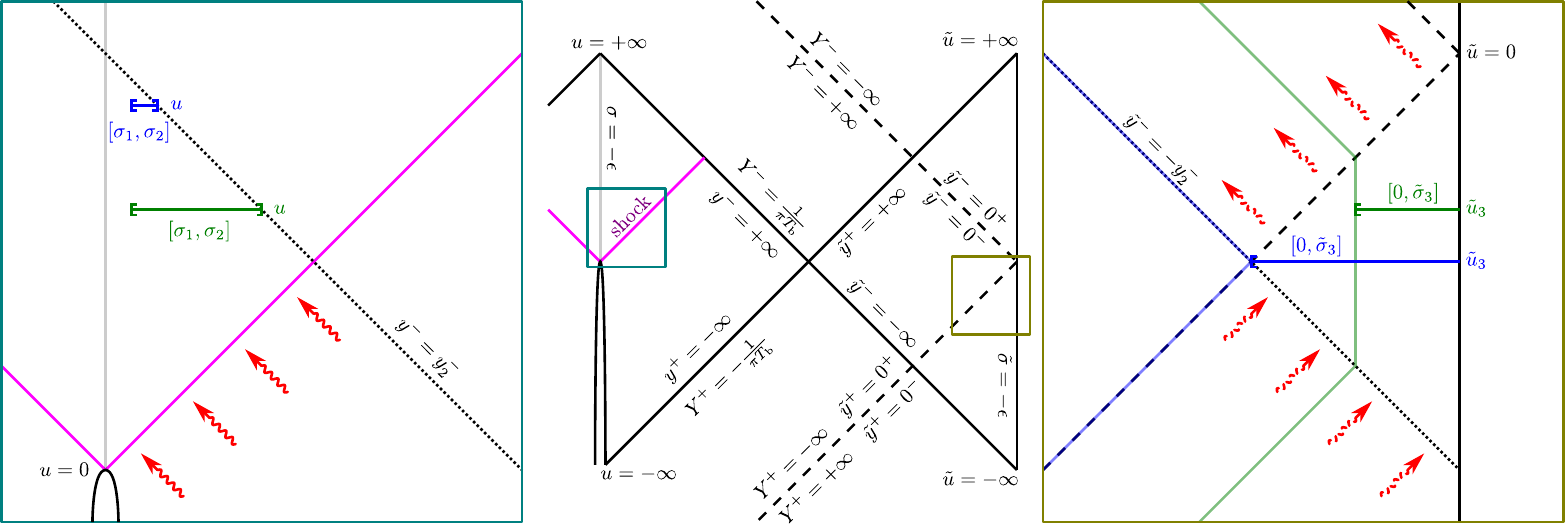}
  \caption{The bath and purifier subsystems. The central panel shows a Penrose diagram of various coordinate patches of the bath and purifier subsystems. The left panel shows two examples, sharing the same $y_2^-$, of an interval $[\sigma_1,\sigma_2]$ of the bath system after the Page time: the shorter blue interval is just barely above the critical length $\Delta_\turn$ needed to recover the black hole interior; the green interval is much longer. Red wavy lines show thermal radiation leaving the bath prior to $y^-=y_2^-$. The right panel shows the corresponding intervals $[0,\sigmapure_3]$ needed in conjunction with the bath intervals (plus $\QM_{\mt{L}}$) to recover the black hole interior. The phase boundaries of $\sigmapure_3$ for recoverability is shown in light blue and green. The dashed wavy lines show the thermal quanta of the purifier that are most entangled with the radiation marked in the left panel.}
  \label{fig:duloxetine}
\end{figure}

A natural follow-up question is whether the black hole interior can be recovered when one can only access a finite portion of the bath's purifier at various times. We shall take the joint system of the bath and its purifier to be in a thermofield double state. Furthermore, we introduce a new set of coordinates $\ypure^\pm = \upure \pm \sigmapure$ for the purifier of the bath, where $\upure$ and $\sigmapure$ are analogous to the $u$ and $\sigma$ coordinates of the bath. These coordinates for the purifier are related to the coordinates we have been using thus far by
\begin{align}\label{wacked}
	Y^\pm
	=& -\frac{1}{\pi T_1}\coth(\pi T_1 \ypure^\pm).
\end{align}
Note that while $\pi T_1 Y^\pm \in (-1,1)$ provides a coordinate chart which includes the bath, the coordinate chart of $(\pi T_1 Y^\pm)^{-1} \in (-1,1)$ includes the purifier of the bath. Moreover, we have
\begin{align}
	\pi T_1 \ypure
	=& -\pi T_1 y + \frac{i\pi}{2},
	&
	d\ypure^+ d\ypure^-
	=& dy^+ dy^-
\end{align}
so that there is no additional Weyl transformation that must be applied to our entropy formulas when endpoints are moved from the bath to its purifier. (Specifically, in \eqref{eq:jackDaniels}, \eqref{eq:equil1}, and \eqref{eq:equil2}, end-points in the purifier of the bath should be treated as though they were simply in the bath and to the past of the shock.) The coverage of the $Y,y,\ypure$ coordinates in the bath and purifier subsystems are summarized in the middle panel of Figure \ref{fig:duloxetine}. 

We may then repeat the analysis of Section \ref{sec:lonely}, now pushing the IR endpoint to a point $\ypure_3^\pm$ in the bath's purifier. That is, we consider whether the black hole interior can be recovered from QM$_{\mt{L}}$, an interval of the bath with endpoints $y_1^\pm=u\mp\sigma_1$ and $y_2^\pm = u\mp \sigma_2$, and an interval of the bath's purifier stretching from an endpoint $\ypure_3^\pm=\upure_3\pm\sigmapure_3$ to the boundary $\sigmapure=0$ --- see figure \ref{fig:duloxetine}.  In general, we shall find that the size of the purifier interval required to recover the black hole interior will depend on the time $\upure_3$ at which the interval is selected.

From Sections \ref{sec:depression} and \ref{sec:lonely}, we see that, to have a chance of recovering the black hole interior, we should take $y_1^+ > y_\Page^+$ and $\sigma_2-\sigma_1>\Delta_\turn$ with $y_\Page^+$ and $\Delta_\turn$ given in \eqref{eq:ImDrunk} and \eqref{eq:areYouSure}. For simplicity, we take $y_2^\pm$ to the future of the shock. The relevant generalized entropies for the R- and N-channels are,
\begin{align}
	S_{\mt{R}}
	=& S^\gen_{\QES-1} + S_{2-3}\,,
	&
	S_{\mt{N}}
	=& S^\gen_\bif + S_{1-2} + S_3\,,
\end{align}
as illustrated in Figure \ref{fig:desvenlafaxine}d. Relating to the problem treated in Section \ref{sec:lonely}, where $\ypure_3^\pm$ is pushed to the IR, in comparing $S_{\mt{R}}$ and $S_{\mt{N}}$, we have the extra contribution
\begin{align}
	S_{\mt{R}}
	- S_{\mt{N}}
	-\left[S_{\mt{R}} - S_{\mt{N}}\right]_{\sigmapure_3\to +\infty}
	=& \frac{c}{6}\log\left[
	\frac{2\pi T_1 x_3^+ (-x_2^- + x_3^-)}{(1-\pi T_1 x_2^-)(x_3^+ - x_3^-)}
	\right].
\end{align}
From \eqref{eq:sociopath}, we see that
\begin{align}
	\left[S_{\mt{R}} - S_{\mt{N}}\right]_{\sigmapure_3\to +\infty}
	\approx&
	-\frac{2 \pi c T_1}{3}(\sigma_2-\sigma_1-\Delta_\turn).
\end{align}
Applying the approximation \eqref{eq:alcohol} for $x_2^-$, we find that the condition $S_{\mt{R}} \le  S_{\mt{N}}$ for the recoverability of the black hole interior translates to
\begin{align}
	\pi T_1 - (x_3^-)^{-1}
	\lesssim& \frac{
	\pi T_1 \left[1-2e^{-4\pi T_1(\sigma_2 -\sigma_1- \Delta_\turn)}\right]
	- (x_3^+)^{-1}
	}{
	e^{2\pi T_1(u-\sigma_2+2\sigma_1+2\Delta_\turn)}
	}.
	\label{eq:christmas}
\end{align}
The RHS, giving the maximal separation of $(x_3^-)^{-1}$ from past null infinity line $(x^-)^{-1}=\pi T_1$ of the bath's purifier, is largest when $(x_3^+)^{-1}$ sits on the future null infinity line $(x^+)^{-1}=-\pi T_1$. Here,
\begin{align}
	\left.\pi T_1 - (x_3^-)^{-1}\right|_{(x_3^+)^{-1} = -\pi T_1}
	\lesssim& \frac{
	2\pi T_1 \left[1-e^{-4\pi T_1(\sigma_2 - \sigma_1-\Delta_\turn)}\right]
	}{
	e^{2\pi T_1(u-\sigma_2+2\sigma_1+2\Delta_\turn)}
	}.
	\label{eq:friendless}
\end{align}
We see that even this is exponentially suppressed (note $u-\sigma_2+2\sigma_1+2\Delta_\turn \ge 2\Delta_\turn$ with $\Delta_\turn$ given in \eqref{eq:areYouSure}). In contrast, with appropriate $\sigma_2-\sigma_1>\Delta_\turn$, $(x_3^+)^{-1}$ can be pushed far from the future null infinity value $-\pi T_1$; the largest separation is achieved when $(x_3^-)^{-1}$ sits on past null infinity:
\begin{align}
	\pi T_1 + \left.(x_3^+)^{-1}\right|_{(x_3^-)^{-1} = \pi T_1}
	\lesssim& 2\pi T_1\left[1-e^{-4\pi T_1(\sigma_2-\sigma_1-\Delta_\turn)}\right].
	\label{eq:eve}
\end{align}

It is also instructive to consider the condition \eqref{eq:christmas} in terms of the spatial interval length $\sigmapure_3$ taken in the purifier. As we elaborate below, due to the step-like nature of the $\tanh$ function in $f$, the constraint \eqref{eq:christmas} becomes a piece-wise linear constraint on $\sigmapure_3$ as a function of $\upure_3$ with interpolation between the pieces on scales of order $\pi T_1^{-1}$.

Let us consider first the case $\sigma_2-\sigma_1-\Delta_\turn \ll (4\pi T_1)^{-1}$. Then, both \eqref{eq:friendless} and \eqref{eq:eve} are small so that we are in the regime
\begin{align}
	\pi T_1 - (x_3^-)^{-1}
	\approx& 2\pi T_1 e^{2\pi T_1 \ypure_3^-},
  &\left( \ypure_3^- \ll -\frac{1}{2\pi T_1} \right)
    \label{eq:abergine}
	\\
	\pi T_1 + (x_3^+)^{-1}
	\approx& 2\pi T_1 e^{-2\pi T_1 \ypure_3^+}
	&\left( \ypure_3^+ \gg \frac{1}{2\pi T_1}\right)
	\label{eq:white}
\end{align}
Note that the bounds given by the RHS's of \eqref{eq:friendless} and \eqref{eq:eve} are complementary in the following sense: if $\pi T_1-(x_3^-)^{-1}$ is much smaller than the RHS of \eqref{eq:friendless}, then \eqref{eq:christmas} reduces to the constraint that $\pi T_1+(x_3^+)^{-1}$ is less than approximately the RHS of \eqref{eq:eve}; on the other hand, if $\pi T_1+(x_3^+)^{-1}$ is much smaller than the RHS of \eqref{eq:eve}, then \eqref{eq:christmas} reduces to $\pi T_1-x_3^-$ being smaller than approximately the RHS of \eqref{eq:friendless}. Considering \eqref{eq:abergine} and \eqref{eq:white}, the interpolation between these two cases occurs on scales of order $(\pi T_1)^{-1}$ in $\ypure_3^\pm$. We thus find a piecewise null phase boundary for $\ypure_3^\pm$: 
\begin{align}
	\begin{split}
	\sigmapure_3 \gtrsim
	- \frac{1}{2\pi T_1}\log[4\pi T_1(\sigma_2-\sigma_1-\Delta_\turn)]+
	\begin{cases}
		-\upure_3
		&\text{if $\upure_3 \lesssim -\frac{y_2^-}{2}$}
		\\
	\upure_3 + y_2^-
	& \text{if $\upure_3 \gtrsim -\frac{y_2^-}{2}$}
\end{cases}
,
\end{split}
  \label{eq:levomilnacipran}
\end{align}
with interpolation between the pieces occurring on scales of order $(\pi T_1)^{-1}$ in $\upure_3$. We see that as $\sigma_2-\sigma_1$ approaches the minimum interval length $\Delta_\turn$ of the bath required for recovery of the black hole interior, $\sigmapure_3$ diverges logarithmically. 

Next, we consider the case where $\sigma_2-\sigma_1-\Delta_\turn \gg (4\pi T_1)^{-1}$. Now, the RHS of \eqref{eq:eve} need not be small, opening the possibility for a new regime where
\begin{align}
	\pi T_1 - (x_3^+)^{-1}
	\approx& 2\pi T_1 e^{2\pi T_1 \ypure_3^+}
  &\left( \ypure_3^+ \ll -\frac{1}{2\pi T_1}\right)
    \label{eq:venlafaxine}
\end{align}
but
\begin{align}
  -\ypure_3^+ \lesssim \sigma_2-\sigma_1-\Delta_\turn
  \label{eq:milnacipran}
\end{align}
so that the bound \eqref{eq:eve} is not yet saturated. Inserting \eqref{eq:venlafaxine} into \eqref{eq:christmas}, we obtain the phase boundary in an intermediate regime between \eqref{eq:friendless} and \eqref{eq:eve}. This phase boundary, at the conclusion of this intermediate regime, \ie{} when \eqref{eq:milnacipran} is saturated, ends deep in the region where \eqref{eq:venlafaxine} holds. Finally, plugging \eqref{eq:venlafaxine} into \eqref{eq:eve}, we obtain a phase boundary in the complement of \eqref{eq:milnacipran}. Altogether, we find
\begin{align}\label{merry4}
	\begin{split}
	\sigmapure_3
	\gtrsim
	\begin{cases}
		-\upure_3 -2(\sigma_2-\sigma_1-\Delta_\turn)
		& \text{if $\upure_3 +\frac{y_2^-}{2}<-(\sigma_2-\sigma_1-\Delta_\turn)$}
		\\
		\frac{y_2^-}{2}-(\sigma_2-\sigma_1-\Delta_\turn)
		&\text{if $\left|\upure_3+\frac{y_2^-}{2}\right|<\sigma_2-\sigma_1-\Delta_\turn$}
		\\
		\upure_3 + y_2^- - 2(\sigma_2-\sigma_1-\Delta_\turn)
		& \text{if $\upure_3+\frac{y_2^-}{2}>\sigma_2-\sigma_1-\Delta_\turn$}
	\end{cases},
      \end{split}
\end{align}
with interpolation on the thermal scale.\footnote{Eq.~\eqref{merry4} can be condensed into 
	\beq\label{merry4a}
	2\tilde{\sigma}_3 \gtrsim   |\tilde{y}_3^+|  +|y_2^-+\tilde{y}_3^-| + y_2^-  - 4 \(\sigma_2-\sigma_2-\Delta_\turn\) \,,
	\eeq 
	which is similar to the general $T_\bath$ result in eq.~\eqref{eq:sigma3optimallinear} with $\delta \sigma_2 \to \sigma_2-\sigma_1-\Delta_\turn$ and $T_\bath = T_{\rm eff} = T_1$. As noted around~\eqref{eq:sigma3optimallinear}, the assumptions leading to the general result are more constraining.} We see that the intermediate case gives the smallest possible region of the purifier needed for reconstruction of the black hole interior.

For the blue and green bath intervals illustrated in the left panel of figure \ref{fig:duloxetine}, the approximate phase boundary \eqref{merry4} is highlighted respectively in light blue and light green in the right panel of figure \ref{fig:duloxetine} and examples of minimal-length purifier intervals are shown in opaque blue and green. The blue case illustrates the phase boundary given by \eqref{merry4} for a bath interval just large enough for \eqref{merry4} to be valid (as opposed to \eqref{eq:levomilnacipran}) --- in this limit, the intermediate piece of \eqref{merry4} vanishes. The green case, on the other hand, has a much larger bath interval. As illustrated in Figure \ref{fig:duloxetine}, \eqref{merry4} has the interpretation of giving the interval of the purifier needed to capture quanta entangled with out-going thermal bath radiation emitted between times $y^-=0$ and $y^-=y_2^-$. When the bath interval length $\sigma_2-\sigma_1$ is barely a few thermal lengths greater than the critical value $\Delta_\turn$ (blue case), nearly all of these quanta must be accessible in the purifier. For much longer bath intervals (green case), fewer purifier quanta are necessary. We shall comment further on this in Section \ref{discuss}.


%% file: sections/nonzeroTb.tex

In the previous section, we analyzed the two-dimensional black hole coupled to a bath system with temperature $T_\bath =T_1$, which together formed a system in thermal equilibrium as soon as the Page time was reached. Compared to the results from the evaporating black hole with zero temperature, \eg \cite{Almheiri:2019hni,Almheiri:2019psf,Chen:2019uhq}, we found qualitatively different behavior in the evolution of the generalized entropy -- and, of course, the role of purification of the bath. In this section, we consider coupling the black hole with temperature $T_1$ after the shock is absorbed into a thermal bath with a general temperature $T_\bath$. The black hole and bath evolve to reach an equilibrium where the black hole temperature matches $T_\bath$. However, the black hole will decrease or increase in size (and entropy) depending on whether $T_\bath < T_1$ or $T_\bath > T_1$. The evolution of the (effective) black hole temperature is shown in eq.~\reef{Teffu} for several cases.


\begin{figure}[htbp]
	\centering\includegraphics[width=4.5in]{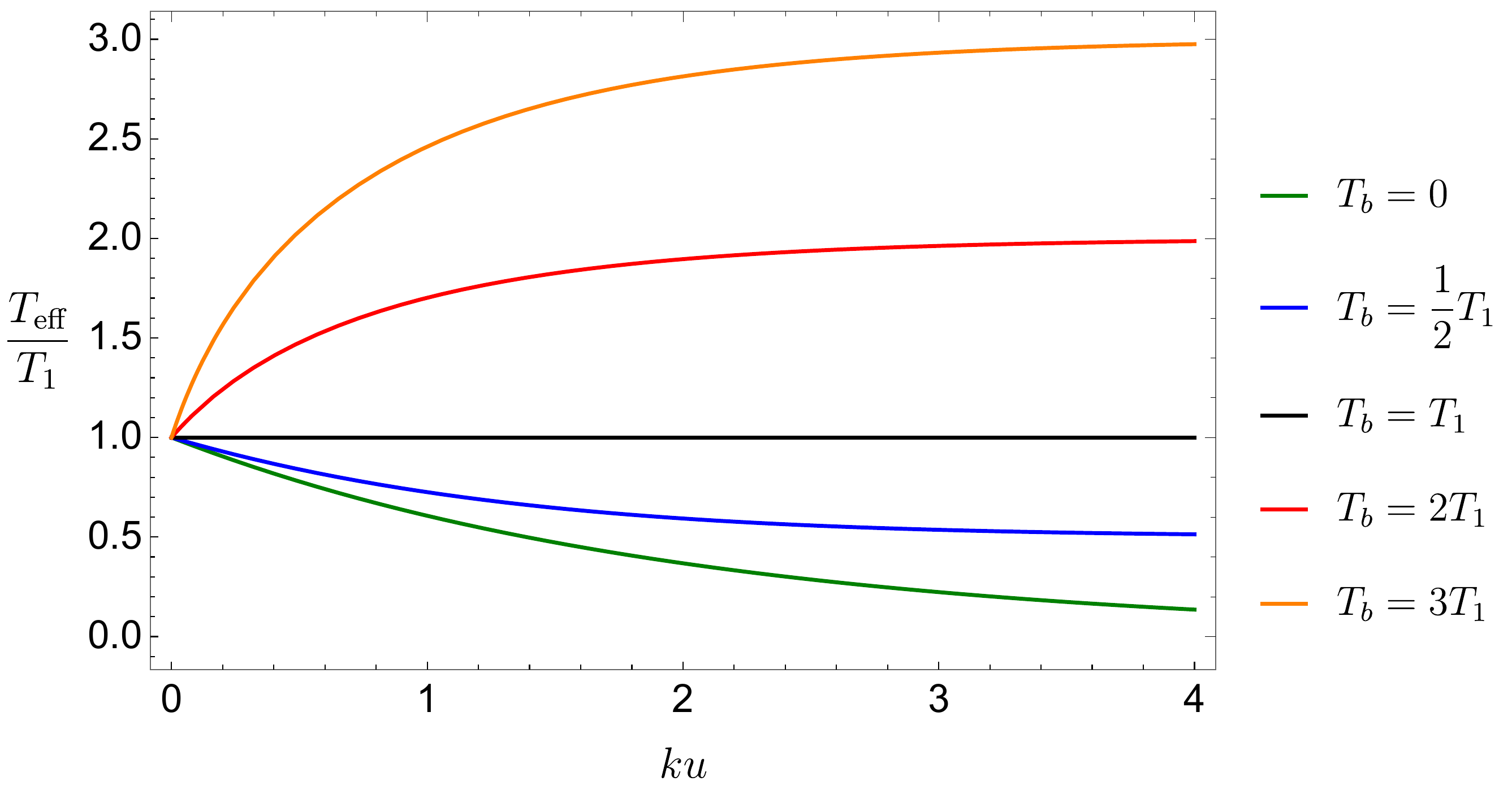}
	\caption{The time dependence of effective temperature of black hole, which simply parametrizes the dynamical behavior of black hole.}\label{Teffu}
\end{figure}
As was explained in section~\ref{sec:quenching}, the Schwarzian equation~\eqref{eq:chocolate} can be solved for arbitrary bath temperature to find the time-map function $f(u)$ in eq.~\eqref{eq:fmap} which reduces to the $\Tb=0$ result of~\cite{Almheiri:2019psf} by taking $\nu =0$.\footnote{We note that for numerical purposes, the following form of the time-map function
\begin{equation}
f(u, T_{b}) =
\frac{2}{k}
\frac{I_\nu(a)\, K_\nu(a e^{-\frac{k u}{2}}) - K_\nu(a)\,I_\nu(a e^{-\frac{k u}{2}})}{ I_\nu (a e^{-\frac{k u}{2}})\(a K_{\nu-1}(a)+ \nu K_{
		\nu}(a)\)  + K_\nu (a e^{-\frac{k u}{2}})\(a I_{\nu-1}(a)- \nu I_{
		\nu}(a)\)  } \,,
\end{equation}
is easier to deal with than the expression in eq.~\eqref{eq:fmap}.
Note that similar expressions appear in \cite{Hollowood:2020cou}.} Taking the limit $ku \to \infty$, one can also define the end of the proper time
\begin{equation}
t_{\infty} = \frac{2}{k} \frac{I_{\nu}(a)}{a I_{\nu-1}(a)- \nu I_{\nu}(a)} = \frac{2}{ka} \frac{I_\nu(a)}{I_\nu'(a)} \,,
\end{equation}
which is also the final position of the QES, \ie $x^{\pm}_{\QES}|_{u\to \infty}= t_{\infty}$. We now stress some important facts about the above map function from coordinate time $t$ to proper (physical) time $u$. First, the function $f(u,\Tb)$ is well defined and real for $\Tb \le T_1$  and also $\Tb\ge T_1$.  Secondly, it is also invariant under the following rescaling
\begin{equation}
T_1 \to \alpha T_1 \,, \quad \Tb \to \alpha \Tb\,, \quad k \to \alpha k\,,\quad u \to \frac{u}{\alpha} \,,\quad\frac{\bar{\phi}_r}{\GN} \to \frac{\bar{\phi}_r}{\alpha^2\GN} \,.
\end{equation}
In other words, the independent dimensionless parameters in the model are
\begin{equation}
T_1 L_{\text{AdS}}\,, \quad  \frac{\Tb}{T_1} \,,\quad  ku \,, \quad \frac{k}{T_1}\,,
\end{equation}
besides of $\frac{\phi_r}{\GN}$. We simply take the radius of AdS as the standard scale by choosing $L_{\AdS}=1$ and all other parameters can be considered to be normalized by $T_1$. From an energetic point of view, it is clear that $\Tb=T_1$ is a critical temperature for the thermal bath, where the black hole will neither lose nor absorb energy. From the energy flow equation \eqref{eq:energy} and Schwarzian equation \eqref{eq:chocolate}, we can define an effective temperature
\begin{equation}
\label{eq:Teff}
T_{\rm{eff}}(u; \Tb) = \sqrt{\Tb^2 + (T_1^2 - \Tb^2)\,e^{- k u} } \,,
\end{equation}
which parametrizes the ADM mass of the dynamical black hole at time $u$ by 
\begin{equation}
E(u) = \frac{\bar{\phi}_r \pi }{4 \GN} T_{\rm{eff}}^2(u) \,.
\end{equation}
 Recalling the energy flux \eqref{eq:stresstensor} on the physical boundary $x^-\approx t$, \ie 
\begin{equation}
	\langle T_{x^- x^-}\rangle
= E_S\, \delta(t) + \frac{c \pi }{12} \frac{1}{(f'(u))^2} \left(
 T_\bath^2 -T_{\rm{eff}}^2(u) \right) \,,
\end{equation}
we can explain the above three terms as the contributions from the shock wave, thermal radiation from the coupled bath system at temperature $T_\bath$, and Hawking radiation escaped from the dynamical black hole. As expected, we can also understand the effective temperature as the measure for the temperature of Hawking radiation at time $u$. For later use, we also show the numerical plot for the time evolution of effective temperature with various $T_\bath$ in figure \ref{Teffu}. 

For $\Tb < T_1$, the black hole loses energy via the absorption of Hawking radiation by the bath and evaporates to a smaller black hole with lower temperature $\Tb$, which is similar to the model with $\Tb =0$ as described in \cite{Almheiri:2019hni,Almheiri:2019psf,Chen:2019uhq}. Conversely, a black hole coupled with a higher temperature bath $\Tb > T_1$ absorbs radiation from the bath and approaches another equilibrium state with temperature $\Tb$ when $k u \gg 1$. In both cases, for $ku \to \infty$, the system is thermalized and shows similar qualitative features to the equilibrium case $\Tb=T_1$. In summary, we have the three different dynamical behaviors in the two-dimensional gravity setup: an evaporating black hole when $\Tb < T_1$; a growing black hole when $\Tb > T_1$; and equilibrium when $\Tb = T_1$. Note that these outcomes are independent of the temperature of the original black hole, \ie $T_0$.

However, diving into the details of the QES and the flow of information, we will see there are different critical temperatures determining the position of the QES relative to the final event horizon, as will be explained in section~\ref{sec:approach-equil}. For this analysis, we approximate the equations for the generalized entropy and find the approximate solutions for QES. Making a small $k$ expansion with {\bf fixed $ku$}, one can find the following approximation of $f(u)$:\footnote{In order to do this analytically, we first note the series expansions of Bessel functions~\cite{abramowitz+stegun} 
\begin{equation}
\label{eq:assymptotic}
\begin{split}
 K_{\nu}(\nu z) & \underset{\nu \to \infty}{\sim} \sqrt{\frac{\pi}{2\nu}} \( \frac{e^{-\nu \eta}}{(1+z^2)^{1/4}} + \mathcal{O}(\frac{1}{e^{\eta \nu} \nu}) \),\quad
 I_{\nu}(\nu z) \underset{\nu \to \infty}{\sim} \sqrt{\frac{1}{2\pi\nu}} \( \frac{e^{\nu \eta}}{(1+z^2)^{1/4}} + \mathcal{O}(\frac{e^{\eta \nu}}{ \nu}) \)\\
& \text{with}\qquad \eta= \sqrt{1+z^2} + \log \frac{z}{1+\sqrt{1+z^2}} \,.
\end{split}
\end{equation}
}
\begin{equation}\label{app_fu}
\frac{f(u)}{t_{\infty}} \approx  \tanh \( \frac{2\pi }{k} \( T_1 - T_{\rm{eff}} - \Tb \log \( \frac{T_1 +\Tb}{\Tb + T_{\rm{eff}}} \) + \frac{ku}{2}\Tb\)\)  \,,
\end{equation}
which reduces to the equilibrium case with $f(u)=\frac{1}{\pi T_1} \tanh \( \pi T_1 u\)$ after taking $\Tb =T_1$. Hence the above simplified form approximates the map-function $f(u)$ even for $\Tb \ge T_1$. From the asymptotic expansion in eq.~\eqref{eq:assymptotic}, one can also obtain the approximation for the upper bound of physical time
\begin{equation}
t_{\infty}  \approx  \frac{1}{\pi T_1} + \frac{k}{4\pi^2 T_1^4}\(T_1^2- \Tb^2 \) + \mathcal{O}(k^2) \,.
 \end{equation}
Let us remark that one can further derive several simpler and useful approximations
\begin{equation}
\label{eq:approx-simple}
\begin{split}
\log\( \frac{t_{\infty}-f(u)}{2t_{\infty}} \) &\sim -\frac{4\pi }{k} \( T_1 - T_{\rm{eff}} - \Tb \log \( \frac{T_1 +\Tb}{\Tb + T_{\rm{eff}}} \)  + \frac{ku}{2}\Tb\) \,,\\
f'(u)&\sim 2 \pi T_{\rm{eff}} \( t_{\infty}-f(u) \) \,,\\
\{u, f(u) \}&= -\frac{1}{(f'(u))^2} \{f(u),u\} \sim  \frac{1}{2(t_{\infty} -f(u))^2} \,.
\end{split}
\end{equation}
which will be used many times in the following analysis.  It is also easy to find that all the above approximations reduce to the same forms used in \cite{Almheiri:2019psf,Chen:2019uhq} upon taking $\Tb=0$. The above approximations are still complicated due to the appearance of $T_{\rm{eff}}(u,\Tb)$, we can further simplify the above results if we focus on times at the order of the Page time by taking the early-time limit $k u \ll 1$ ({\bf linear region}) \footnote{In the later, we will find that our most analytical approximations (at leading order) in the linear region present linear behaviors in time. One can consider some transition point $ku \sim \frac{1}{\#}$ as the endpoint for the linear region. \label{footnote_linear}}. In the linear regime, the effective temperature $T_{\rm{eff}} \sim T_1$ and we find the following linear approximations
\begin{equation}\label{app_linear}
\begin{split}
\log\( \frac{ t_{\infty}-f(u)}{2t_{\infty}} \) &\sim -2\pi T_1 u + {\cal O}\(ku^2\)\,
\\
 \log \(  \frac{1}{f'(u)}\) &\sim 2\pi T_1 u -\log \( 4\pi T_1 t_{\infty} \) \,, \\
\end{split}
\end{equation}
where the leading-order contributions are not sensitive to the temperature of the bath $\Tb$ because the black hole does not evaporate very much in this phase.
We are also interested in the {\bf late-time region} with $e^{ku} \gg 1$ where we need the following approximations \footnote{Note that the $ku\to \infty$ and the $\Tb \to 0$ limits of eq.~\eqref{eq:assymptotic} do not commute.}
\begin{equation}\label{app_late}
\begin{split}
\log\( \frac{ t_{\infty}-f(u)}{2t_{\infty}} \) &\sim \frac{4\pi}{k} \( \Tb -T_1 +\Tb \log\( \frac{T_1+\Tb}{2\Tb}\) - \frac{ku}{2}\Tb + \frac{T_1^2-\Tb^2}{4\Tb}e^{-ku}  \) \,, \\
\log \(  \frac{1}{f'(u)}\) &\sim \frac{1}{k} \( 2\pi \Tb ku - 4 \pi \(\Tb -T_1 +\Tb \log\( \frac{T_1+\Tb}{2\Tb}\) \)\)-\log \( 4\pi \Tb t_{\infty} \) \,. \\
\end{split}
\end{equation}
Lastly, we note that the coefficient of the linear term changes to $2\pi \Tb$ whereas, which is expected because the temperature of the black hole at late time ($e^{ku}\gg1$) is close to $\Tb$. Instead of analytical approximations, we also performed numerical calculations for all the results as the double-check for these approximations in the following analysis. 
For convenient comparisons with the results at $T_\bath=0$, all numerical plots are done by choosing the numerical parameters listed in table \ref{tab:baseline}, which are the same as those chosen in \cite{Chen:2019uhq}.

\begin{table}[t]
	\centering
	\begin{tabular}{l|c|c|c|c|c|c|c|c}
		\hline
		\hline
		& & &    & & & &\\[-2ex]
		\textbf{Parameter} & $L_{\mt{AdS}}$ &  $k$ & 	$T_1$ & $T_0$ &c &$\epsilon$ &	$\phi_0$ &	$\bar{\phi}_r$  \\ \hline
		& & &    & & & &\\[-2ex]
		\textbf{Value} & 1	& $\frac{1}{4096}$  & $\frac{1}{\pi}$ &$\frac{63}{64\pi}$ & 4096	& $\frac{1}{4096}$ & 0 & $\frac{1}{4096^2}$\\[0.5ex] \hline \hline
	\end{tabular}
	\caption{Baseline parameters for all numerical plots in this paper.}
	\label{tab:baseline}
\end{table}

\subsection{QES and Page curve}\label{subsec:QES_Page}
In the following section, we first consider the generalized entropy of the subsystem QM$_{\mt{R}}$, which is the same with that of subsystem consisting of QM$_{\mt{L}}$, the thermal half-line and another half-line containing the purification. That is, we would like to find the position of the QES, \ie  $x^{\pm}_{\QES}$, in the late-time phase when we anchor  the endpoint $x^{\pm}_1 $ on the boundary between AdS and flat spacetime. The generalized entropy reads
\begin{equation}\label{Sgen_QESx1}
\begin{split}
S_{\gen,\rm{late}}(\Tb) =&\frac{\bar{\phi}_r}{4 \GN} \left[ 2\frac{1-(\pi T_1)^2 x^+_{\QES}x^-_{\QES} + \frac{k}{2}\,I\!\(x^+_{\QES},x^-_{\QES};x^-_{\QES}\)}{x^+_{\QES}-x^-_{\QES}} \right.\\
&\left.+ 2k \log\( \frac{2}{\epsilon} \frac{\sinh\(\pi \Tb\(y^-_1 -\ym\)\)}{\pi \Tb}  \frac{(x^+_{\QES}-x^+_1)}{(x^+_{\QES}-x^-_{\QES})}\sqrt{\frac{f'(y^-_{\QES})}{f'(y^+_1)}} \) \right] \,, \\
\end{split}
\end{equation}
with the integral term defined as
\begin{equation}\label{hausp}
\begin{split}
I(x^+,x^-;x)&= \int_0^{x} \, \(x^+-t\)\(x^--t\) \(\{u,t\} -2 \(\frac{\pi \Tb}{f'(u)}\)^2\)dt\,,\\
{\rm with}\qquad&\{u,t\} - 2\(\frac{\pi \Tb}{f'(u)}\)^2= \(\frac{1}{f'(u)}\)^2 2\pi^2 \( T_1^2 -\Tb^2\) e^{-ku} \,.
\end{split}
\end{equation}
In order to minimize the generalized entropy, we need to solve the differential equations $\partial_{\pm} S_{\gen}=0$. Explicitly, we have
\begin{equation}\label{QES_equations}
\begin{split}
0&=2\( \pi T_1 x^-_{\QES} \)^2 -2 -k I\(x^-_{\QES }, x^-_{\QES};x^-_{\QES }\) +2k \(x^+_{\QES}- x^-_{\QES}\)^2  \( \frac{1}{x^+_{\QES}-x^+_1}  - \frac{1}{x^+_{\QES}-x^-_{\QES}} \) \,,\\
0&= 2-2\( \pi T_1 x^+_{\QES} \)^2  +k I\(x^+_{\QES }, x^+_{\QES};x^-_{\QES }\)  \\
&+2k \(x^+_{\QES}- x^-_{\QES}\)^2  \(\frac{\pi \Tb}{\tanh\( \pi \Tb(y^-_{\QES}-y_1^-) \)} \frac{1}{f'\(\ym \)}+\frac{1}{x^+_{\QES}-x^-_{\QES}}  +\frac{1}{2}\frac{f''(y^-_{\QES})}{\(f'(y^-_{\QES})\)^2}\) \,.
\end{split}
\end{equation}
To solve these equations, we will need the approximation for the time-map function $f(u)$ in eq.~\eqref{app_fu} (and it's subsequent limits in eqs.~\eqref{eq:approx-simple}, \eqref{app_linear} and \eqref{app_late}), but we still need to carefully deal with the integral term that originates from the backreaction of the dilaton in the JT gravity.  From the late time limit of eq.~\eqref{QES_equations} we find
\begin{equation}
I_\infty \equiv I(t_\infty,t_\infty;\tinf) = \frac{2}{k}\(\( \pi T_1 t_\infty\)^2-1\)\,,
\end{equation}
which is the leading-order contribution to the integral at late times because the position of QES should be located near $t_{\infty}$, \ie $x^+_{\QES} \sim t_{\infty} \sim t\sim x^-_{\QES} $. As before, we start from considering the generalized entropy of the subsystem consisting of QM$_\mt{L}$ and the whole bath (with its purification) by taking $x_1$ on the conformal boundary of AdS
\begin{equation}
x^+_1\approx t \approx x^-_1 \qquad \(\ie y^+_1\approx u \approx y^-_1\)\,,
\end{equation}
where we ignored the correction at the order $\mathcal{O}(\epsilon f'(u))$.\footnote{Recall the point on AdS boundary at proper time $u$ is defined by $t=f(u)= \frac{x^+_1+x^-_1}{2}, s=\frac{x^+_1-x^-_1}{2}\approx \epsilon f'(u)$.}

\subsubsection{Turn on the temperature of bath}
From the intuition derived from studying the $T_\bath =0$ case in refs.~\cite{Almheiri:2019psf,Chen:2019uhq}, we expect the position of the QES after the Page time to satisfy
\begin{equation}\label{ass_order}
0< x^+_{\QES} - \tinf  <   \tinf-t \ll \tinf- x^-_{\QES}  \ll \tinf  \,.
\end{equation}
We will therefore solve the extremal equations~\eqref{QES_equations} by expanding around $t_\infty$. With the help of the approximations in eqs.~\eqref{eq:approx-simple}, we can approximate the integral
\begin{equation}\label{app_Ixm}
\begin{split}
&I\(x^-_{\QES }, x^-_{\QES};x^-_{\QES }\) \sim I_{\infty} - (\tinf -x^-_{\QES}) \partial_- I \(x^-_{\QES }, x^-_{\QES};x^-_{\QES }\)\\
&\sim \frac{2}{k}\(\( \pi T_1 t_\infty\)^2-1\) + (\tinf -x^-_{\QES}) \log\(\frac{\tinf -x^-_{\QES}}{\tinf}\)\(1-\frac{\Tb^2}{T^2_{\rm{eff}}(v_{\QES})}\)  \,.\\
\end{split}
\end{equation}
where for $\partial_- I\(x^-_{\QES }, x^-_{\QES};x^-_{\QES }\) \equiv \frac{d I \(x^-_{\QES }, x^-_{\QES};x^-_{\QES } \)}{d x^-_{\QES}}$, we used the approximation 
\begin{equation}
\begin{split}
& \int_0^{x^-_{\QES}} \, 2\(x^-_{\QES}-t\) \(\{u,t\} -2 \(\frac{\pi \Tb}{f'(u)}\)^2\)dt
\approx  \int_0^{x^-_{\QES}} \, \frac{\(\tinf-t+x^-_{\QES}-\tinf\)}{(\tinf -t)^2} \(1- \frac{ T^2_b}{T^2_{\rm{eff}}(u)}\)dt \,.
\end{split}
\end{equation}
However, it is not easy to perform the above integral of $t$ due to the appearance of time $u$. Instead, we apply the middle value theorem and find
\begin{equation}\label{eq:middlevalue}
\begin{split}
\partial_- I \(x^-_{\QES }, x^-_{\QES};x^-_{\QES }\)&\approx  \int_0^{x^-_{\QES}} \, \frac{1}{(\tinf -t)} \(1- \frac{ T^2_b}{T^2_{\rm{eff}}(u)}\)dt\\
&\approx -\(1- \frac{\Tb^2}{T^2_{\rm{eff}}(v_{\QES})}\) \log\(\frac{\tinf- x^-_{\QES}}{\tinf}\) \,,\\
\end{split}
\end{equation}
where $v_{\QES} \in \[0,y^-_{\QES} \]$ is referred to as the middle value for the $t$ integral from $0$ to $x^-_{\QES}$. Similarly, we can obtain the other integral $I\(x^+_{\QES }, x^+_{\QES};x^-_{\QES }\) $ by
\begin{equation}\label{app_Ixp}
\begin{split}
&\int_0^{x^-_{\QES}} \(  (\tinf -x^+_{\QES})^2 -2 (\tinf -x^+_{\QES})(\tinf-t) + (\tinf -t)^2 \) \(\{u,t\} -2 \(\frac{\pi \Tb}{f'(u)}\)^2\)dt\,,\\
&\sim (\tinf - x^+_{\QES}) \log\(\frac{\tinf -x^-_{\QES}}{\tinf} \)\(1-\frac{\Tb^2}{T^2_{\rm{eff}}(v_{\QES})}\) +  I_{\infty} -\frac{1}{2}(\tinf -x^-_{\QES}) \(1-\frac{\Tb^2}{T^2_{\rm{eff}}(y_{\QES}^-)}\) \,,
\end{split}
\end{equation}
where we have ignored the first integral at the order $\mathcal{O}((x^+_{\QES}-\tinf)^2)$, used mean value theorem for the second integral again with the {\bf same} middle value $v_{\QES}$ as before and considered the third integral as a function of $x^-_{\QES}$ with its Taylor expansion around $x^-_{\QES} \sim \tinf$ as
\begin{equation}
\begin{split}
\int^{x^-_{\QES}}_0 (\tinf -t)^2  \(\{u,t\} -2 \(\frac{\pi \Tb}{f'(u)}\)^2\)dt
\sim I_{\infty} - \frac{1}{2}(\tinf -x^-_{\QES}) \( 1 - \frac{\Tb^2}{T^2_{\rm{eff}}(y^-_{\QES})}\)\,.
\end{split}
\end{equation}
Although we can not decide the middle value for any $y^-_{\QES}$, it is easy to find $T_{\rm{eff}}(v_{\QES}) \sim T_1$ at the linear region with $ku \ll 1$.

Combining our assumptions \eqref{ass_order} and the approximations of the integrals in eqs.~\eqref{app_Ixm} and \eqref{app_Ixp}, we can approximate the equations for QES by much simpler forms
\begin{equation}\label{QES_simpler}
\begin{split}
&\frac{4\pi T_1}{\tinf -x^-_{\QES}} \( \pi T_1 \tinf + \frac{k}{4\pi T_1} \log\(\frac{\tinf -x^-_{\QES}}{\tinf} \)\(1-\frac{\Tb^2}{T^2_{\rm{eff}}(v_{\QES})}\) \) \approx \frac{2k}{x^+_{\QES}-t} \,,\\
& 4\pi T_1 \(\xp-\tinf\)\( \pi T_1 \tinf + \frac{k}{4\pi T_1} \log\(\frac{\tinf -x^-_{\QES}}{\tinf} \) \(1-\frac{\Tb^2}{T^2_{\rm{eff}}(v_{\QES})}\)   \) \approx \frac{k}{2}\(\tinf -x^-_{\QES}\)\Gamma_{\rm{eff}}(\ym)\,,\\
\end{split}
\end{equation}
where we have defined 
\begin{equation}
\begin{split}
\Gamma_{\rm{eff}} (\ym)&\equiv\(  1-\frac{\Tb}{T_{\rm{eff}}(y_{\QES}^-)} \)^2 \,,
\end{split}\label{loop1}
\end{equation}
and only keep the leading-order contributions. 
The non-negative coefficient $\Gamma_{\rm{eff}}$ approaches zero as the black hole reaches thermal equilibrium with the bath.
For later use, we also present the numerical plot for $\Gamma_{\rm{eff}}$ for various temperature in figure \ref{fig:Deltaeff}.

With the above equations, it is straightforward to find the solutions, \ie the location of QES
\begin{equation}\label{QES_xpxm}
\begin{split}
\xp&\approx  \tinf + \frac{\Gamma_{\rm{eff}} }{4- \Gamma_{\rm{eff}} }\(\tinf -t \) \,,\\
\xm&\approx \tinf -\frac{8\pi T_1}{k(4-\Gamma_{\rm{eff}} )}\( \tinf-t\)   \( \pi T_1 \tinf + \frac{k}{4\pi T_1} \log\(\frac{\tinf -x^-_{\QES}}{\tinf} \)\(1-\frac{\Tb^2}{T^2_{\rm{eff}}(v_{\QES})}\) \) \,.
\end{split}
\end{equation}
Assuming the time delay $u-\ym$ is not large (\ie $k(u-\ym) \ll 1$), one can
use the approximation
\begin{equation}
\begin{split}
\log \( \frac{\tinf - f(\ym)}{\tinf -f(u)}  \) &\approx -\frac{4\pi}{k} \( \Teff (u)- \Teff(\ym) - \Tb \log\(\frac{\Tb +\Teff(u)}{\Tb+\Teff(\ym)}\) -\frac{k}{2}(u-\ym)\)  \\
&\sim 2\pi \Teff (\ym) (u-\ym) 
\sim 2\pi \Teff (u) (u-\ym) \,,
\end{split}
\end{equation}
and further simplify the position of the QES to
\begin{equation}\label{QES_ym}
\begin{split}
\ym &\approx u - u_{\HP} \\
u_{\HP}&=\frac{1}{2\pi \Teff(u)} \log\[\frac{8 \pi T_1}{k(4-\Gamma_{\rm{eff}} (u))}  \( 1 + \frac{k}{4\pi T_1} \log\(\frac{\tinf -t}{\tinf} \)\(1-\frac{\Tb^2}{T^2_{\rm{eff}}(v_{\QES})}\)  \) \]\,,\\
\end{split}
\end{equation}
where the second term can be understood as the Hadyen-Preskill time, as will be explained later. It is noted that the time scale $u_{\HP}$ is not constant in general because the black hole is also dynamical. For $\Tb=0$, the solutions reduce to
\begin{equation}
\begin{split}
\ym (\Tb=0)&\approx u - \frac{1}{2\pi T_1 e^{-\frac{k}{2}u}} \log\( \frac{8\pi T_1 e^{-\frac{k}{2}u}}{3k}  \)\,,
\end{split}
\end{equation}
which is in agreement with the results of \cite{Almheiri:2019psf}.\footnote{Here we explicitly write the right side as a function of time $u$ which should be understood as the leading-order contribution. To be more precise, we can also use $e^{\frac{k}{2}\ym}$ rather than $e^{\frac{k}{2}u}$.} Similarly to the zero temperature bath case, the QES moves towards the horizon at $x^+=\tinf$. However, we want to stress the importance of the role of the non-zero factor $\Gamma_{\rm{eff}}$ that captures the speedup of the equilibration process because the thermal bath also emits radiation to the AdS region when $\Tb \neq 0$.

Furthermore, we can also compare our new solutions with the explicit and linear solution found in \cite{Chen:2019uhq}. Focusing on the Page transition within the linear region, we can further simplify the results for the position of QES and explicitly obtain
\begin{equation}
\begin{split}
\xp&\approx  \tinf + \frac{\Gamma_0}{4- \Gamma_0}\(\tinf -t \) \,, \qquad \Gamma_0=\( 1-\frac{\Tb}{T_1} \)^2 \,, \\
\ym &\approx
u -\frac{1}{2\pi T_1} \log\( \frac{8\pi T_1}{k(4-\Gamma_0) }\) \,,\qquad  ku \ll 1\ (\text{linear region}) \,,
\end{split}
\end{equation}
where we can rewrite the time delay
\begin{equation}
u_{\mt{HP}}(ku \ll 1)=\frac{1}{2\pi T_1} \log\( \frac{8\pi T_1}{k\(4-(1-\Tb/T_1)^2\)} \) \,,
\end{equation}
 as the Hadyen-Preskill time in linear region. It is also easy to check that the above results are reduced to the linear results presented in \cite{Chen:2019uhq} by setting $\Tb=0$, \ie $\Gamma_{\rm{eff}}=1$.

 After finding the position of the QES, we are able to consider the evolution of the generalized entropy \eqref{Sgen_QESx1}. The generalized entropy is dominated by the classical area term from the dilaton
 \begin{equation}
 \begin{split}
\phi & \approx 2\bar{\phi}_r  \left( \frac{1-(\pi T_1)^2 x^+_{\QES}x^-_{\QES} + \frac{k}{2}I\(\tinf,x^-_{\QES};x^-_{\QES}\)}{\tinf-x^-_{\QES}} \right)\(1 - \frac{\xp -\tinf}{\tinf -\xm } \)\\
&\sim \frac{2\bar{\phi}_r}{\tinf - \xm}   \left[   1-\big(\pi T_1 \tinf\big)^2 - (\pi T_1)^2 \tinf \big(\xp -\tinf \big) - (\pi T_1)^2 \tinf (\xm -\tinf)  \right.\\
&\quad \left. + \frac{k}{2}\(  I_{\infty} +\frac{(\tinf -\xm)}{2}\(1-\frac{\Tb^2}{T^2_{\rm{eff}}(v_{\QES})}\) \log\( \frac{\tinf -\xm}{\tinf}\)  \)  \right] \\
&\sim   \bar{\phi}_r \( 2(\pi T_1)^2 \tinf  +\frac{k}{2}\(1-\frac{\Tb^2}{T^2_{\rm{eff}}(v_{\QES})}\) \log\( \frac{\tinf -\xm}{\tinf}\)  - \frac{k \Gamma_{\rm{eff}}(y_{\QES}^-)}{4}\) \,,
 \end{split}
 \end{equation}
which is approximated by the value of dilaton on the horizon at $\xp = \tinf$. Recall that $\Gamma_{\rm{eff}}(y_{\QES}^-)$ is given in eq.~\reef{loop1}.  Comparing the area term (without divergences associated with short range entanglement)
 \begin{equation}\label{Sphi}
 S_{\phi_{{\QES}}} = \frac{\phi}{4 \GN} \sim  \frac{c}{12 k}\( 2\pi T_1 +\frac{k}{2}\(1-\frac{\Tb^2}{T^2_{\rm{eff}}(v_{\QES})}\) \log\( \frac{\tinf -\xm}{\tinf}\)  \) \,,
 \end{equation}
 with the time delay in position of QES, \ie eq.~\eqref{QES_ym}, we can rewrite the time shift as
 \begin{equation}\label{eq:HPtime}
 u_{\HP} \approx \frac{1}{2\pi T_{\mt{eff}}(u) } \log \( \frac{S_{\phi_{\QES}}}{c}  \)  + \mathcal{O}(1) \approx  \frac{1}{2\pi T_{\mt{eff}}(u) } \log \( \frac{S(u)-S_0}{c}  \) \,,
 \end{equation}
 where we have restored the extremal entropy $S_0 \equiv \frac{\phi_0}{ 4 \GN}$ (see \eqref{eq:action}) in the complete entropy  $S(u)$ of our dynamical black hole. Here we can explain the entropy $S(u)$ as the density of state at time $u$ and take $S_0$ as the ground state entropy associated with the value $\phi_0$. As discussed in \cite{Almheiri:2019psf}, the time delay $u_{\mt{HP}}$ appearing in $y_{\QES}^-$ can be understood as the Hayden-Preskill time. 

 \subsubsection{Page transition}
 
 \begin{figure}[htbp]
 	\centering\includegraphics[width=5.6in]{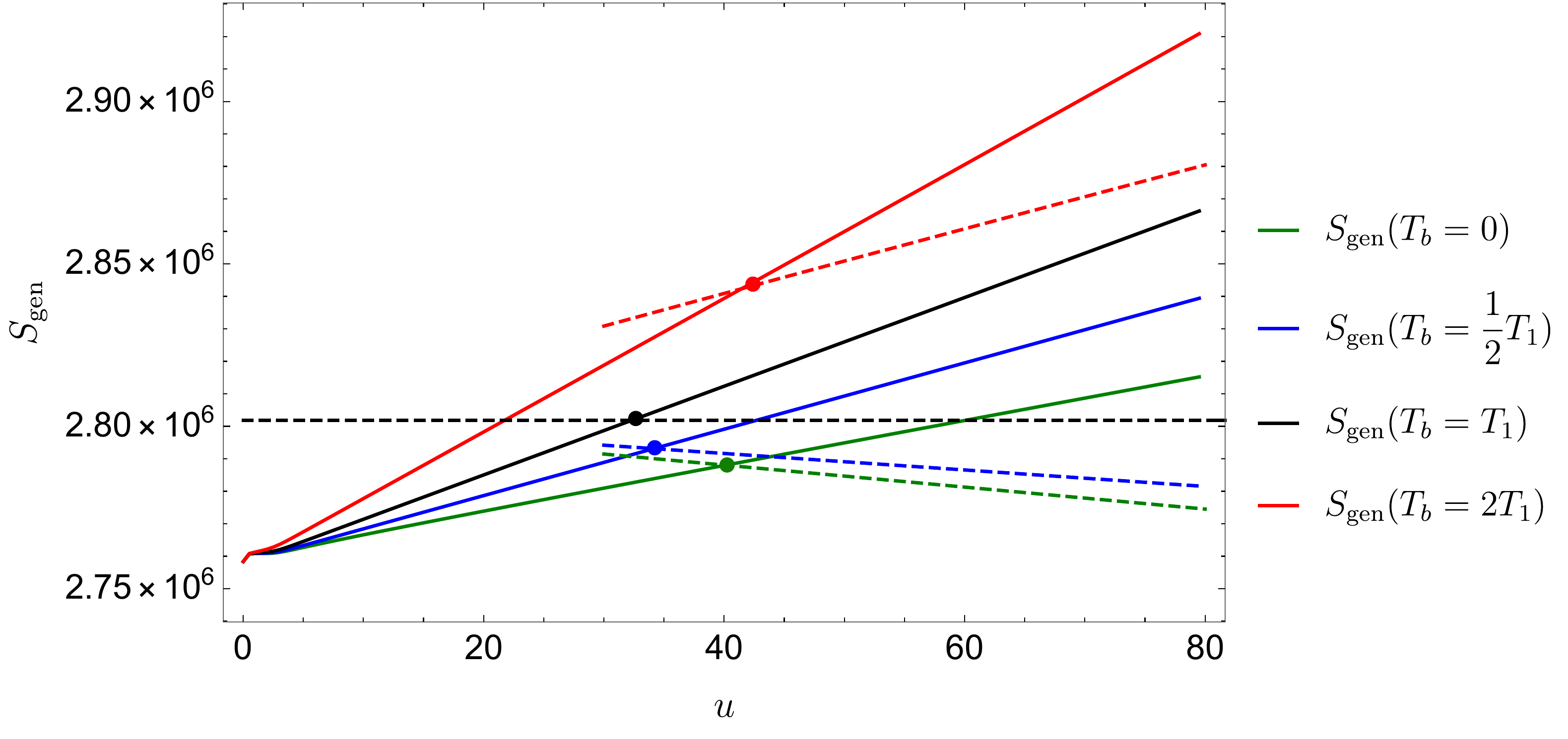}
 	\caption{The Page curve of generalized entropy around Page transition from scrambling phase to late-time phase for different bath temperatures. The solid lines represent the analytical results at the scrambling phase and the dashed lines indicate the numerical results for the late-time phase which are also approximated by solutions \eqref{QES_xpxm} and their approximate generalized entropy \eqref{Sgen_late_linear}. Note that the black dashed line shows the generalized entropy at equilibrium case, which is the constant given in eq.~\eqref{SgenT1}. }\label{fig:Sgen3T1_pagetime}
 \end{figure}

 The subleading term of the generalized entropy is the bulk entropy
 \begin{equation}
 \begin{split}
  \frac{4\GN}{\bar{\phi}_r}S_{\rm{bulk}} &=\( 2k \log\( \frac{2}{\epsilon} \frac{\sinh\(\pi \Tb\(u-y_{\QES}^-\)\)}{\pi \Tb}  \frac{(x^+_{\QES}-t)}{(x^+_{\QES}-x^-_{\QES})}\sqrt{\frac{f'(y^-_{\QES})}{f'(u)}} \) \) \\
  &\sim 2k \( \log \( \frac{8}{(4-\Gamma_{\rm{eff}})\epsilon} \frac{\sinh\(\pi \Tb u_{\mt{HP}}\)}{\pi \Tb} \)  - \pi T_{\rm{eff}} u_{\HP}+ \frac{k u_{\HP}}{4 T_{\rm{eff}}^2} (T_1^2- \Tb^2) e^{-ku}  \) \,,
 \end{split}
 \end{equation}
 which remains constant in the linear region with $ku \ll 1$.
As expected, it also reproduces the results in \cite{Chen:2019uhq} after fixing $\Tb=0$. In order to derive the Page time, we can also explicitly write generalized entropy at late time phase in the {\bf linear region}
\begin{equation}\label{Sgen_late_linear}
\begin{split}
 S_{\gen,\rm{late}} (ku \ll 1)&\approx \frac{\bar{\phi}_r}{4\GN}  \left[ 2\pi T_1 -k \pi T_1\(1-\frac{\Tb^2}{T_1^2}\) (u-u_{\HP})   \right. \\
& \left.+2k \log \( \frac{8}{(4-\Gamma_0)\epsilon} \frac{\sinh\(\pi \Tb u_{\mt{HP}}\)}{\pi \Tb}\)- 2k\pi T_{\rm{eff}} u_{\HP}   + \mathcal{O}(k^2 \log k)  \right]\,,
\end{split}
\end{equation}
which displays linear decrease (increase) of the generalized entropy after the Page transition for $\Tb < T_1$ ($\Tb > T_1$). Given the fact that the time delay $u_{\HP}$  is a constant when $T_\bath=T_1$, it is obvious that the entropy $ S_{\gen,\rm{late}} $ also reduces to a constant when $T_\bath = T_1$ (see eq.~\eqref{QES_ym}), which is the same as the result derived in eq.~\eqref{SgenT1} for the equilibrium case. Shortly before the Page time, we can obtain the generalized entropy in the scrambling phase with $1 \ll \pi T_1 u $ and $ku \ll 1$ by 
 \begin{equation}\label{Sgen_scrambling_linear}
 \begin{split}
&S_{\gen,\rm{scrambling}} \approx \frac{\bar{\phi}_r}{4\GN}  \( 2\pi T_0 + 2k \log\( \frac{24\pi E_s}{\epsilon c} \frac{\sinh \pi (\Tb u)}{\pi \Tb} \frac{1}{\sqrt{f'(u)}}\) + \kappa \) \\
&\approx \frac{\bar{\phi}_r}{4\GN}   \( 2\pi T_0 +2k \pi (\Tb+T_1) u    + 2k \log\( \frac{12 E_s}{\epsilon c \Tb}\)  + \frac{k^2 u}{2}\(1- \frac{\Tb^2}{T_1^2}\)(1-u\pi T_1)  + \kappa \) \,.
 \end{split}
 \end{equation}
 where we put all other contributions in $S_{\rm{Bulk}}$ into  $\kappa$ which approaches a constant when $t =f(u) \to t_{\infty}$.\footnote{The analysis for the scrambling phase is similar to that in \cite{Chen:2019uhq}. See the section 2.2.2 for more details.}
The leading-order terms of the generalized entropy in the scrambling phase~\eqref{Sgen_scrambling_linear} are a constant related to the entropy of the original black hole and two linearly increasing terms, \ie $2k \pi (T_1 + T_\bath)$, due to the entanglement of radiation escaping from the non-zero temperature bath and black hole, respectively. Because the temperature of black hole is approaching $T_\bath$, we will show later that the linear increase is replaced by $2k \pi (T_1 + T_1)$ term.\footnote{Technically, this is due to the approximations for $\log \frac{1}{\sqrt{f'(u)}}$ for different time regions, see \eqref{app_linear} and \eqref{app_late}.}
As a result, the generalized entropy at the scrambling phase increases indefinitely while that of the late time phase asymptotes to the entropy of a black hole with temperature $\Tb$, we expect there is a phase transition (Page transition) between them when $S_{\gen,\rm{scrambling}} =S_{\gen,\rm{linear}}$. This transition occurs at the Page time
 \begin{equation}
 \label{eq:pagetime}
 u_{\Page}(\Tb) \approx \frac{2}{4-\Gamma_0} \frac{T_1- T_0}{kT_1} + \frac{1- \frac{\Tb^2}{T_1^2}}{4-\Gamma_0} u_{\HP}+ \mathcal{O}(1)\,,
 \end{equation}
which decreases with the increase of $\Tb$ for $\Tb < T_1$, reaches a minimum at $\Tb=T_1$ and then increases for larger $\Tb$. In contrast, the generalized entropy at the Page time
 \begin{equation}\label{Sgen_Page}
 S_{\gen}(u_{\Page}) \approx \frac{\bar{\phi}_r}{4\GN}  2\pi \( T_0 + \frac{2(T_1-T_0)}{3- \frac{\Tb}{T_1}} + \cdots\)\,,
 \end{equation}
 increases with the increase of $\Tb$. These linear behaviors are explicitly shown in the figure \ref{fig:Sgen3T1_pagetime}. As a comparison, we represent the Page transition at linear region with $\Tb \ge T_1$.

Lastly, we add that the expressions for the Page time in eq.~\eqref{eq:pagetime} and the Page entropy in eq.~\eqref{Sgen_Page} diverge for $\Tb\to 3T_1$. This is an artifact of approximating the coefficient $\Gamma_{\rm{eff}}$ defined in eq.~\reef{loop1} by $\Gamma_0= \( 1- \frac{T_\bath}{T_1} \)^2$. However, if we include the subleading terms in $\Gamma_\mathrm{eff}$, we find that both of these quantities remain finite. We return to discuss this point in section~\ref{sec:overheat}.

 \begin{figure}[htbp]
 	\centering\includegraphics[width=5.5in]{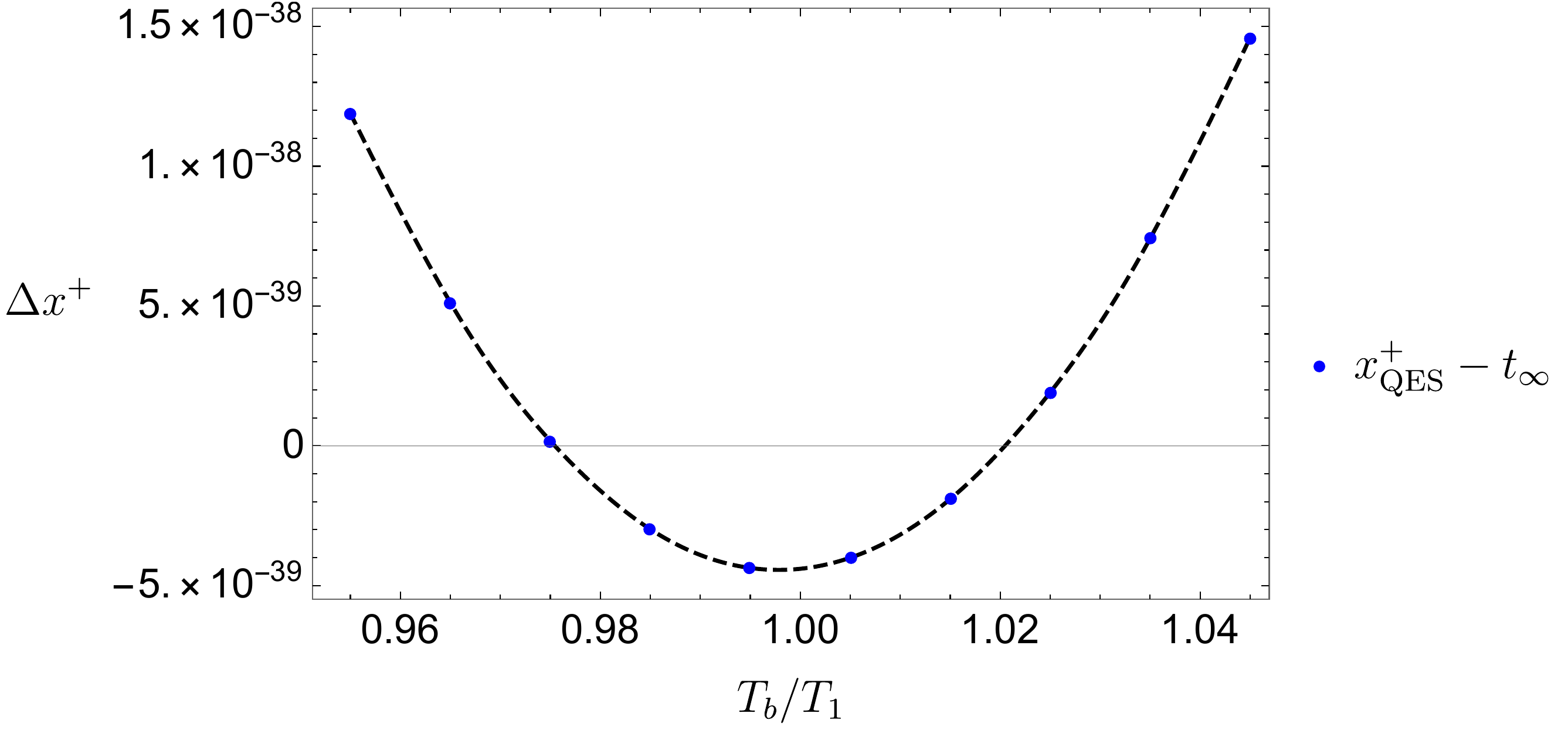}
 	\caption{The numerical results from solving QES equations for the deviation of QES from horizon, \ie $\( \xp -\tinf\)$, at a fixed time slice $u=40$ (after Page transition) with different bath temperatures $\Tb$.}\label{criticalT01}
 \end{figure}

 \subsubsection{Approach Equilibrium}\label{sec:approach-equil}
For the equilibrium situation studied in \cite{Almheiri:2019yqk}, the QES sat outside of the horizon, and resulting in part of the quantum extremal island being located outside the black hole.  The same behaviour was found in section \ref{sec:equilibrium} -- see eq.~\reef{eq:soot} -- where the bath temperature matches that of the black hole after it has absorbed the shockwave. Therefore in the present case where the two temperatures do not match, we still expect that as the black hole approaches its final equilibrium, \ie in the late time phase with $T_{\rm{eff}}\approx \Tb$,  the QES  will move outside of horizon at some critical temperature.\footnote{Similar behaviour was found in \cite{Hollowood:2020cou}.}

Ultimately, we wish to track the position of the QES for a black hole as a function of boundary time $u$ starting with a temperature $T_1$ and the bath at some fixed temperature $\Tb$. However, the analysis is simplified by asking how with fixed $u$ and $T_1$, the position of the QES moves as we vary the bath temperature $\Tb$.
With this approach, we can find different phases according to the position of QES as a function of the  boundary time $u$
 \begin{itemize}
 	\item Inside horizon \,\qquad $T_{c_1}(u) < \Tb < T_{c_2}(u)$\,,
 	\item On the horizon \qquad $\Tb = T_{c_1}(u)$ or $\Tb = T_{c_2}(u)$\,,
 	\item Outside horizon \qquad $\Tb < T_{c_1}(u)$ or $\Tb >T_{c_2}(u)$ \,,
 \end{itemize}
where the critical temperatures $T_{c_1}(u)$ and $T_{c_2}(u)$ will be derived in the following -- see eq.~\eqref{eq_Tc}.

If we extend the position \eqref{QES_xpxm} of the QES  to the equilibrium case with $\Gamma_{\rm{eff}}=0$, we find the QES is located on the horizon at $x^+=\tinf$, which is not what we found in section~\ref{sec:equilibrium}. Recalling the simplified solutions for QES \eqref{QES_simpler}, it is obvious that the non-negative term on the right-hand side, \ie
\begin{equation}
\frac{k}{2}\(\tinf -x^-_{\QES}\)\Gamma_{\rm{eff}}(\ym)\,,
\end{equation}
 implies we always have $\xp \ge  \tinf$. The solution to this puzzle is simple: all the approximations used in the previous analysis for the QES are based on the assumptions in eq.~\eqref{ass_order}, which are invalid when $\Tb$ is extremely near $T_1$. Technically speaking, it is traced back to the fact that
$\Gamma_{\rm{eff}}=\(1-\frac{\Tb}{T_{\rm{eff}}}\)^2$ around this narrow region suppresses the leading-order contribution. In order to find the critical temperature for the transition of QES, we need to track the (some) sub-leading contributions which compete with the leading-order terms when $\Gamma_{\rm{eff}} \sim k$. Although it is not easy to perform the integral $I$ to next order, we can determine these corrections by perturbing from the equilibrium case at $\Tb=T_1$ because the critical temperature should satisfy $T_1 - T_{c} \sim \sqrt{k}$. In other words, we can approach the critical temperature from regular $\Tb$ and from $T_1=\Tb$ and look for all necessary corrections.

Instead of directly solving the QES equations for the equilibrium case, we can approximate the two equations \eqref{eq_QES_equilibrium}  with $x^\pm_1=t$ by
\begin{equation}\label{eq_QES_equilibrium02}
\begin{split}
4\pi T_1 (\tinf -\xm)\( 1 + \frac{\xm- \tinf}{\tinf} \) &\approx 2k \left(\tinf-\xm\right)^2 \( \frac{1}{\xp-t} - \frac{1}{\tinf-\xm} \) \,,\\
\longrightarrow \qquad \xp -t &\sim \frac{k}{2\pi T_1} \( \tinf -\xm\) \,,
\end{split}
\end{equation}
and
\begin{equation}
\begin{split}
 4\pi T_1 \(\xp -\tinf \) &\approx 2k\left(\xp- \xm\right)^2 \left(\frac{1}{\tinf-x^-} -\frac{\xp -\tinf}{(\tinf-\xm)^2}-\frac{1}{\tinf-x^-} + \frac{t-\tinf}{(\tinf-\xm)^2}\right)  <0  \\
 &\approx 2k\( t -\xp \) + 4k \frac{\xp-\tinf}{\tinf-\xm}(t-\xp) \\
\longrightarrow \qquad \xp -\tinf &\sim \frac{k}{2\pi T_1} \( t -\xp\) \longrightarrow \xp  \sim \tinf -\frac{k}{2}\( \tinf -t\) \,, \\
\end{split}
\end{equation}
where the leading-order contribution $\frac{(\tinf -\xm)^2}{\tinf -\xm}$ (positive) vanishes and we have to keep the next order correction $t-\xp$ (negative). The fact that the sub-leading term has the opposite sign to the (almost vanishing) leading term is what positions the QES outside the horizon, in contrast with the cases when $\Tb$ is not perturbatively close to $T_1$. From this lesson, we also need to keep that correction for $\Gamma \sim k$ where the leading order is competing with the sub-leading order. Adding this correction to \eqref{QES_simpler}, we need to correct the right side of the second equation by
\begin{equation}
\frac{k}{2}\(\tinf -x^-_{\QES}\)\Gamma(\ym) \quad \longrightarrow \frac{k}{2}\(\tinf -x^-_{\QES}\)\Gamma(\ym) + 2k\(t-\xp \) \,,
\end{equation}
and arrive at
\begin{equation}\label{QES_corrections}
\frac{2}{\xp -t}\(\xp-\tinf \) \approx  \frac{1}{2} \( 1- \frac{\Tb}{\Teff (\ym)} \)^2 - \frac{k}{\pi T_1} \,,
\end{equation}
which now agrees with the results of section~\ref{sec:equilibrium}.

Comparing with the results of section~\ref{sec:equilibrium}, we can interpret the $\Gamma$ term as a correction from equilibrium results, which is reinforced by the fact that $\Gamma$ approaches zero as the system thermalizes. We can expect that further corrections we missed should be only at the order $\mathcal{O}(\Gamma\times k) \sim k^2$.
The critical temperatures of $\Tb$ for which the QES changes position with respect to the event horizon are given by
\begin{equation}\label{eq_Tc}
\begin{split}
T_{c_1} (u) &\approx \(1 - \sqrt{\frac{2k}{\pi T_1}} \) \Teff(\ym)\,,\\
T_{c_2} (u) &\approx \(1 + \sqrt{\frac{2k}{\pi T_1}} \) \Teff(\ym) \,,
\end{split}
\end{equation}
which define a small region of temperatures where the QES is located outside the horizon. 

Lastly, we mention that since $T_\mt{eff}$ approaches $\Tb$ as the system thermalizes, even when $\Tb$ is far from $T_1$, the QES will eventually cross the event horizon for late enough times. This is to be expected since, as we claimed before, for $ku\to \infty$, the system behaves as the equilibrium case studied in section~\ref{sec:equilibrium}. Indeed, when 
\begin{equation}\label{eq:outhorizon01}
ku \gtrsim\log\left(\left|1-\frac{T_1^2}{\Tb^2}\right| \sqrt{\frac{\pi T_1}{8k}}\right)\,,
\end{equation}
the QES is located outside the event horizon. By these times, the effective temperature is very close to the bath temperature $T_\mathrm{eff} \approx \Tb \(1 \pm \sqrt{\frac{2k}{\pi T_1}}\)$, where the sign is determined by whether $\Tb$ is greater or smaller than $T_1$, and the correction parameter  $\Gamma_\mathrm{eff} \approx \frac{2k}{\pi T_1}$ is perturbatively small. For bath temperatures that are very close to $T_1$,
\begin{equation}\label{eq:outhozion}
\frac{|\Tb-T_1|}{T_1}  \lesssim  \sqrt{\frac{2k}{\pi T_1}} e^{\frac{T_1-T_0}{2T_1}}\,,
\end{equation}
the QES is already outside of the event horizon by the Page time in eq.~\eqref{eq:pagetime}.

\subsubsection{Overheated black holes}\label{sec:overheat}
\begin{figure}[htbp]
	\centering\includegraphics[width=5.0in]{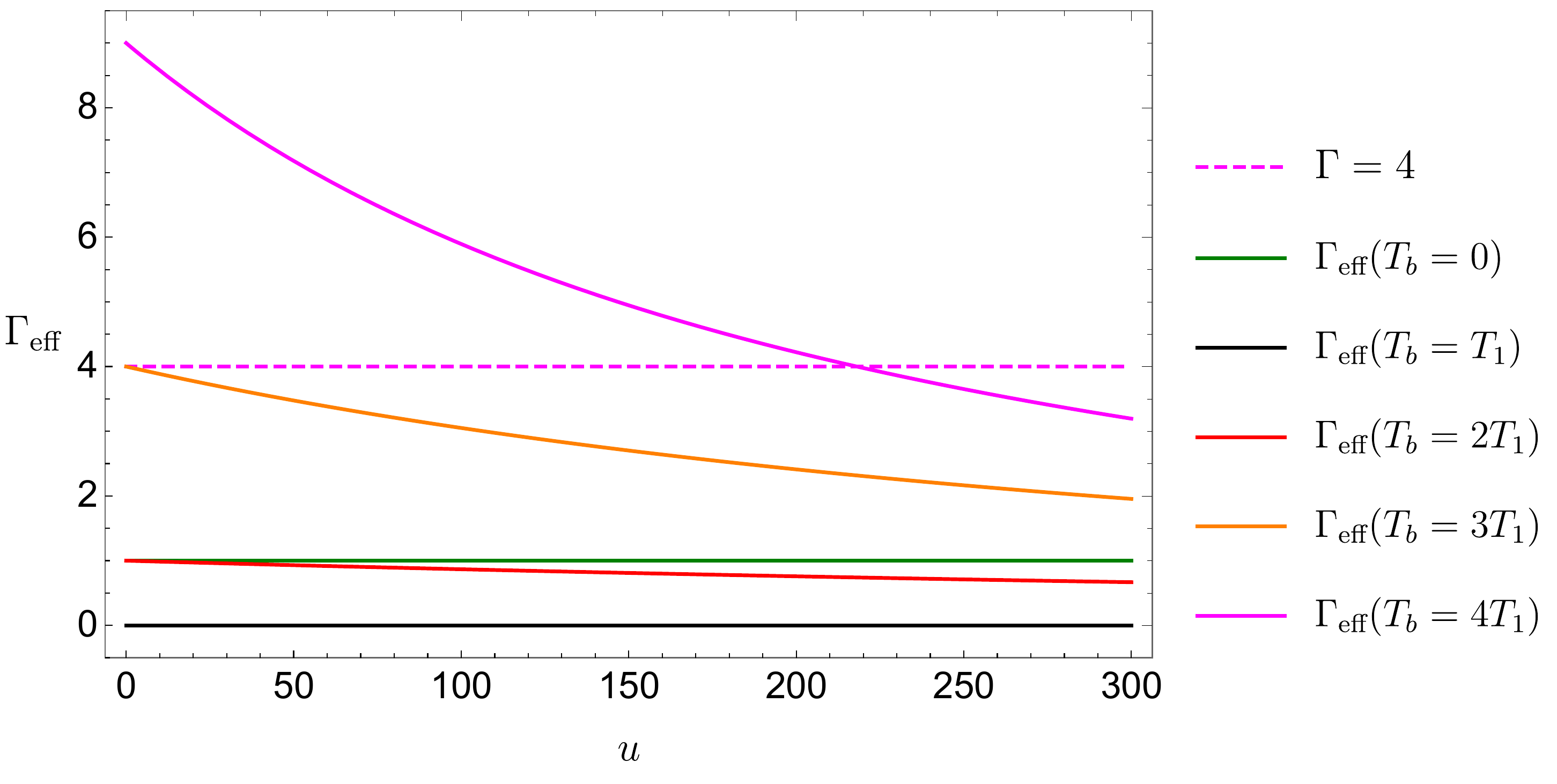}
	\caption{The time evolution of function $\Gamma_{\rm{eff}}(u)$ for various bath temperature.}\label{fig:Deltaeff}
\end{figure}

In the previous subsections, we derived the leading order expressions of the position of QES and discussed the importance of the subleading corrections when the bath temperature approaches $T_1$ with $\Gamma \sim k \sim 0$ because $\xp -\tinf$ changes its sign after the transition point at $T_{c_1}$ and $T_{c_2}$. Although we claim our previous approximations apply for arbitrary temperatures, it is obvious that our solution \eqref{QES_xpxm} appears singular at $\Gamma =4$ and further it appears the sign of $\xp-\tinf$ changes. It may appear that we have to consider next order corrections at another ``critical temperature'', \ie
\begin{equation}
\Tb = 3 T_1 \,,\quad \text{with} \qquad \Gamma_0 = \(1- \frac{\Tb}{T_1}\)^2=4\,.
\end{equation}
However, this is incorrect. The next order of correction can not help to solve this problem. Aside from $\xp$, the solutions for $\ym, \xm$ (see eqs.~\eqref{QES_xpxm} and \eqref{QES_ym}) show more problems because they are not well-defined when $\Gamma \ge 4$. At linear order, the generalized entropy in the late-time phase of the overheated black holes increases very rapidly, as can be seen by the coefficient of the linear term (see \eqref{Sgen_late_linear})
\begin{equation}
k\pi T_1 \( \frac{\Tb^2}{T_1^2} -1 \) \,.
\end{equation}
This rate of increase in generalized entropy may appears larger than that in the scrambling phase where the speed is dominated by linear term (see \eqref{Sgen_scrambling_linear})
\begin{equation}\label{eq:Sgenscra_slope}
2k \pi (T_1 +\Tb) u  \quad  (ku \ll 1)\,, \qquad \text{or}  \quad 4k \pi \Tb u    \quad (e^{ku} \gg 1)\,,
\end{equation}
where one contribution of ($2k\pi \Tb u$) comes from the radiation from bath and the other ($2k\pi T_1u$ and $2k\pi \Tb u$) from the black hole (for which $T_{\mt{eff}} \sim T_1$ at early times and  $T_{\mt{eff}} \sim \Tb$ for late times). One may wonder whether that means we can find a critical temperature $T_c$ above which a Page transition doesn't occur because the generalized entropy in late-time phase increases faster than that from the scrambling phase. The answer is again no.

\begin{figure}[htbp]
	\centering\includegraphics[width=5.6in]{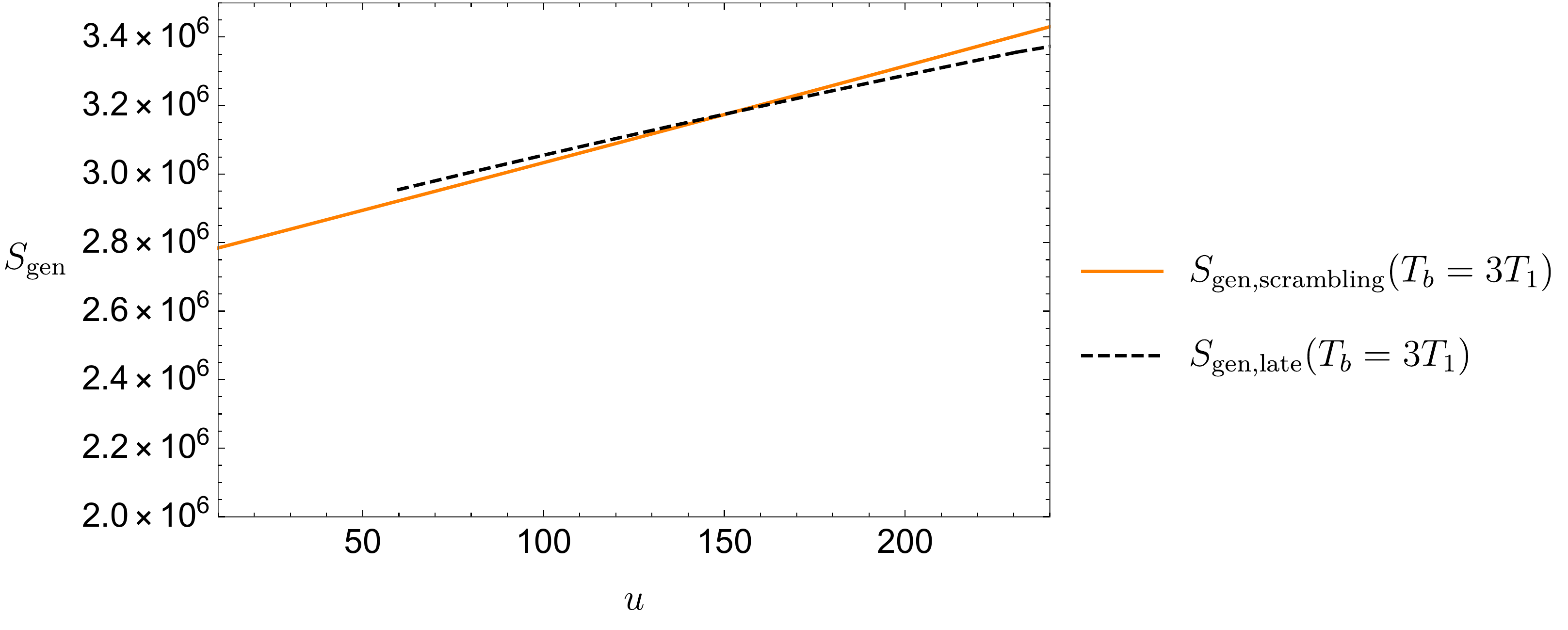}
	\caption{The Page transition with ``critical" bath temperature at $\Tb=3T_1$}\label{fig:Sgen3T1}
\end{figure}

All of the above questions or puzzles are actually due to the invalidity of the leading-order contributions in the linear region for overheated black holes. Our complete solutions are valid for arbitrary temperatures $\Tb$ outside of the critical region close to $T_1$ discussed in the previous section where subleading terms become important. One key ingredient to consider is that $\Gamma_{\rm{eff}}$ approaches zero with time, see the figure \ref{fig:Deltaeff}. For example, we always have $\Gamma_{\rm{eff}} (\Tb =3T_1) < 4$ for $u>0$. So there is no such new critical temperature at $\Tb =3T_1$. Another important fact is the delay of Page time with an increase of $|\Tb-T_1|$. Compared to the Page time at $\Tb=T_1$, the Page time with $\Tb > \#\,T_1$ is pushed to a later time that guarantees we have $\Gamma_{\rm{eff}}\(u=u_{\Page}\) < 4$. One might also wonder whether this time delay is really physical and why we should have a  restriction on the initial time for the solutions at late-phase. Let's remark this restriction is reminiscent of what we have seen in the zero bath-temperature case and also the equilibrium case. More explicitly, the equilibrium case also presents this similar restriction on time $u$, \ie the inequality \eqref{equi_inequalities}. The final ingredient that prevents the late time solutions in eq.~\eqref{QES_xpxm} from becoming singular is the high bath-temperature itself because it creates a new and large coefficient ${\Tb^2}/{T_1^2}$ that enhances the next-order corrections to the linear region. For example, we can see those effects from the expansion of $\Gamma_{\rm{eff}}$, \ie
\begin{equation}
\begin{split}
\Gamma_{\rm{eff}} (u)&=\(  1-\frac{\Tb}{T_{\rm{eff}}(u)} \)^2
\approx \( 1-\frac{\Tb}{T_1} \)^2 -\frac{ \Tb (T_1-\Tb)^2 (T_1+\Tb)}{T_1^4}ku + \cdots \,,
\end{split}
\end{equation}
where the second order correction cannot be simply ignored for large $\Tb/T_1$. To verify that there is no divergence, we show the Page transition using numerics for the ``critical temperature'' $\Tb=3T_1$ in figure \ref{fig:Sgen3T1}. We also compare the position of the QES using our approximation \eqref{QES_xpxm} with numerical results and they fit well as for the small $\Tb$ cases. 
\subsubsection{Page Curve and Thermalization}

\begin{figure}[htbp]
	\centering\includegraphics[width=4.0in]{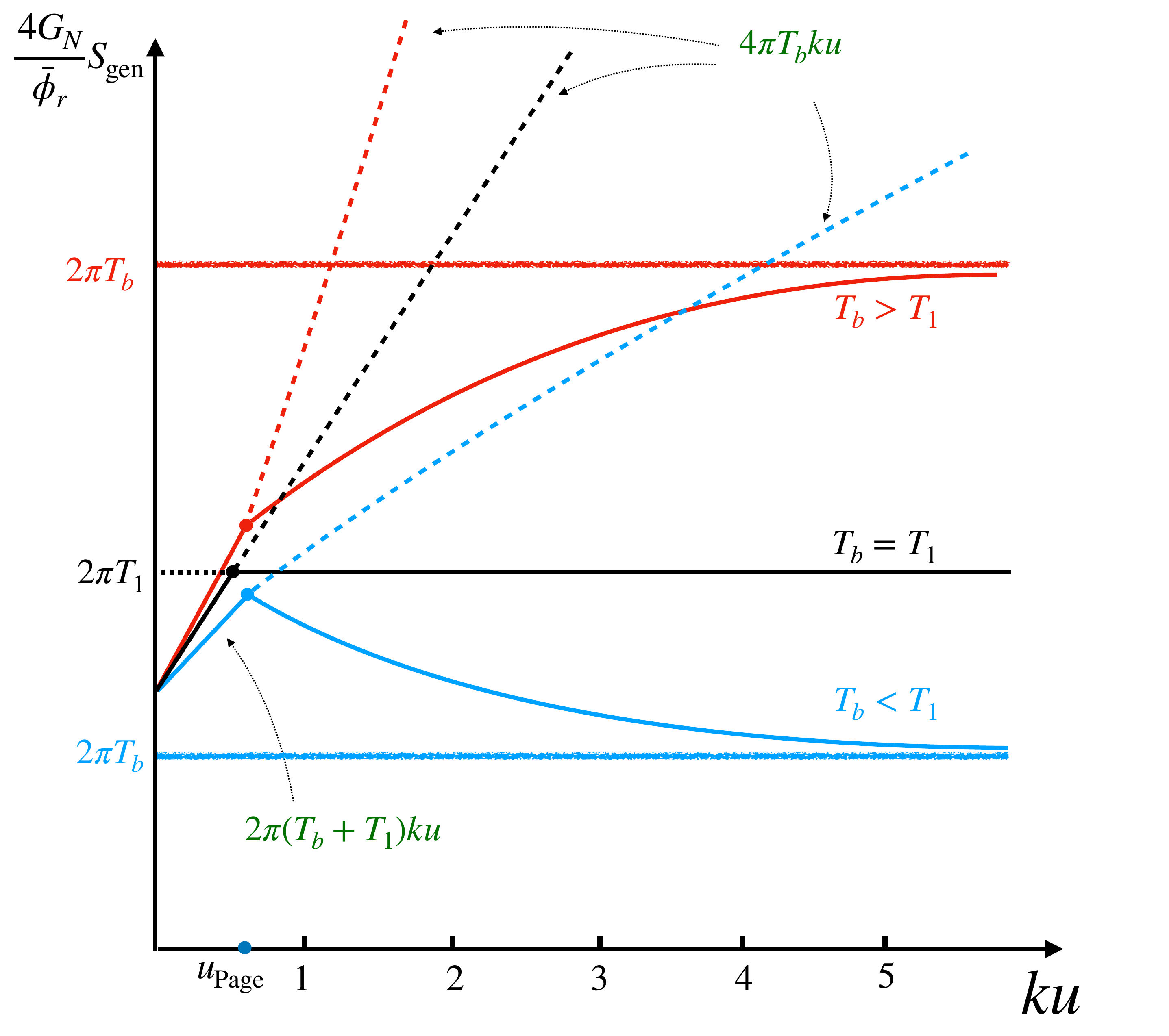}
	\caption{The schematic diagram for Page curve of black hole coupled with a thermal bath at different temperatures. The red, black, and blue solid lines show the Page curve for a growing black hole with $\Tb> T_1$, an external black hole at equilibrium status with $T_\bath = T_1$, and an evaporating black hole with $\Tb < T_1$, respectively. The corresponding dashed lines present the generalized entropy at the late-time region, whose behavior is dominated by the linear term $4\pi T_\bath ku$ as discussed around \eqref{eq:Sgenscra_slope}.}\label{fig:Page_curve}
\end{figure}

So far, we have focused on the evolution of the generalized entropy of the evaporating black hole up to times comparable with the Page time. As we will now show, we can also use the position of the QES in eq.~\eqref{QES_xpxm}, to find a full Page curve from $u=0$ to the late time regime with $e^{ku} \gg 1$. The expected behavior of the generalized entropy $S_{\gen,\rm{late}}$ at late times is that the subleading corrections slow down the linear decrease ($\Tb < T_1$) or increase ($\Tb>T_1$) of the generalized entropy, which will eventually approach a constant $S_{\gen,\rm{late}}(\Tb)$ corresponding to the entropy of a black hole with temperature $\Tb$, as derived in eq.~\eqref{SgenT1}.
However, we cannot simply substitute solution into the definition of generalized entropy to derive its time evolution due to the absence of approximation for the middle value $v_{\QES}$ at late times.

Instead of considering the generalized entropy itself, we can take the time derivative of $S_{\gen,\rm{late}}\(u, \xp,\xm\)$ as defined in \eqref{Sgen_QESx1}
\begin{equation}
\label{eq:derivative}
\begin{split}
\frac{4\GN}{\bar{\phi}_r}\frac{d S_{\gen,\rm{late}}}{du} = 2k\(-\pi \Tb\frac{\cosh\(\pi \Tb(\ym-u)\)}{\sinh \( \pi \Tb(\ym -u)\)}  - \frac{1}{2}\partial_u \( \log f'(u) \)  + \frac{f'(u)}{t-\xp}\) \,,\\
\end{split}
\end{equation}
where we have used the facts
\begin{equation}
 \frac{\partial S_{\gen,\rm{late}}}{\partial \xp} = 0\,, \qquad \frac{\partial S_{\gen,\rm{late}}}{\partial \xm} = 0 \,,
\end{equation}
from the definition of QES. The time derivative in eq.~\eqref{eq:derivative} can be further simplified by taking the limits
\begin{equation}
\begin{split}
\pi \Tb \coth \( \pi \Tb (u-\ym) \) &\approx \pi \Tb \,,\\
-\frac{1}{2}\partial_u \( \log f'(u) \)&\approx \pi \Teff(u) + k \frac{\Teff^2 - \Tb^2}{4\Teff^2} \,,\\
\frac{f'(u)}{t-\xp} &\approx -\frac{f'(u)}{\tinf -t} \frac{4-\Gamma_{\rm{eff}}}{4} \approx \pi \Teff(u) \frac{\Gamma_{\rm{eff}}-4}{2} \,,
\end{split}
\end{equation}
to obtain
\begin{equation}
\label{eq:derivapprox}
\frac{d S_{\gen,\rm{late}}}{du}  \approx - \frac{\bar{\phi}_r}{4\GN}  \( 1- \frac{\Tb^2}{\Teff^2(u)}\) k\pi\Teff(u)  \,.
\end{equation}
Taking a linear approximation of eq.~\eqref{eq:derivapprox} agrees with the results found in the previous section -- see eq.~\eqref{Sgen_late_linear}. Furthermore, since $\Teff$ approaches $\Tb$ for late times ($e^{ku} \gg 1$) the time derivative obviously decays to zero in this limit, implying the generalized entropy in the late-time region is indeed approaching a constant. We can then rewrite our generalized entropy at time $u$ in late time phase in integral form
\begin{equation}
S_{\gen,\rm{late}}(u) \approx S_{\gen}(u_{\Page}) - \frac{\bar{\phi}_r}{4\GN}  \int^u_{u_{\Page}}  \( 1- \frac{\Tb^2}{\Teff^2(\tilde{u})}\) k\pi\Teff(\tilde{u}) d\tilde{u} \,,
\end{equation}
where the start point is the generalized entropy at Page time $S_{\gen}(u_{\Page}) $ that has been derived at  \eqref{Sgen_Page} in the linear region. Fortunately, the above integral is fully analytical and can be performed to yield 
\begin{equation}\label{Sgen_late_analytical}
\begin{split}
S_{\gen,\rm{late}}(u) &\approx S_{\gen}(u_{\Page}) + \frac{\bar{\phi}_r}{4\GN}  (2\pi\, \Teff(u) -2\pi\, \Teff\!\(u_{\Page}\) )\,,\\
2\pi \Teff\(u_{\Page}\) &\approx 2\pi T_1 -k\pi T_1\( 1- \frac{\Tb^2}{T_1^2}\)u_{\Page}\,.
\end{split}
\end{equation}
The dominant term is nothing but the black hole entropy with temperature $\Teff$, \ie 
\begin{equation}\label{area}
S_{\gen,\rm{late}}(u) \sim \frac{\phi(u)}{4\GN} \sim \frac{2\pi \Teff(u) \bar{\phi}_r}{4\GN}    \,,
\end{equation}
 and the extra contributions from the leading order of the bulk entropy are all encoded in the value at Page time. Finally, combining with the generalized entropy at scrambling phase
 \begin{equation}
\begin{split}
&S_{\gen,\rm{scrambling}} (u\gg 1)\approx \frac{\bar{\phi}_r}{4\GN}  \( 2\pi T_0 + 2k \log\( \frac{24\pi E_s}{\epsilon c} \frac{\sinh \pi (\Tb u)}{\pi \Tb} \frac{1}{\sqrt{f'(u)}}\) + \kappa \) \,,\\
\end{split}
\end{equation}
we found the expected Page curve by taking eq.~\eqref{Sgen_late_analytical} as the generalized entropy for the late-time phase. After quench-phase, the generalized entropy is decided by that in the scrambling phase and then jumps to the late-time phase after Page time. Finally, we remark that the generalized entropy at scrambling phase $S_{\gen,\rm{scrambling}}$ represents the fine-graining entropy because its increase is dominated by the increase of entanglement entropy from these thermal radiations emitting from the thermal bath and black hole itself. Whereas, the generalized entropy at the late-time phase obviously denotes the coarse-graining entropy, \ie the area of the dynamical black hole, as shown in \eqref{area}.
As a summary, we show a diagram to presents the information about Page curve derived in the last several subsections in figure \ref{fig:Page_curve}.

In this subsection, we focused on the QES and generalized entropy of the subsystem consisting of QM$_{\mt{L}}$, the complete thermal bath, and its purification. Similar to the analysis in section \ref{sec:depression} for $T_1= T_\bath$ (see section 3.1 in \cite{Chen:2019uhq} for the case with zero bath temperature), we can consider a smaller subsystem by cutting a bath interval $[0, \sigma_1]$, corresponding to shifting the anchor point $x_1^\pm$ away from AdS$_2$ boundary into the bath with choosing $y_1^\pm = u \mp \sigma_1$. However, it is not easy to perturbatively solve the QES in general because our order assumption \eqref{ass_order} may break. Instead, we can begin with assuming another order condition
\begin{equation}
0< x^+_{\QES} - \tinf  <   \tinf-x_1^+ \ll \tinf- x^-_{\QES}  \ll \tinf  \,.
\end{equation}
Naively,  the above condition requires that we do not put the anchor point near the shock wave in order to guarantee $x_1^+ = f(u-\sigma_1) \approx \tinf$. In other words, we can generalize the approximations in this subsection to the case with $x_1$ near AdS$_2$ boundary. In most places, we only need to change $u, t$ to $u-\sigma_1, x_1^+$. Finally, one can obtain the corresponding QES as 
\begin{equation}\label{QES_xpxm_bath}
\begin{split}
\xp&\approx  \tinf + \frac{\Gamma_{\rm{eff}} }{4- \Gamma_{\rm{eff}} }\(\tinf -x_1^+ \) \,,\\
\xm&\approx \tinf -\frac{8\pi \Teff}{k(4-\Gamma_{\rm{eff}} )}\( \tinf-x_1^+\) \,,
\end{split}
\end{equation}
from which we can find the $y_{\QES}^-$ is shifted in the way of 
\begin{equation}
\ym \approx  u-\sigma_1 - u_{\HP} \,.
\end{equation}
Further, it is consistent with our numerical results and also the zero bath temperature case which is studied in section 3.1 of \cite{Chen:2019uhq}. 


\subsubsection{Simpler derivation of QES} 
In the above, we followed the analysis in \cite{Almheiri:2019psf,Chen:2019uhq} to derive the position of QES as shown in eq.~\eqref{QES_xpxm}. However, there is one undetermined middle value $v_{\QES}$ appearing in many expressions due to integral over the dilaton profile \eqref{eq:dilatonbefore}. Comparing the results in eqs.~\eqref{area} and \eqref{Sphi}, one finds the identify
\begin{equation}\label{identification}
2\pi \Teff(u)  \sim 2\pi \Teff(y^-_\QES) \sim   2\pi T_1 +\frac{k}{2}\(1-\frac{\Tb^2}{T^2_{\rm{eff}}(v_{\QES})}\) \log\( \frac{\tinf -\xm}{\tinf}\) \,, 
\end{equation}
where corrections of order $k$ are ignored. With the above approximation, we can simplify our results, \eg by using $\Teff (y^-_\QES)$ rather than $v_{\QES}$. To confirm our result, we can derive the position of QES in a more direct way. It is based on the observation in \cite{Moitra:2019xoj,Hollowood:2020cou} that the dilaton profile can be expressed without any integrals as\footnote{We thank Tim Hollowood and Prem Kumar for bringing this simple result to our attention. This expression is also discussed in detail in Section 3.9 in their recent paper \cite{Hollowood:2020cou}.}
\begin{equation}
\phi (x^\pm) = \phi_r\(  \frac{2f'(y^-)}{x^+-x^-} + \frac{f''(y^-)}{f'(y^-)} \)\,, 
\end{equation}
when the components of the stress tensor, $\langle T_{x^+x^+} \rangle$ and $\langle T_{x^+x^-} \rangle$ vanish.  Correspondingly, we can rewrite the solution \eqref{QES_equations} from extremizing the generalized entropy as 
\begin{equation}\label{QES_equations02}
\begin{split}
0&= \frac{1}{\phi_r}\frac{\partial \phi (x^\pm_\QES) }{\partial x^+_\QES} +2k\( \frac{1}{x^+_{\QES}-x^+_1}  - \frac{1}{x^+_{\QES}-x^-_{\QES}} \) \,,\\
0&= \frac{1}{\phi_r}\frac{\partial \phi (x^\pm_\QES) }{\partial x^-_\QES}+2k \(\frac{\pi \Tb}{\tanh\( \pi \Tb(y^-_{\QES}-y_1^-) \)} \frac{1}{f'\(\ym \)}+\frac{1}{x^+_{\QES}-x^-_{\QES}}  +\frac{1}{2}\frac{f''(y^-_{\QES})}{\(f'(y^-_{\QES})\)^2}\) \,.
\end{split}
\end{equation}
Noting the small $k$ expansion leads to the approximation $f'(y) \approx 2\pi \Teff (y) \( \tinf - f(y)\)$ and our ordering condition \eqref{ass_order}, it is straightforward to find the derivatives of dilaton are approximated by 
\begin{equation}
\begin{split}
\frac{1}{\phi_r}\frac{\partial \phi (x^\pm) }{\partial x^+_\QES} &\approx \frac{4\pi \Teff(y^-_\QES)}{\tinf -x^-_\QES} \,,\\
\frac{1}{\phi_r}\frac{\partial \phi (x^\pm) }{\partial x^-_\QES} &\approx
-4\pi \Teff(y^-_\QES) \frac{x^+_\QES -\tinf}{(\tinf -x^-_\QES)^2} + \frac{ T_{\mt{eff}}'(y^-_\QES)}{(\tinf -x^-_\QES)\Teff(y^-_\QES)} \,,
 \\
\end{split}
\end{equation} 
where we note the fact that $T_{\mt{eff}}'(y) =\partial_y \Teff (y)= -k\frac{T^2_{\mt{eff}} -T_\bath^2}{2 \Teff}$. Combining the above expressions with the approximations for the bulk entropy at $x_1^\pm=t$, one can find the QES is determined by 
\begin{equation}
\begin{split}
 4\pi \Teff(y^-_\QES)(x^+_\QES-t) \approx 2k(\tinf -x^-_\QES)\,,\\
4\pi \Teff(y^-_\QES)(x^+_\QES -\tinf) \approx \frac{k}{2} \Gamma_{\rm{eff}}(y^-_\QES) (\tinf -x^-_\QES)\,.
\end{split}
\end{equation}
which is exactly equivalent to our result in eq.~\eqref{QES_simpler} after substituting eq.~\eqref{identification}. Finally, we can find the position of QES  as
\begin{equation}
\begin{split}
 x^+_\QES &\approx   \tinf + \frac{\Gamma_{\rm{eff}} }{4- \Gamma_{\rm{eff}} }\(\tinf -t \) \,,\\
 x^-_\QES  &\approx \tinf -\frac{8\pi \Teff}{k(4-\Gamma_{\rm{eff}} )}\( \tinf-t\)  \,,
\end{split}
\end{equation}
which is the same as \eqref{QES_xpxm}. 
One can also easily find an approximation for dilaton profile
\begin{equation}
\phi (x^\pm_{\QES}) \approx \phi_r\(  \frac{2f'(y^-_\QES)}{\tinf-x^-} - 2\pi \Teff (y^-_\QES)\) \approx 2\pi \Teff(y^-_\QES)\phi_r \,, 
\end{equation}
where we ignored the derivative term $T_{\rm{eff}}'(y^-_\QES)$ as being order $k$.


\subsection{Information flow}\label{subsec:flow}
In the previous section, we studied the generalized entropy of QM$_\mt{L}$ plus the whole bath and its purification. (Of course, since the entire system is in a pure state, we could also think of this more simply as the entropy of \QMR.) In this section, we chop parts of the (purified) bath and discuss which intervals, together with QM$_\mt{L}$, are essential to having the ability to recover the information in the interior of the black hole. In contrast to the semi-infinite interval case studied in the previous section where the bulk entropy is described by the two-point function on the UHP, the generalized entropy instead has one more endpoint, \ie $x^\pm_2$ (or $y^\pm_2 = u\mp \sigma_2$) as the right end-point of the finite bath interval which can be used with QM$_\mt{L}$ to recover the black hole interior. According to the position of $x_2^\pm$ after or before the shock, we can divide the generalized entropy into two cases.

We begin by examining the case where the end-point $x_2^\pm$ is located after the shock, \ie $x^+_2 >0$ or $y^+_2=u-\sigma_2 >0$. Similarly to the equilibrium case studied in section~\ref{sec:equilibrium}, we have two competing channels. The N-channel (where the black hole interior is non-recoverable) has the QES at the bifurcation point $x^\pm_{\QES''}=\pm\frac{1}{\pi T_0}$, and the generalized entropy for this channel showing in figure \ref{fig:desvenlafaxine}c is given by 
\begin{equation}\label{SN_after}
\begin{split}
\frac{4\GN}{\bar{\phi}_r}S_{\mt{N}}(y^+_2 \ge 0) &\equiv \frac{4\GN}{\bar{\phi}_r} \( S^{\gen}_{\QES''} +S_{1-2} \) \\
&= 2\pi T_0 + 2k \log 2 + 2k \log\( \frac{1}{\epsilon^2}\frac{\sinh\( \pi \Tb(y^-_2 -y_1^-)\)}{\pi \Tb} \frac{x^+_1 -x_2^+}{\sqrt{f'(y^+_1) f'(y^+_2)} }\) \,, \\
\end{split}
\end{equation}
where $\phi_{\QES''}= 2\pi T_0\bar{\phi}_r$. When this channel is preferred, the entanglement wedge of the bath interval plus QM$_\mt{L}$ does not contain the interior of the black hole. The R-channel (where the interior is recoverable) instead has the QES at the same location as the late-time phase QES. Correspondingly, the generalized entropy for this R-channel corresponding to figure \ref{fig:desvenlafaxine}c reads
\begin{equation}\label{SR_after}
\begin{split}
\frac{4\GN}{\bar{\phi}_r}S_{\mt{R}}(y^+_2 \ge 0) &\equiv \frac{4\GN}{\bar{\phi}_r} \( S^{\gen}_{\QES-1} +S_{2} \) \\
&= \frac{\phi_{\QES}}{\bar{\phi}_r} + 2k \log\( \frac{2}{\epsilon} \frac{\sinh\(\pi \Tb(\ym -y^-_1) \)}{\pi \Tb}  \frac{x_1^+- \xp}{\xp-\xm}  \sqrt{\frac{f'(\ym)}{f'(y^+_1)}}\)\\
&\qquad \quad + 2k \log\( \frac{12 \pi E_s}{ c \epsilon} \frac{\sinh\(\pi \Tb y^-_2\)}{\pi \Tb}  \frac{x^+_2}{\sqrt{f'(y^+_2)}}\) \,.
\end{split}
\end{equation}
Evidently, when the R-channel is preferred, the entanglement wedge of the corresponding bath region plus the QM$_\mt{L}$ system includes the interior of the black hole. To find the transition where the bath interval (plus QM$_\mt{L}$) is able to reconstruct the black hole interior, we require  \ie
\begin{equation}
\frac{4\GN}{\bar{\phi}_r}\( S_{\mt{N}} - S_{\mt{R}} \) \ge 0\,,
\end{equation}
or equivalently
\begin{equation}\label{conditionAB_aftershock}
2k \log\(  \frac{c}{6\pi E_s \epsilon} \frac{(x^+_1 -x^+_2) \sinh \(\pi \Tb(y^-_2 - y^-_1)\)}{x^+_2\sqrt{f'(y^+_1)} \sinh(\pi \Tb y^-_2) }\) \ge \frac{4 \GN}{\bar{\phi}_r} S_{\gen,\rm{late}}(\sigma_1) -2\pi T_0 \,,
\end{equation}
where we rewrite the left side as the simple part because $S_{\gen,\rm{late}} $ with $\sigma_1=0$ has been discussed in the last section. We will focus on analyzing the left-hand side of eq.~\eqref{conditionAB_aftershock} in the following calculations. 

When the right end-point $y^+_2 \le 0$ is before the shock, we have two similar competing channels for the generalized entropy
\begin{equation}\label{eq:SNym}
\begin{split}
\frac{4\GN}{\bar{\phi}_r}S_{\mt{N}}(y^+_2 \le 0) &\equiv \frac{4\GN}{\bar{\phi}_r} \( S^{\gen}_{\QES''} +S_{1-2} \) \\
&= 2\pi T_0 + 2k \log 2 + 2k \log\( \frac{12 \pi E_s}{c\epsilon^2}\frac{x^+_1 \sinh\(\pi \Tb (-y^+_2)\)\sinh\( \pi \Tb (y^-_2 -y^-_1) \)}{(\pi \Tb)^2 \sqrt{f'(y^+_1)}} \) \,, \\
\end{split}
\end{equation}
and also
\begin{equation}
\begin{split}
\frac{4\GN}{\bar{\phi}_r}S_{\mt{R}}(y^+_2 \le 0) &\equiv \frac{4\GN}{\bar{\phi}_r} \( S^{\gen}_{\QES-1} +S_{2} \) \,.
\end{split}
\end{equation}
with 
\begin{equation}
\frac{4\GN}{\bar{\phi}_r} S_{2} = 2k \log\( \frac{1}{ \epsilon} \frac{\sinh\(\pi \Tb( y^-_2- y^+_2)\)}{\pi \Tb}  \) \,.
\end{equation}
The condition for the bath interval to have ability to reconstruct the interior of black hole when the right end-point is located before the shock is then given by
\begin{equation}\label{conditionAB_preshock}
2k \log\(  \frac{24\pi E_s }{c\epsilon} \frac{x^+_1 \sinh \(\pi \Tb(- y^+_2)\) \sinh \(\pi \Tb (y^-_2 -y^-_1)\)}{\sqrt{f'(y^+_1) }\pi \Tb\sinh \(\pi \Tb(y_2^- -y_2^+)\)}\) \ge \frac{4 \GN}{\bar{\phi}_r} S_{\gen,\rm{late}}(\sigma_1) -2\pi T_0 \,.
\end{equation}
Lastly, we remark that the N-channel and R-channel show the same divergence $2k \log \(\frac{1}{\epsilon^2}\)$, so the AdS cutoff $\epsilon$ is exactly canceled in the comparison and does not play an important role in the following calculations.

\subsubsection{Regularization of Shock Wave} \label{laxpair}
\begin{figure}[htbp]
	\includegraphics[width=0.49\textwidth]{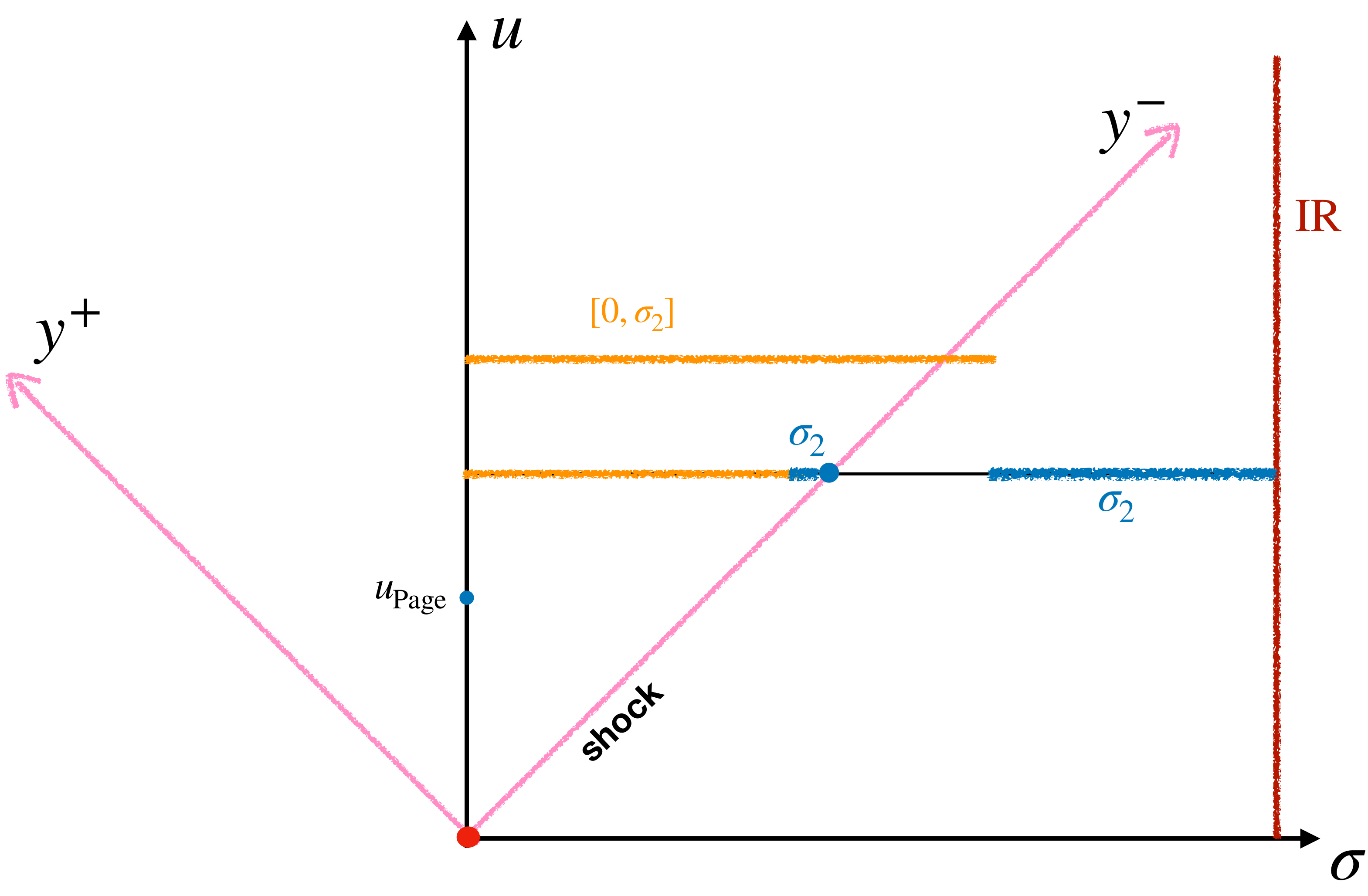} \hspace{0.01\textwidth}
	\includegraphics[width=0.49\textwidth]{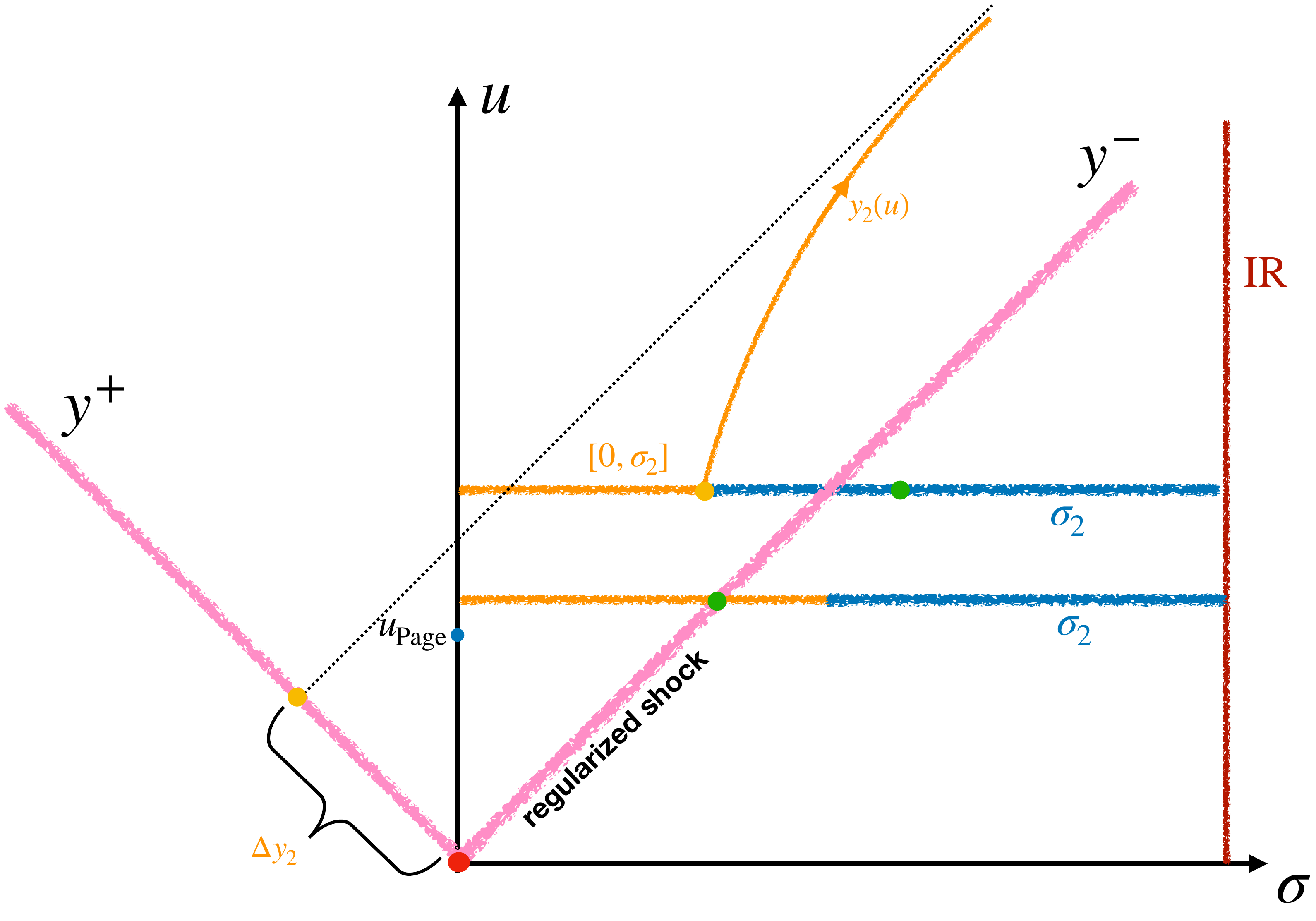}
	\caption{The yellow lines show the finite bath interval with $T_\bath \le T_p$ at a fixed time slice $u$ which has the ability to reconstruct the black hole interior with only including QM$_{\mt{L}}$ but not the purifier. The blue shadow region presents the expected region where we can put the endpoint of the finite bath interval, \ie $y_2$, and make the subsystem recover the information of black hole.
	Left:  The simple shock wave as a line. Right: The regularized shock wave as a small region indicated by the pink shadow. The yellow curve presents the endpoint  $y^+_2$ of the minimal bath interval approaches a constant $\Delta y_2$ derived in \eqref{finaly2plus} with the evolution of time.}
	\label{fig:bath_intervals}
\end{figure}

Before we discuss the condition for the finite bath-interval plus QM$_\mt{L}$ (as shown in the figure \ref{fig:desvenlafaxine}c) to reconstruct the black hole interior, we can roughly estimate the region for $y_2$ that makes the above equalities hold by looking at the divergence structure of $S_{\mt{N}}-S_{\mt{R}}$ with endpoint $y_2$ at special points. We will encounter an apparent paradox that will require a careful regularization for the region of the shock wave with the help of parameter $E_s/c$.

Explicitly, we can take the endpoint $y_2$ to the IR cut-off, \ie the limit $\sigma_2 \to \infty$.  It not hard to find the two competing channels show similar divergence
\begin{equation}\label{divergence_IR}
\begin{split}
\frac{4\GN}{\bar{\phi}_r}S_{\mt{N}}(y^+_2 \le 0) \quad &\longrightarrow \quad 4k \pi \Tb \sigma_2 + 4k \log \(\frac{1}{\epsilon \pi \Tb}\) \,,\\
\frac{4\GN}{\bar{\phi}_r}S_{\mt{R}}(y^+_2 \le 0) \quad &\longrightarrow \quad 4k \pi \Tb \sigma_2 + 4k \log \(\frac{1}{\epsilon \pi \Tb}\)\,,\\
\end{split}
\end{equation}
for thermal bath with nonzero $\Tb$. However, with the bath at zero temperature, the divergence takes a different form in these two channels
\begin{equation}
\begin{split}
\frac{4\GN}{\bar{\phi}_r}S_{\mt{N}}(y^+_2 \le 0) \quad &\longrightarrow \quad 4k \log\(  \frac{ \sigma_2}{\epsilon}\) \,,\\
\frac{4\GN}{\bar{\phi}_r}S_{\mt{R}}(y^+_2 \le 0) \quad &\longrightarrow \quad 2k \log\(  \frac{ 2\sigma_2}{\epsilon}\) +2k \log\(  \frac{\ell}{\epsilon}\) \,,\\
\end{split}
\end{equation}
where $\ell$ is some finite length-scale. This difference in the divergence structure makes the R-channel preferred when $y_2$ is around the IR cut-off in the bath and guarantees the purification of the thermal bath is not necessary for the interior reconstruction.

Now let us consider the limit of taking $y_2$ near the shock wave at $\sigma_{\mt{shock}}=u$, \ie $ y^+=0$. We have to consider approaching the shock from the region before the shock or after the shock. Under the limit $y^+_2 \to 0^-$, one can find
\begin{equation}
\begin{split}
\frac{4\GN}{\bar{\phi}_r}S_{\mt{N}}(y^+_2 \le 0) \quad &\longrightarrow \quad 2k \log\( \frac{-y^+_2}{\pi \Tb\epsilon^2}\) \,,\\
\frac{4\GN}{\bar{\phi}_r}S_{\mt{R}}(y^+_2 \le 0) \quad &\longrightarrow \quad 2k \log\( \frac{\sinh \( 2\pi \Tb \sigma_2\)}{(\pi \Tb\epsilon)^2 } \) \,,\\
\end{split}
\end{equation}
which implies the N-channel is preferred $(\log(-y^+_2) \to -\infty)$\footnote{Taken at face value, the generalized entropy in the N-channel becomes negative for sufficiently small $|y_2^+|$. This is another hint that the shockwave needs to be regularized.} when $y_2^+ < 0$ is located in the region around the shock. On the other hand, the limit $y^+_2 \to 0^+$ (or $x^+_2 \to 0^+$) leads us to
\begin{equation}
\begin{split}
\frac{4\GN}{\bar{\phi}_r}S_{\mt{N}}(y^+_2 \ge 0) \quad &\longrightarrow \quad 2k \log\( \frac{\bar{\ell}^2}{\epsilon^2}\) \,,\\
\frac{4\GN}{\bar{\phi}_r}S_{\mt{R}}(y^+_2 \ge 0) \quad &\longrightarrow \quad 2k \log\( \frac{x_2^+}{\epsilon } \) +2k \log\( \frac{1}{\pi \Tb\epsilon } \)\,,\\
\end{split}
\end{equation}
where $\bar{\ell}$ is some finite length-scale.For the zero bath temperature case, the divergence structure is the same.\footnote{The $\frac{1}{\pi \Tb}$ in the last logarithm is replaced by some finite length scale $\tilde{\ell}$ when $\Tb = 0$.} From the simple lesson coming from these divergences, it is obvious that the R-channel is preferred when $y^+_2$ is located in the region near the shock but after the shock wave. This region for $y_2$ which allows the bath interval plus QM$_\mt{L}$ to recover the interior of the black hole is shown in the left diagram in figure \ref{fig:bath_intervals}.
This conclusion meets an obvious paradox because we can contain a larger part of the bath by moving the right end-point of the interval from the after-shock region to the pre-shock region. The above analysis implies that adding more bath interval surprisingly makes one lose the ability to recover the black hole interior.

However, the above paradox appears just because we consider the shock-wave as a line located at $\sigma_{\rm{shock}}=u$. More precisely, the generalized entropy for the R and N-channel around the shock wave is disconnected. In order to solve this problem, we have to regularize the region of the shock wave and simultaneously make the generalized entropy a continuous function. In other words, we need to consider the entropy from one-point and two-point functions
\\
\\
{\setlength{\arrayrulewidth}{0.5mm}
	\setlength{\tabcolsep}{18pt}
	\renewcommand{\arraystretch}{1.5}
\begin{tabular}{ |p{0.1cm}|p{6cm}|p{6cm}| }
	\hline
	 & before-shock $y^+_2 <0$ &after shock $y^+_2 >0$ \\
	\hline
	$S_2$ & $\log\( \frac{1}{ \epsilon} \frac{\sinh\(\pi \Tb( y^-_2- y^+_2)\)}{\pi \Tb}  \)$ &  $\log\( \frac{12 \pi E_s}{ c \epsilon} \frac{\sinh\(\pi \Tb y^-_2\)}{\pi \Tb}  \frac{x^+_2}{\sqrt{f'(y^+_2)}}\)$\\
    $S_{12}$ & \tiny{$\log\( \frac{12 \pi E_s}{c\epsilon^2}\frac{x^+_1 \sinh\(\pi \Tb (-y^+_2)\)\sinh\( \pi \Tb (y^-_2 -y^-_1) \)}{(\pi \Tb)^2 \sqrt{f'(y^+_1)}} \) $ } &  \tiny{$\log\( \frac{1}{\epsilon^2}\frac{\sinh\( \pi \Tb(y^-_2 -y_1^-)\)}{\pi \Tb} \frac{x^+_1 -x_2^+}{\sqrt{f'(y^+_1) f'(y^+_2)} }\)$} \\
	\hline
\end{tabular}
}
\\
\\
The identifications
\begin{equation}
\begin{split}
S_{2,\rm{before}} (y_2^+ \to -0) &= S_{2,\rm{after}} (y_2^+ \to +0)\,, \\
S_{12,\rm{before}} (y_2^+ \to -0) &= S_{12,\rm{after}} (y_2^+ \to +0)\,, \\
\end{split}
\end{equation}
fixes the two boundaries of shock-wave region as
\begin{equation}
\begin{split}
 1 = \frac{12 \pi E_s}{ c} \frac{x^+_2}{f'(y^+_2)} \,, \qquad y_2^+ \to +0\\
 \frac{12 \pi E_s}{ c} \frac{\sinh (-\pi \Tb y^+_2)}{\pi \Tb} =1 \,,\qquad y_2^+ \to -0\,.\\
\end{split}
\end{equation}
 Recalling the property of $f(u)$ such as $f'(0)=1, x=f(y)\sim 0 \sim y$, we can take the energy of shock wave $E_s$ as a regulator and regularize the shock wave as a small region defined by
 \begin{equation}
 y^+_{\rm{shock}} \equiv \[-\frac{c}{12 \pi E_s} ,\frac{c}{12 \pi E_s} \]= \[ -\frac{k}{(T_1^2 - T_0^2)\pi^2} ,\frac{k}{(T_1^2 - T_0^2)\pi^2}\] \,,
 \end{equation}
which is independent of the temperature of bath as expected.

After identifying this small region as the shock-wave, we can take out this part from the bath interval and then make the generalized entropy connected when we move the endpoint $y_2$ from the after-shock region to the pre-shock region. The connectivity guarantees that we do not have the paradox about the ability of bath interval to recover the information of the black hole interior anymore. More explicitly, we will discuss this problem in detail in the next subsections.
\subsubsection{Need for the Purification}
\label{sec:purif}

The conditions for a finite bath interval plus {\QML} in a zero temperature bath to reconstruct the interior of the black hole were discussed in~\cite{Chen:2019uhq}. Moreover, we found in section \ref{sec:equilibrium} that the equilibrium case with $\Tb=T_1$, even the whole semi-infinite bath interval with {\QML} does not contain the appropriate information to reconstruct the interior of the black hole. Rather we had to also include its purification (or at least a portion of the latter). In the previous section, we have seen the difference in divergence structure between a non-zero temperature bath and that with zero temperature when $y_2$ approaches the IR cut-off between the bath and its purification. The smaller divergence of the leading term \eqref{divergence_IR} in a zero-temperature bath guarantees we can use the whole bath interval $y_2 \to \infty$ with {\QML} to reconstruct the interior of the black hole. Obviously, this is expected because this subsystem as one part of the bipartite pure system will be able to reconstruct its complementary part, \ie the black hole interior, after Page transition. A natural question is whether all bath intervals with non-zero temperature $\Tb$ require a part of the purification in order to reconstruct the black hole interior. In this subsection, we show that only the bath interval with a temperature higher than the critical temperature $T_p$ in eq.~\eqref{eq:tp} requires its purification.

To this end, we consider large bath intervals by putting the left end-point after the shock and the right end-point before the shock and focus on the inequality in eq.~\eqref{conditionAB_preshock}. The left-hand side of the inequality monotonically increases with $y^+_1=u-\sigma_1$ ($u\ge u_{\Page}$) and the right-hand side decreases, so the weakest condition for that inequality is choosing $\sigma_1=0$, \ie anchoring the initial point of bath interval at AdS boundary, which satisfies our physical expectation. Let's move on to the condition for $y^+_2$ by considering  figure \ref{fig:desvenlafaxine}c  and the corresponding conditions in eq.~\eqref{conditionAB_preshock}, \ie 
\begin{equation}\label{conditionAB_preshock2}
2k \log\(  \frac{24\pi E_s }{c\epsilon} \frac{x^+_1 \sinh \(\pi \Tb(- y^+_2)\) \sinh \(\pi \Tb \sigma_2\)}{\sqrt{f'(y^+_1) }\pi \Tb\sinh \(2\pi \Tb\sigma_2\)}\) \ge \frac{4 \GN}{\bar{\phi}_r} S_{\gen,\rm{late}} -2\pi T_0 \,,
\end{equation}
with $\sigma_1=0$. Again, it is straightforward to show
\begin{equation}
\partial_{\sigma_2} \( \frac{ \sinh \(\pi \Tb(- y^+_2)\) \sinh \(\pi \Tb \sigma_2\)}{\pi \Tb\sinh \(2\pi \Tb\sigma_2\)} \) = \frac{1}{2}\frac{\cosh (\pi \Tb u)}{\cosh^2 (\pi \Tb \sigma_2)} >0\,,
\end{equation}
which implies the maximum of the left-hand side in above inequality is the value at the limit $\sigma_2 \to \infty$. As expected, the weakest condition for the bath interval plus \QML\ to have enough information about the black hole interior is if we consider the entirety of the bath with $\sigma_1=0$ and $\sigma_2 \to \infty$. The condition to recover the interior of the black hole is then given by
\begin{equation}
\label{eq:recover}
\text{Max}=2k \log\(  \frac{12 E_s }{c\epsilon \Tb} \frac{f(u) e^{-\pi \Tb u}}{\sqrt{f'(u) }}\) \ge \frac{4 \GN}{\bar{\phi}_r} S_{\gen,\rm{late}} -2\pi T_0 \,.
\end{equation}
It is useful to notice that the right-hand side decreases with time $u$ and the left-hand side increases for $\Tb < T_1$ and decreases for $\Tb >T_1$.\footnote{It is easy to show that from the approximation \eqref{app_linear} and \eqref{app_late} because the dominated term for $k\log\(\frac{1}{f'(u)}\)$ involve from $2k\pi T_1u$ to $2k \pi \Tb u$. } As a result, the inequality cannot hold for $\Tb >T_1$ because the maximum of the left-hand side is bounded by $2k\, \log\!\( \frac{12 E_s \tinf}{c\epsilon \Tb}\)$. This implies that in the setup where the bath heats up the black hole ($\Tb>T_1$), the bath and QM$_\mt{L}$ systems are never able to reconstruct the black hole interior. We now focus on the evaporating black hole model ($\Tb < T_1$) and take the the late-time approximation \eqref{app_late} at $e^{ku} \gg 1$ for which the LHS gives a constant
\begin{equation}
2k \log\(  \frac{12 E_s \tinf}{c\epsilon \Tb}\) -k \log\( 4\pi \Tb \tinf\)  + 4 \pi \(T_1 -\Tb -\Tb \log\( \frac{T_1+\Tb}{2\Tb}\) \) \,.
\end{equation}
Correspondingly, the RHS reaches its minimum at the same late-time limit
\begin{equation}
\frac{4 \GN}{\bar{\phi}_r} S_{\gen,\rm{late}} -2\pi T_0  \approx 2\pi \( \Tb- T_0\) + 2k \log\( \frac{1}{\pi T_b \epsilon } \) \,.
\end{equation}
The inequality from generalized entropy in eq.~\eqref{eq:recover} then yields the condition for the temperature of bath
\begin{equation}
\Tb  \lesssim \frac{2T_1 +T_0}{3}-\frac{2\Tb}{3} \log\( \frac{T_1+\Tb}{2\Tb}\) +\frac{k}{3\pi} \log\( \frac{6E_s}{c T_1} \sqrt{\frac{T_1}{\Tb}} \) \,.
\end{equation}
Finally, we can find the critical temperature of bath is
\begin{equation}
\label{eq:tp}
T_p \approx T_1 -\frac{1}{2}\( T_1 - T_0 \)+\frac{k}{2\pi} \log\( \frac{6E_s}{c T_1}\) \,,
\end{equation}
which defines the lowest bath temperature for which the purification of the bath is needed to reconstruct the interior of the black hole. It is interesting to note that the critical temperature is also near $T_1$ due to the ansatz $T_0 \sim T_1$. However, it is different from the critical temperatures $T_{c_1}$ and $T_{c_2}$ in eq.~\eqref{eq_Tc} because the former depends on $T_0$, the temperature of original black hole, while $T_{c_1}$ and $T_{c_2}$ are independent of $T_0$.

To summarize, a bath with a temperature $\Tb < T_p$ admits finite bath intervals plus {\QML} to reconstruct the interior of the black hole. When the temperature of the bath increases beyond $T_p$, even the whole semi-infinite bath interval plus {\QML} does not have enough information for interior reconstruction if part of the purification is not included.
Lastly, we can also change the left end-point $\sigma_1$ rather than taking it to the AdS boundary ($\sigma_1\to 0$) and do a similar late-time approximation to obtain the constrains on bath temperature.
As expected, we find a stronger condition and get a smaller critical temperature
\begin{equation}
T_p (\sigma_1)\approx \frac{1}{2}\( T_1 + T_0 \) -2kT_1 \sigma_1+\frac{k}{2\pi} \log\( \frac{6E_s}{c T_1}\) \,, \text{with} \quad k\sigma_1 \ll 1\,,
\end{equation}
for the reconstruction of information in black hole. We should also note the chopping off too much of the bath interval by taking the initial point from $\sigma_1=0$ to a finite one also may make the thermal bath interval plus \QML\  lose
essential information to reconstruct the black hole interior if $\sigma_1$ is too large. The size of the bath interval we can ignore is also restricted by 
\begin{equation}
\sigma_1 \lesssim \frac{1}{4k T_\bath} \( 2T_1+T_0 - 3T_\bath - T_\bath \log\(  \frac{T_1+T_\bath}{2T_\bath}\)\)  + \frac{1}{4\pi T_\bath} \log\( \frac{6E_s}{c T_1 }\sqrt{\frac{T_1}{T_\bath}}\) \,,
\end{equation}
where we have assumed the bath temperature is not too small. 

\subsubsection{Finite Bath Interval}\label{sec:lowerTem}
In this subsection, we will assume the bath temperature is lower than the critical $T_p$ and discuss how much bath interval needed in order to reconstruct the black hole interior and in particular, what is the closest we can bring the right end-point $\sigma_2$ to the AdS boundary and still reconstruct the black hole interior. The two competing channels are described in figure \ref{fig:desvenlafaxine}c.This analysis can be understood as an extension of the late-time protocol of~\cite{Chen:2019uhq} to the thermal bath model.
Let's first assume we only need the bath interval after the shock to which the radiation of black hole escapes. Taking the time slice at $u$ after the Page time and putting the left end-point of the bath interval at the AdS boundary ($\sigma_1=0$), the bath interval we are looking for satisfies eq.~\eqref{conditionAB_aftershock}
\begin{equation}\label{conditionAB_aftershock2}
2k \log\(  \frac{c}{6\pi E_s \epsilon} \frac{(t -x^+_2) \sinh \(\pi \Tb\sigma_2 \)}{x^+_2\sqrt{f'(u)} \sinh(\pi \Tb (u+\sigma_2)) }\) \ge \frac{4 \GN}{\bar{\phi}_r} S_{\gen,\rm{late}}-2\pi T_0 \,,
\end{equation}
which imposes a constrain on the size of the bath interval, \ie the value of $\sigma_2(u)$. Assuming we can have $\pi T_1 y_2^+\gg 1 $ (or $f(y^+_2) \approx \tinf$) and still stay at the linear-region with $ku <1$,
we can recall the approximation again
\begin{equation}\label{app_conditionAB}
\begin{split}
2k\log\( \frac{\sinh \(\pi \Tb\sigma_2 \)}{\sinh(\pi \Tb (u+\sigma_2)) } \) &\approx -2k\pi \Tb u  \\
2k\log\(  \frac{t-x^+_2}{x^+_2} \) &\approx \log\(  \frac{\tinf-f(y^+_2)}{\tinf} \)
\sim  -4k\pi T_1 (u-\sigma_2) \,,\\
2k\log\( \frac{1}{\sqrt{f'(u)}}\) &\approx 2k\pi T_1 u  - k\log\(4\pi T_1 \tinf \) \,,\\
\frac{4 \GN}{\bar{\phi}_r} S_{\gen,\rm{late}} &\approx 2\pi T_1 - k\pi T_1(u-u_{\HP}) \(1-\frac{\Tb^2}{T_1^2} \) + \mathcal{O}(k\log(\cdots))\,.
\end{split}
\end{equation}
Then the above inequality leads us to
\begin{equation}\label{linear_sigma2}
\sigma_2 (u) \gtrsim \frac{T_1- T_0}{2k T_1} + \frac{u}{4}\( 1 + \frac{\Tb}{T_1}\)^2 + \frac{u_{\HP}}{4}\(1-\frac{\Tb^2}{T_1^2} \) \,,
\end{equation}
or equivalently
\begin{equation}\label{linear_latepro_y2p}
y^+_2 (u)\equiv u- \sigma_2 \lesssim \frac{u}{4} \( 3 - 2\frac{\Tb}{T_1} -\frac{\Tb^2}{T_1^2} \) -\frac{T_1 - T_0}{2k T_1} -\frac{u_{\HP}}{4}\(1-\frac{\Tb^2}{T_1^2} \) \,,
\end{equation}
which constrains the size of the bath interval able to reconstruct the black hole interior. By setting $\Tb=0$, we recover the results reported in~\cite{Chen:2019uhq} (see eq.~(3.61) and eq.~(3.62)). It is also clear that the thermal bath with $\Tb \gtrsim T_1$ obviously breaks the that inequality, implying we cannot find a bath interval with only {\QML} able to recover the information in the black hole. This conclusion is consistent with that found in the previous subsection. However, we also want to stress the validity of the condition~\eqref{linear_latepro_y2p}. One can find the critical value
around the Page time is not physical, \ie
\begin{equation}
y^+_2\big|_{u_{\rm{Page}}} \approx \frac{T_1- T_0}{2k T_1} \( \frac{3 - 2\frac{\Tb}{T_1} -\frac{\Tb^2}{T_1^2}  }{3 + 2\frac{\Tb}{T_1} -\frac{\Tb^2}{T_1^2} } -1\)  \lesssim 0 \,.
\end{equation}
This invalidity implies the condition \eqref{linear_latepro_y2p} is only valid for time slices after the Page time with  $\exp\( \pi T_1  (u -u_{\Page})\)\gg 1$.
To be precise, the reason is we can only find a small $  y^+_2 \ll \frac{1}{\pi T_1}$ instead of $y^+_2 \gg \frac{1}{\pi T_1}$ as a solution around Page time. However, the value is so small that it is actually located in the shock-wave region. As a result, it means we cannot find a bath interval able to reconstruct the black hole interior with only the region after the shock wave at the Page time. One can look at the eq.~(3.59) in~\cite{Chen:2019uhq} as an example of this. The calculation is similar and we do not repeat it here because the value of small $y^+_2$ in that region is not really physical after the regularization of the shock-wave. After the Page time, the critical $y^+_2$ will exponentially increase and move quickly to the linear region as shown in eq.~\eqref{linear_latepro_y2p}. The allowed region for the endpoint of $y_2^+$ is shown in the right plot in figure \ref{fig:bath_intervals}.

\begin{figure}[htbp]
	\includegraphics[width=0.499\textwidth]{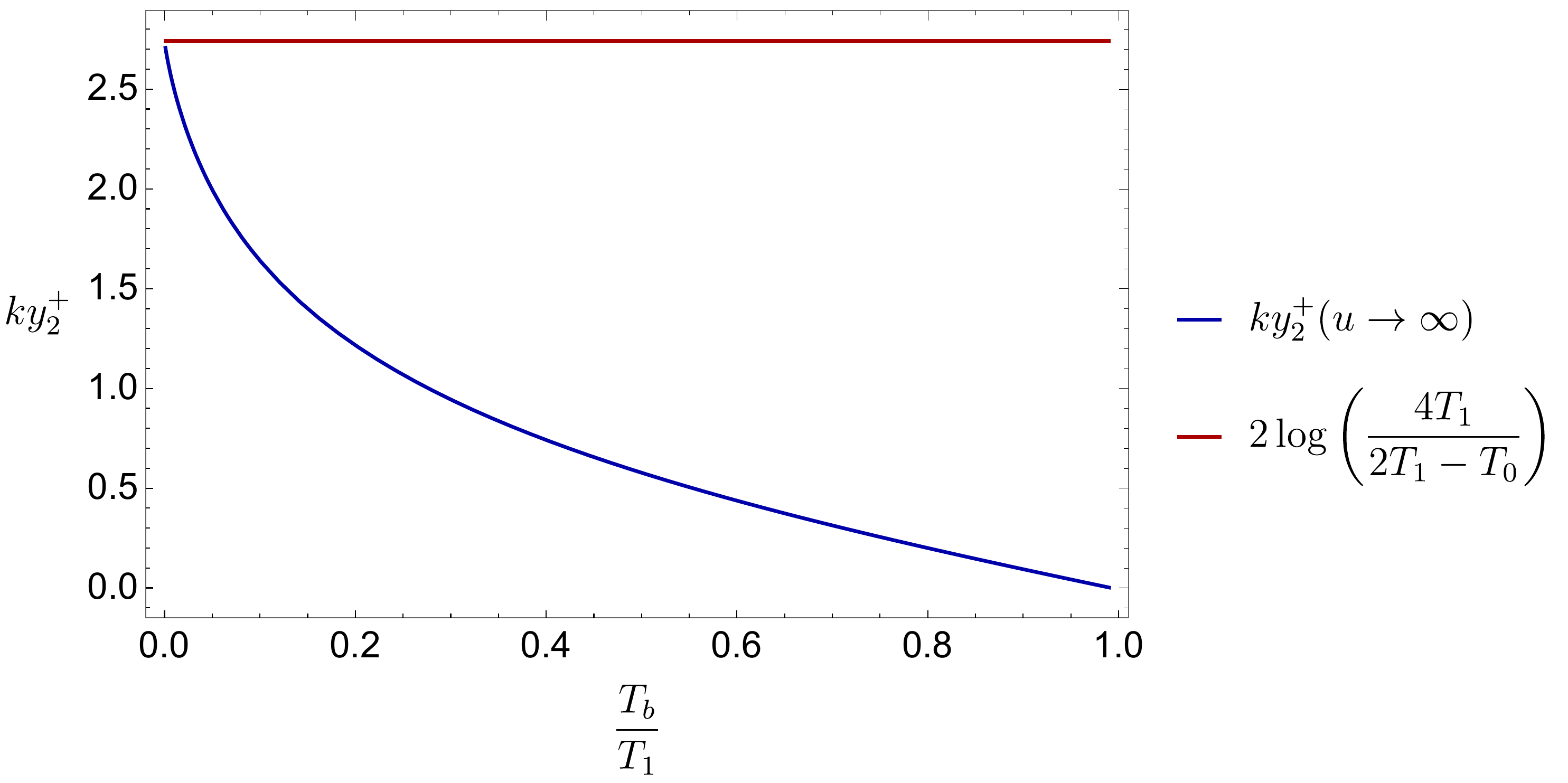} \hspace{0.01\textwidth}
	\includegraphics[width=0.499\textwidth]{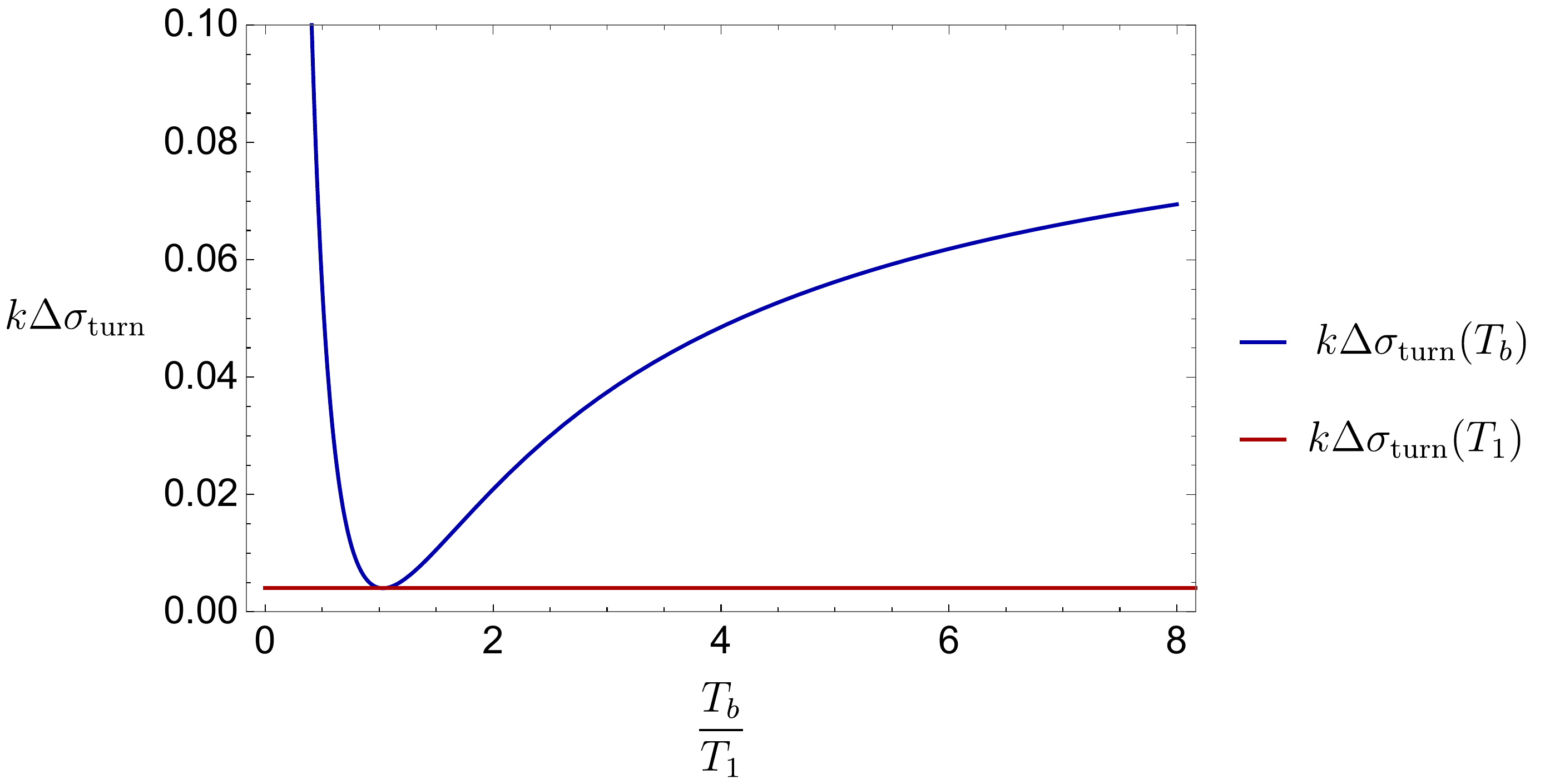}
	\caption{Left: the final position of the null surface $y^+_2$, \ie \eqref{finaly2plus} as the endpoint of bath interval with the ability to reconstruct the information of the interior of black hole. Right: the bath temperature dependence of the minimal length $k \Delta \sigma_{\turn}$, \ie \eqref{sigmaturn},  that is necessary for the reconstruction of the interior of black hole. }\label{fig:y2plus}
\end{figure}

Taking the lesson from the zero-temperature case, we can expect that the linear growth of $y^+_2$ is suppressed with the time evolution and finally $y^+_2(u)$ will approach a null surface. In order to show that explicitly, we should take the late-time ($e^{ku} \gg 1$) approximation in eq.~\eqref{app_late} and use the following approximations
\begin{equation}\label{eq:applate}
\begin{split}
2k\log\(  \frac{t-x^+_2}{x^+_2} \) &\approx \log\(  \frac{\tinf-f(y^+_2)}{\tinf} \)
\sim  -4k\pi \Tb y^+_2  + 8\pi (\Tb - T_1)\(1-e^{-ky^+_2/2}\) \,,\\
\frac{4 \GN}{\bar{\phi}_r} S_{\gen,\rm{late}}\(e^{ku} \gg 1\) &\approx \frac{4 \GN}{\bar{\phi}_r} S_{\gen,\rm{late}} (\Tb)\approx 2\pi \Tb + 2k \log\( \frac{1}{\pi T_\bath \epsilon} \)\,,
\end{split}
\end{equation}
where we also keep the second-order contribution $e^{-ky^+_2/2}$ because the condition $e^{ky^+_2} \gg 1$ is not guaranteed.  Combining all these approximations, the condition~\eqref{conditionAB_aftershock2} becomes
\begin{equation}\label{null_y2p}
k\Tb y^+_2 +2\( \Tb - T_1\) e^{-ky^+_2/2} \lesssim -\( T_1 - \frac{T_0+\Tb}{2} + \Tb \log\(\frac{T_1 +\Tb}{2\Tb} \) \) -\frac{k}{2\pi} \log\( \frac{6E_s}{c T_1}\sqrt{\frac{T_{\mt{eff}}}{T_1}}\) \,.
\end{equation}
When $\Tb=0$, we get the final null surface for critical $y^+_2$ as
\begin{equation}
y^+_2\(e^{ku}\gg 1\) \big|_{\Tb=0}\approx \frac{2}{k} \log\( \frac{4T_1}{2T_1 -T_0} \) \,,
\end{equation}
which agrees with the result in \cite{Chen:2019uhq}. For non-zero $\Tb$, the analytical solution is written as
\begin{equation}\label{finaly2plus}
y^+_2\(e^{ku}\gg 1\)  \approx \frac{X + 2\Tb \ W\!\( \frac{e^{-\frac{X}{2\Tb}}(T_1-\Tb)}{k\Tb} \)}{k\Tb} = \Delta y_2\,,
\end{equation}
where $X$ represents the right side in \eqref{null_y2p} and $W(z)$ is the Lambert W-function or product logarithm defined by $z=W(z)e^{W(z)}$. As a summary, the time dependence of $y^+_2(u)$ is shown in the right plot in figure \ref{fig:bath_intervals}. It is clear that the constant $\Delta y_2$ indicates how much early radiation is not necessary in the reconstruction of black hole interior. 
Lastly, we show the numerical plot for position of the final null surface as a function of $\Tb/T_1$ in figure \ref{fig:y2plus}. As expected, it decays with the increase of $\Tb$ and stops at a point extremely near $T_1$ because  the value at $\Tb =T_1$, \ie  $-\frac{T_1-T_0}{T_1}$ is smaller than zero.
\subsubsection{The Role of Purification}
\label{sec:purif2}

\begin{figure}[htbp]
	\includegraphics[width=0.499\textwidth]{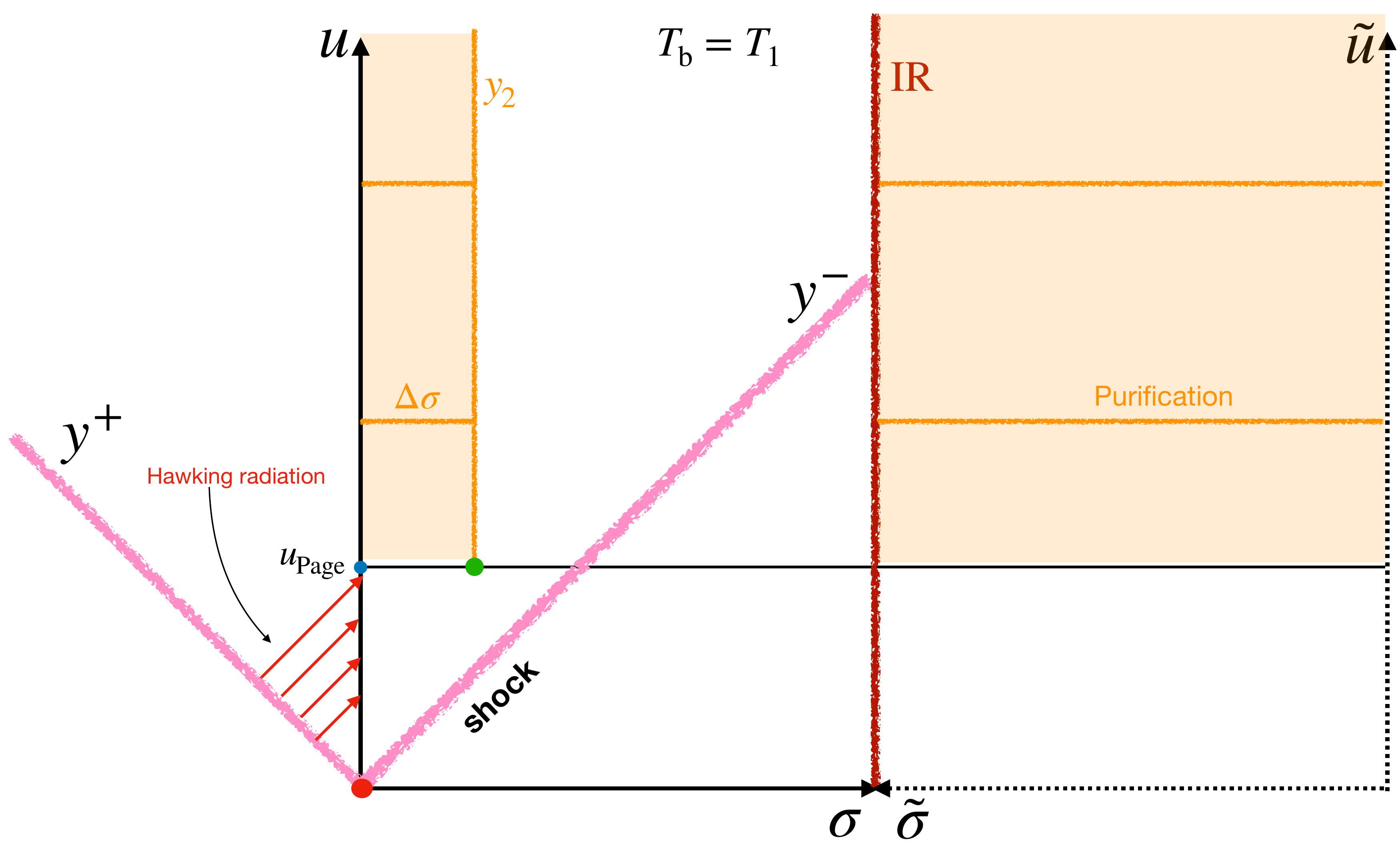} \hspace{0.01\textwidth}
	\includegraphics[width=0.499\textwidth]{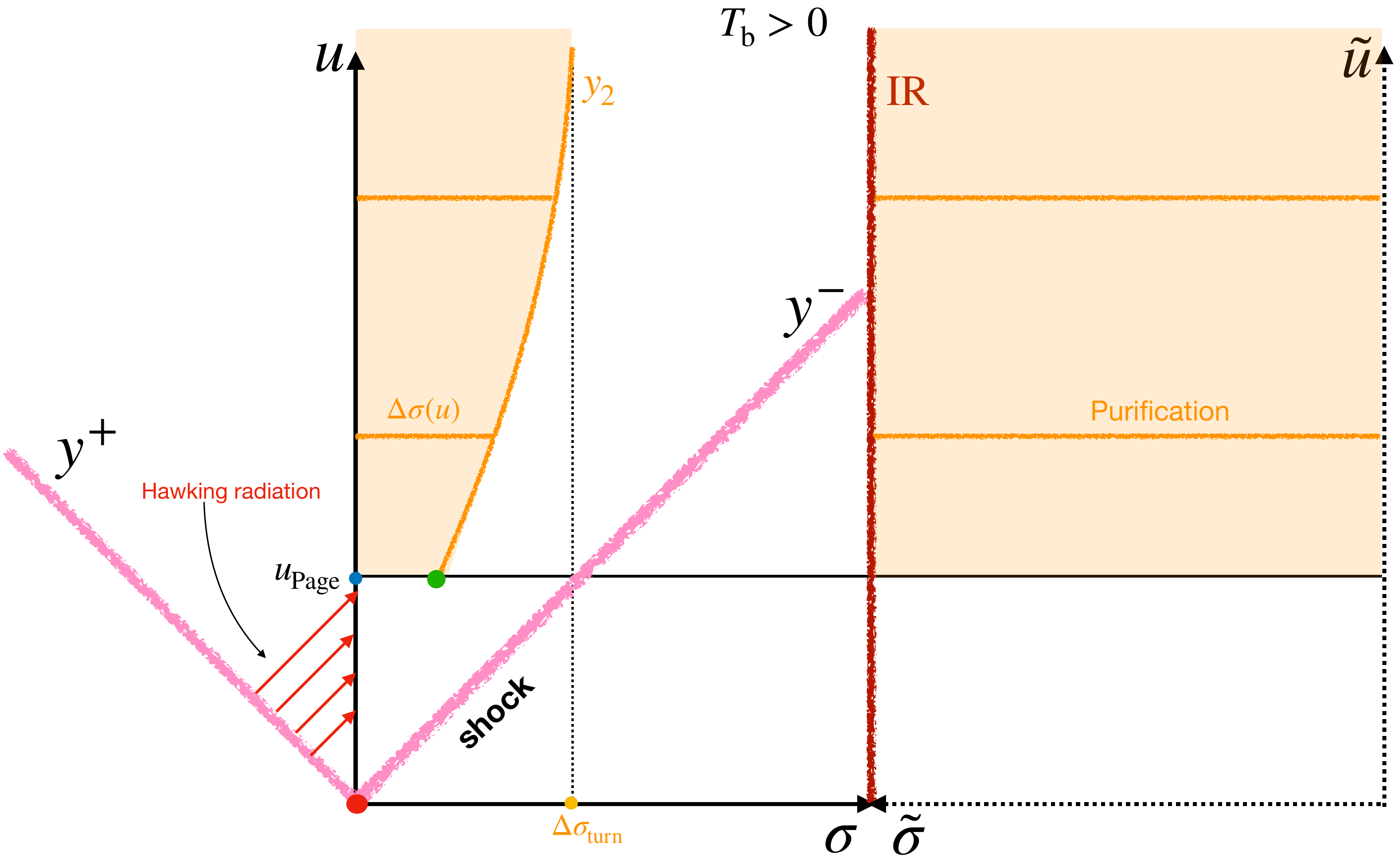}
	\caption{The yellow shadow denotes the minimal bath region including a full half-line as the purification of thermal bath for reconstructing the interior of the black hole. The yellow lines represent the necessary bath region at a fixed time slice after Page transition. Left: the equilibrium case with $T_\bath= T_1$. Right: Non-equilibrium case where $k\, \delta \sigma(u)$ increases with the time evolution and approaches a constant $\Delta \sigma_{\turn}$ defined in eq.~\eqref{sigmaturn}.}\label{fig:sigmaturn}
\end{figure}

In the previous subsection, we focused on a finite bath interval with the bath temperature lower than the critical temperature $T_p$ derived in eq.~\eqref{eq:tp} because we wanted to omit the purification of the bath itself. To reconstruct the black hole interior at higher bath temperatures, \ie to observe a Page transition, we need to include (a portion of) the purification. For simplicity, we take one endpoint of the finite bath interval on AdS boundary with $\sigma_1=0$ and ask how large the bath interval $[0,\sigma_2]$ needs to be to reconstruct the black hole interior. Concretely, we consider the purified bath interval with temperature $T_b >0$ and discuss the condition for QM$_{\mt{L}}$,  a finite bath interval $[0,\sigma_2]$ (where partial Hawking radiations reside) and the full purification to reconstruct the interior of the black hole.\footnote{The $T_b\to 0$ limit has some subtleties here. The ``purification'' of the bath with zero temperature is a pure state coupled to the bath system by direct product. Thus, the purification of the bath does not help with interior reconstruction because it is a fully unentangled region and the corresponding R-channel is defined by $S_{\mt{R}}= S^\gen_{\QES-1} +S_2+ S_{\halfLine}$, which cannot be derived from \eqref{eq:SRN_4b} by taking the limit $T_\bath \to 0$. The reason is that, for $T_b=0$, the holographic (3D) spacetime for bath interval and its purification is defined by two separated regions rather than a smooth and connected spacetime, as in the $T_b\neq 0$ case, where the entanglement between two regions glues the spacetime. However, note that the naive $T_b\to 0$ limit would leave one IR divergent term since $S_{\mt{N}}-S_{\mt{R}} \sim \frac{c}{6}\log \frac{\bar{l}}{\sigma_\IR}$.} Similar to the equilibrium case shown in eq.~\eqref{eq:beer}, the two competing channels showing to figure \ref{fig:desvenlafaxine}b  are defined as 
\begin{equation}\label{eq:SRN_4b}
S_{\mt{R}}
=S^\gen_{\QES-1} + S_{2-\IR}\,, \qquad 
S_{\mt{N}}
= S^\gen_\bif + S_{1-2} + S_\halfLine \,.
\end{equation}
where the condition for reconstruction is decided by $S_{\mt{N}} - S_{\mt{R}}\ge 0$. 
Most pieces in the above two equations have been discussed in the last subsection (see eqs.~\eqref{SN_after} and \eqref{SR_after}) in details except for 
\begin{equation}\label{S2_extra}
\begin{split}
S_\halfLine&=\frac{c}{6} \log \(\frac{\sinh\(2\pi T_\bath \sigma_{\IR}\)}{\pi T_\bath\epsilon} \) \approx 
\frac{c}{6}\(2\pi T_\bath \sigma_\IR
+ \log\left(\frac{1}{2\pi \epsilon T_\bath}\right)  \)\,, \\
S_{2-\IR}&=  \frac{c}{6}\log\( \frac{12 \pi E_s}{c \(\epsilon\pi T_\bath\)^2} \frac{x^+_2 \sinh\(\pi T_\bath y^+_{\IR}\) \sinh\(\pi T_\bath (y_2^--y^-_{\IR})\) }{\sqrt{f'(y_2^+)}}  \) \,.
\end{split}
\end{equation}
In the limit $\sigma_{\IR} \to \infty$, we can rewrite that extra term as
\begin{equation}\label{S2IR_app}
S_{2-\IR}\approx   \frac{c}{6} \(  2\pi T_\bath \sigma_\IR
+ \log\left(\frac{1}{2\pi \epsilon T_\bath}\right) -2\pi T_\bath y^-_2 +     \log \( \frac{12 \pi E_s}{c\epsilon } \frac{x^+_2 \sinh\(\pi T_\bath y^-_2\) }{\pi T_\bath\sqrt{f'(y_2^+)} }   \)     \)\,,
\end{equation}
where we pick up a time slice after the Page transition, the first two terms denoting the thermal entropy of a half-line can compensate the same divergence appearing in $S_\halfLine$ and the last term is the same as the one-point function contribution $S_2$ appearing in the case without purification, \ie eq.~\eqref{SR_after}. Physically, we can explain the third term $2\pi T_\bath y^-_2$ as the entanglement from the thermal radiation generated by the thermal bath. That extra term is traced back to the inclusion of the purification and the negative sign reflects the fact that the bath interval is entangled with its purification. As a result, we can expect the introduction of purification can help fulfill the condition for reconstruction as we will show below. 

First, let's work on the simple linear region with $ku\ll 1$. Adding the new contributions $\eqref{S2_extra}$ in eq.~\eqref{conditionAB_aftershock2} and taking the linear  approximations in eq.~\eqref{app_conditionAB} again,  the condition $S_{\mt{N}} - S_{\mt{R}} \ge 0$ can be rewritten as a restriction on the length of the finite bath interval $\Delta \sigma = \sigma_2$, \ie  
\begin{equation}\label{linear_sigma2_purification}
\sigma_2(u) \gtrsim \frac{T_1- T_0}{2k \(T_1+T_\bath\)} + \frac{T_1}{4(T_1 + T_\bath)}\(u\( 1 - \frac{\Tb}{T_1}\)^2 + u_{\HP}\(1-\frac{\Tb^2}{T_1^2} \) \) + \frac{\log\(\frac{6E_s}{cT_1}\)}{2\pi (T_1 + T_\bath)} + \cdots \,,
\end{equation}
where we also have assumed $\pi T_1 y^+_2 \gg 1$ (or $f(y_2^+) \approx t_{\infty}$). Comparing to the finite bath interval without purification, \ie eq.~\eqref{linear_sigma2}, we see that introducing the purification decreases the minimal length of the necessary bath interval for reconstructing the interior of the black hole, and also slows down the speed of its linear increase with time. More importantly, it also makes the subsystem consisting of  QM$_{\mt{L}}$, a finite bath interval $[0,\sigma_2]$ (with only a fraction of the Hawking radiation) and the purification have the ability to recover the information of the black hole even when $T_b>T_p$.  As the unreliability of the linear approximations in the overheated case, we should remark that we also need to consider some corrections for the above approximate $\sigma_2(u)$ if the temperature of the thermal bath is too high, \ie $T_\bath \gtrsim 3 T_1$. 

As one might expect, for much larger times, the linear increase of $\sigma_2(u)$ breaks down. Since the black hole eventually equilibrates with the bath, we expect qualitatively similar behavior to the equilibrium case of section~\ref{sec:equilibrium}, that is, we expect $\Delta\sigma = \sigma_2$ to approach half the (equilibrium) Page time as in eq.~\eqref{eq:areYouSure}. To derive this explicitly, we focus on the time derivative of the critical length denoted by $\partial_u \sigma_2^\ast(u)$ directly. Noting the time evolution of the generalized entropy at late-time phase (after Page transition) has been shown in \eqref{eq:derivapprox}, we explicitly start from the approximation of $S_{\mt{R}}-S_{\mt{N}} = 0$ by 
\begin{equation}\label{eq:SRequalsSN}
2k  \(2\pi T_\bath y^-_2+ \log\(  \frac{c}{6\pi E_s \epsilon} \frac{(f(u)-x^+_2) \sinh \(\pi \Tb\sigma_2 \)}{x^+_2\sqrt{f'(u)} \sinh(\pi \Tb (u+\sigma_2)) }\) \)\approx \frac{4 \GN}{\bar{\phi}_r} S_{\gen,\rm{late}}-2\pi T_0 \,,
\end{equation}
and obtain the differential equation 
\begin{equation}\label{eq:derivativsigma2}
\begin{split}
4\( T_{\rm{eff}}(y^+_2) +T_\bath \) \partial_u \sigma^\ast_2 &\approx  -T_{\rm{eff}}(u)\(1- \frac{T^2_\bath}{T^2_{\rm{eff}}(u)}\)- 2T_\bath -2T_{\rm{eff}}(u) + 4 T_{\rm{eff}}(y^+_2) + \mathcal{O}(k)  \\
&\approx T_{\rm{eff}} (u)\(   1- \frac{T_\bath}{T_{\rm{eff}}(u)}\)^2 \,,
\end{split}
\end{equation}
where we denoted the solution of $S_{\mt{R}}-S_{\mt{N}} = 0$ as $\sigma_2^\ast(u)$ and mainly used the approximation $f'(u) \sim 2\pi T_{\rm{eff}}(u) \(\tinf - f(u)\)$ and associated approximations derived in eqs.~\eqref{eq:approx-simple}. To double check, we can focus on the linear region with $T_{\rm{eff}} \approx T_1$ again and obtain 
\begin{equation}
\partial_u \sigma_2^\ast (u) \approx  \frac{T_1}{4\(T_1+ T_\bath\)} \( 1 - \frac{T_b}{T_1} \)^2\,,
\end{equation}
which agrees with eq.~\eqref{linear_sigma2_purification} as expected. On the other hand, it is also obvious that the time derivative at late time region approaches zero, \ie 
\begin{equation}
\partial_u \sigma_2^\ast \big|_{e^{ku} \gg 1} \longrightarrow 0 \,,
\end{equation}
because of the simple approximation $T_{\rm{eff}}\(e^{ku} \gg 1\) \approx T_\bath$ for the effective temperature. In other words, the evolution towards equilibrium pushes the minimal length $\Delta \sigma_2$ to be a constant, which exactly matches what was shown in the equilibrium case in eq.~\eqref{eq:areYouSure}. As a result, we can find the minimal length $\Delta \sigma_2$ with purification should approach a constant whereas the lightcone coordinate $y_2^+ = u- \sigma_2 $ reaches a constant as indicated in eq.~\eqref{finaly2plus} if we do not include the purification. We sketch a plot to illustrate the time dependence of $\Delta \sigma_2(u)$ when the purification is included in figure \ref{fig:sigmaturn}. More explicitly, we apply the late-time approximation with $e^{ku} \gg 1$ on $S_{\mt{N}}- S_{\mt{R}} \ge 0$ and derive the constraint
\begin{equation}\label{sigmaturn}
\sigma_2\({ e^{ku} \gg 1}\)\gtrsim \frac{1}{4k T_\bath} \( 2T_1 -T_\bath -T_0  -2T_\bath \log\( \frac{T_1+ T_\bath}{2T_\bath}  \) \) +\frac{\log\(\frac{6E_s}{cT_1}\sqrt{\frac{T_{\bath}}{T_1}}\)}{4\pi  T_\bath}   \equiv \Delta \sigma_{\turn}\,.
\end{equation}
As expected, it returns to the result shown in eq.~\eqref{eq:areYouSure} for the equilibrium case by setting $T_\bath = T_1$. As a final remark, we point out that the minimum of the dimensionless length scale $ k \Delta \sigma_{\turn}$ (at leading-order) is realized at the near-equilibrium case with $T_\bath = T_1 \( 2\frac{T_1}{T_0} - 1 \)$. A simple numerical plot is also shown in the right figure \ref{fig:y2plus}.  The interesting feature we want to highlight is that when the black hole is evaporating, we need more bath interval to recover the interior of the black hole, while a thermalized black hole requires less bath interval in which less of the outgoing Hawking radiation is located. 

In the above, we have seen how including the entire purification of the thermal bath allows for the reconstruction of the black hole interior. The next natural question is how much of purifier is really necessary for this reconstruction, as was considered in section \ref{sec:smoke} for the equilibrium case. In order to investigate that question, we consider a subsystem with QM$_\mt{L}$, a bath interval $[0,\sigma_2]$ and a finite interval $[0,\tilde{\sigma}_3]$ in the purification (on the time slice $\tilde{u}_3$). As shown in figure \ref{fig:desvenlafaxine}d, the generalized entropy for the two competing channels are defined as 
\begin{equation}
S_{\mt{R}}
=S^\gen_{\QES-1} + S_{2-3}\,, \qquad 
S_{\mt{N}}
= S^\gen_\bif + S_{1-2} + S_3\,,
\end{equation}
where the three endpoints are taken as a point on the AdS boundary with $y_1^\pm = u$ (\ie $\sigma_1=0$), the bath point $y_2^\pm = u \mp \sigma_2$ in the region $\rm{\Rtwo}$ and the point with $\tilde{y}_3^\pm = \tilde{u}_3 \pm \tilde{\sigma}_3$ in the purification region, respectively. As before, the two terms $S^\gen_{\QES-1},\ S^\gen_\bif + S_{1-2} $ are given by eqs.~\eqref{SR_after} and \eqref{SN_after}, respectively. We only need to consider two new ingredients, \ie 
\begin{equation}
\begin{split}
S_3  &=\frac{c}{6} \log \(\frac{\sinh\(2\pi T_\bath \tilde{\sigma}_3\)}{\pi T_\bath\epsilon} \)  \,,\\
S_{2-3} &=\frac{c}{6}\log\( \frac{12 \pi E_s}{c \(\epsilon\pi T_\bath\)^2} \frac{x^+_2 \cosh\(\pi T_\bath \tilde{y}^+_{3}\) \cosh\(\pi T_\bath (y_2^-+\tilde{y}^-_3)\) }{\sqrt{f'(y_2^+)}}  \) \,.\\
\end{split}
\end{equation}
which can be derived from the counterparts with point $y_3^\pm$ in the region $\rm{\Rfour}$ by the map $\pi T_\bath \tilde{y}_3^\pm = \frac{i \pi }{2} - \pi T_\bath y^\pm_3$. First of all, it is easy to find that we can retrieve the results in the last subsection (see \eqref{S2_extra}) where we include the full purification region, by pushing the third point $\tilde{\sigma}_3$ to the IR cut-off surface with $\tilde{\sigma}_3\to\sigma_{\IR} \sim + \infty$ (\ie approaching the null surface in the spacetime of bath's purifier). More explicitly, we can define the difference due to the finite $\tilde{\sigma}_3$, \ie  
\begin{equation}
\(S_{\mt{N}}-S_{\mt{R}} \)  - \(S_{\mt{N}}-S_{\mt{R}} \)\big|_{\tilde{\sigma}_3 \to \infty}= \frac{c}{6} \log \( \frac{\sinh\( 2\pi T_\bath \tilde{\sigma}_3  \)}{ 2 e^{\pi T_\bath y^-_2} \cosh \( \pi T_\bath (\tilde{y}_3^+) \)\cosh (\pi T_\bath (y_2^- + \tilde{y}_3^-) )}  \)\,,
\end{equation}
as $\Delta S_{\mt{NR-NR}}$. Equipped with the above difference for the two configurations in figures \ref{fig:desvenlafaxine}c and \ref{fig:desvenlafaxine}d , we can discuss the result of cutting part of the purification in the reconstruction. Noting that the dependence on $\tilde{\sigma}_3$ only appears on $\Delta S_{\mt{NR-NR}}$,  one can easily find the derivative of $S_{\mt{N}}-S_{\mt{R}}$  satisfies
 \begin{equation}
 \begin{split}
  &\quad \frac{\partial \(S_{\mt{N}}-S_{\mt{R}}\)}{\partial \tilde{\sigma}_3} = \frac{\partial \(\Delta S_{\mt{NR-NR}}\)}{\partial \tilde{\sigma}_3}  \\
 &= \frac{c}{6}  \( 2\coth (2\pi T_\bath \tilde{\sigma}_3) -\tanh\( \pi T_\bath (\tilde{y}_3^+)\) +\tanh \(  \pi T_\bath (y_2^- + \tilde{y}_3^-) \)   \)  \ge  0\,,\\
 \end{split}
 \end{equation}
 due to the simple facts that $\coth x \ge 1$ for $x\ge0$ and $|\tanh  x| \le 1$. The above positive derivative shows that $S_{\mt{N}}-S_{\mt{R}}$ monotonically increases with the increase of $\sigma_2$, implying that it is easier to reconstruct the black hole interior by including a larger interval in the bath. We can then rewrite the condition for this subsystem to reconstruct the black hole interior as 
 \begin{equation}\label{eq:condition_d}
 S_{\mt{N}}-S_{\mt{R}}   =  \(S_{\mt{N}}-S_{\mt{R}} \)\big|_{\tilde{\sigma}_3 \to \infty} + \Delta S_{\mt{NR-NR}} \ge 0\,,
 \end{equation}
 where $\(S_{\mt{N}}-S_{\mt{R}} \)\big|_{\tilde{\sigma}_3 \to \infty} $ is positive if and if the condition in eq.~\eqref{linear_sigma2} or \eqref{sigmaturn} is satisfied. Because the maximum of $\Delta S_{\mt{NR-NR}}$ is defined as $\tilde{\sigma}_3 \to \infty$ and is zero, we always have  $\Delta S_{\mt{NR-NR}} < 0$ for a finite $\tilde{\sigma}_3$, indicating that we need to include more bath interval than the critical length $\sigma_2^{\ast}(u)$ (derived in eq.~\eqref{linear_sigma2} or \eqref{sigmaturn}) in order to make the channel with a finite portion of the purification recoverable. Recalling the $\sigma_2$-dependence of \eqref{eq:SRequalsSN}, we can find the following decomposition 
 \begin{equation}\label{eq:sigma23}
 S_{\mt{N}}-S_{\mt{R}}  \approx \frac{c}{6} \(2\pi T_\bath y^-_2+ \log\(  \frac{c}{6\pi E_s \epsilon} \frac{(f(u)-x^+_2) \sinh \(\pi \Tb\sigma_2 \)}{x^+_2\sqrt{f'(u)} \sinh(\pi \Tb (u+\sigma_2)) }\) \)+ \Delta S_{\mt{NR-NR}}  +   \cdots\,,
 \end{equation}
 where we ignored the extra terms without dependence on $\sigma_2, \tilde{\sigma}_3$. 
  Then we can simply take the results in the above subsection to derive the necessary conditions for $\sigma_2$ and $\tilde{\sigma}_3$. However, it is more convenient to define the length of the finite bath interval beyond the critical value as\footnote{We hide the complicated expressions which are not shown in \eqref{eq:sigma23}  by using $\sigma_2^\ast(u)$. For the equilibrium case discussed in section \ref{sec:smoke}, we considered a more general set-up with $\delta \sigma_2= \sigma_2 -\sigma_1 -\Delta_{\rm{turn}}$ where the critical value is just the constant $\Delta_{\rm{turn}}$. }
 \begin{equation}
 \delta \sigma_2 = \sigma_2 - \sigma_2^\ast (u)\,,
 \end{equation}
 which helps us to show the effect of including more bath interval and cutting part of the bath purifier. It is straightforward to rewrite the necessary condition \eqref{eq:condition_d} to support the recoverable channel for the  linear region ($ku \ll 1$) as 
 \begin{equation}
 \begin{split}
 2\pi \(T_1 + T_\bath\) \delta \sigma_2  +  \log \( \frac{\sinh\( 2\pi T_\bath \tilde{\sigma}_3  \)}{ 2 e^{\pi T_\bath y^-_2} \cosh \( \pi T_\bath (\tilde{y}_3^+) \)\cosh (\pi T_\bath (y_2^- + \tilde{y}_3^-) )}  \) \ge 0 \,, \\
 \end{split}
 \end{equation} 
 by noting the approximation \eqref{app_conditionAB} and its result $\(S_{\mt{N}}-S_{\mt{R}} \)\big|_{\tilde{\sigma}_3 \to \infty} = \frac{c}{3} \pi (T_1 +T_\bath) \delta \sigma_2$.  
 Noticing the other approximation \eqref{app_late} and the simple relation $\(S_{\mt{N}}-S_{\mt{R}} \)\big|_{\tilde{\sigma}_3 \to \infty} = \frac{2c \pi }{3} \pi T_\bath \delta \sigma_2$ in the late-time region, one can find the condition for reconstructing the interior of black hole reads 
 \begin{equation}
 4\pi  T_\bath \delta \sigma_2  +  \log \( \frac{\sinh\( 2\pi T_\bath \tilde{\sigma}_3  \)}{ 2 e^{\pi T_\bath y^-_2} \cosh \( \pi T_\bath (\tilde{y}_3^+) \)\cosh (\pi T_\bath (y_2^- + \tilde{y}_3^-) )}  \) \ge 0 \,, \\
 \end{equation}
 whose further reductions depend on the sign of the terms inside $\cosh$ functions and are similar to what have done in section \ref{sec:smoke}.  For example, if we assume all length scales on the above are larger than $\frac{1}{\pi T_\bath}$, we can simply find the length of extra bath interval $[\sigma_2^\ast(u), \sigma_2]$ is constrained by 
 \begin{equation}\label{eq:sigma3linear}
 \delta\sigma_2 \gtrsim
 \begin{cases} \frac{T_\bath}{2(T_1 + T_\bath)} \(  |\tilde{y}_3^+|  +|y_2^-+\tilde{y}_3^-| + y_2^- -2\tilde{\sigma}_3\)\,,  &\mbox{if} \quad  ku \ll 1 \\ 
 \frac{1}{4} \(  |\tilde{y}_3^+|  +|y_2^-+\tilde{y}_3^-| + y_2^- -2\tilde{\sigma}_3\)\,, & \mbox{if} \quad  e^{ku} \gg 1 \end{cases} \,.
 \end{equation}
 Then it is easy to find that the RHS of the above equation can be reduced to four cases where one of them vanishes, implying we need to consider the regime with $2\pi ( T_1+T_\bath )\delta\sigma_2 \ll 1$,  and other three cases at late-time region retrieve the results derived in \eqref{merry4}. Finally, we also comment the above linear dependence would like appear for the time region between the two limits due to the complicate dependence of entropy on $\sigma_2$. However, if we only focus on a small perturbation with $\delta\sigma_2/\sigma_2^* \ll1, (T_{\rm{eff}}+ T_\bath) \delta\sigma_2 \gg 1 $, we can calculate the derivate of $\(S_{\mt{N}}-S_{\mt{R}} \)\big|_{\tilde{\sigma}_3 \to \infty} $ with respect to $\sigma_2$ and find the following expected result
 \begin{equation}\label{eq:sigma3linearB}
2(T_{\rm{eff}}(y_2^+) + T_\bath) \delta\sigma_2 \gtrsim T_\bath\(  |\tilde{y}_3^+|  +|y_2^-+\tilde{y}_3^-| + y_2^- -2\tilde{\sigma}_3\)\,, 
 \end{equation} 
 where the two terms on the RHS describe the entropy of radiation located on the small region $[\sigma_2^\ast, \sigma_2]$ and emitted from the black hole and the thermal bath, respectively.

 Starting from the subsystem with \QML, bath interval with the critical bath length $\sigma_2^\ast$ and all purification,  the above inequalities in eqs.~\eqref{eq:sigma3linear} and \eqref{eq:sigma3linearB} tell us how much bath interval we need to include if we want to exclude part of the purification in $\tilde{\sigma}=[\tilde{\sigma}_3, \tilde{\sigma}_{\mt{IR}}]$. Needless to say, we can interpret these inequalities in the opposite way, \eg\footnote{Although eqs.~\eqref{merry4} and \eqref{eq:sigma3optimallinear} look very similar, it is important to keep in mind that the assumptions leading to the two results are different. While for eq.~\eqref{merry4} we simply needed to assume that all lengths we are dealing with are larger than the thermal scale, for eq.~\eqref{eq:sigma3optimallinear} we further needed to restrict to the cases where $\delta \sigma_2 \ll \sigma_2$.}
 \begin{equation}\label{eq:sigma3optimallinear}
   2\tilde{\sigma}_3 \gtrsim   |\tilde{y}_3^+|  +|y_2^-+\tilde{y}_3^-| + y_2^-  - 2\frac{T_{\rm{eff}}(y_2^+) + T_\bath}{T_\bath} \delta\sigma_2  \,.
 \end{equation}
Then we can learn how much bath's purifier is necessary for reconstruction for a fixed bath interval $[0, \sigma_2]$ plus \QML. 
In particular, we specify an interval in the purifier by both its length $\tilde{\sigma}_3$ and the time slice $\tilde{u}_3$ on which it is placed in the spacetime of purification region. First, we observe from \eqref{eq:sigma3optimallinear} if $|\tilde{u}_3|$ is very large both of the two expressions with absolute values on the right-hand side would be very large. That is, for very large $|\tilde{u}_3|$, we would need a large interval in the purifier with $\tilde{\sigma}_3\sim|\tilde{u}_3|$ to recover the black hole interior. 
Varying over the time slice $\tilde{u}_3$, we find that the ``optimal purifier'' with smallest length is determined by 
 \begin{equation}\label{eq:optimallinear}
 \tilde{\sigma}_3 \approx \frac{1}{2} y_2^- - \frac{T_{\rm{eff}}(y_2^+)+T_\bath}{2T_\bath} \delta \sigma_2  \,,\quad \text{with} \quad \Big|\tilde{u}_3+\frac12 y_2^-\Big| \le \frac{T_{\rm{eff}}(y_2^+)+T_\bath}{2T_\bath} \delta \sigma_2 \,.
 \end{equation}
We note that this expression simply reduces to the equilibrium case shown in the second case in eq.~\eqref{merry4} after taking either the late-time limit or setting $T_\bath=T_1$. Hence the present results are analogous to those illustrated for the equilibrium case in figure \ref{fig:duloxetine}. That is, from eq.~\reef{eq:optimallinear}, the optimal purifier lies anywhere on a band of time slices centered at $\tilde{u}_3 = -y_2^-/2$ and with width  $\Delta\tilde{u}_3 =\frac{T_{\rm{eff}}(y_2^+)+T_\bath}{2T_\bath} \delta \sigma_2$. In this band, the length of the purifier interval is given by the expression above. Therefore when $\delta \sigma_2/\sigma_2$ is small, the optimal purifier is simply an interval of length $\tilde{\sigma}_3 = y_2^-/2$ on the time slice $\tilde{u}_3 = -y_2^-/2$. 

In this subsection, we have discussed the necessity of the thermal bath's purification when the bath temperature is beyond the critical temperature and also the constrain on the length of the bath interval and its purifier. To complete the explorations on the role of purification, the last question we ask is what is the minimal length of the bath's purifier. Of course, we have shown it is zero when $T_\bath \le T_p$. For a bath system with higher temperature, it is natural to expect that the length of the bath's purifier is minimal when the entire bath interval is included in the subsystem for reconstruction. Making some more efforts, one can find that that expectation is true by showing $\partial_{\sigma_1} \( S_{\mt{N}} -S_{\mt{R}}\) \le 0$ and $\partial_{\sigma_2} \( S_{\mt{N}} -S_{\mt{R}}\) \ge 0$. It means that the best for reconstruction is including all the bath interval with $\sigma \in \[0,\sigma_{\mt{IR}}\]$. In the limit $\sigma_2 \to \sigma_{\mt{IR}} \sim +\infty$, one can read the entropy two completing channels as 
\begin{equation}
S_{\mt{R}}
=S^\gen_{\QES-1} + S_{3-\mt{IR}}\,, \qquad 
S_{\mt{N}}
= S^\gen_\bif + S_{1-\mt{IR}} + S_3\,,
\end{equation}
where the entropy for the two-point function $S_{1-\mt{IR}}$ is defined in  \eqref{eq:SNym} with taking $\sigma_1=0, \sigma_2 =\sigma_{\mt{IR}}$ and the last new ingredient $S_{3-\mt{IR}}$ is derived as
\begin{equation}
S_{3-\mt{IR}}= \frac{c}{3} \log \(  \frac{\sinh \( \pi T_\bath (\tilde{\sigma}_{\mt{IR}}- \tilde{\sigma}_3) \) }{ \pi T_\bath \epsilon} \)\,.
\end{equation}
Similar to the calculations for critical temperature, one can find the condition $S_{\mt{N}}-S_{\mt{R}}$ is rewritten as 
\begin{equation}
\begin{split}
&2k \(  \log\( \frac{6E_s}{c } \frac{x_1^+ \sinh(\pi T_\bath (-y^+_{\mt{IR}})) \sinh(\pi T_\bath (y_{\mt{IR}}-y_1^-) )}{\sqrt{f'(y_1^+)}} \) + 4\pi T_\bath \tilde{\sigma}_3 - 2\pi T_\bath \tilde{\sigma}_{\mt{IR}} \) \\
&\gtrsim   2\pi \( T_{\rm{eff}} \(u\) -T_0 \) \,,\\
\end{split}
\end{equation}
where we can easily see that more purification interval is more helpful for the reconstruction. Taking $\sigma_1 =0$ and late-time limit $e^{ku} \gg 1$, we can finally find the minimal purifier is constrained by 
\begin{equation}
 \tilde{\sigma}_3 \gtrsim  \frac{1}{4k T_\bath } \( 3T_\bath -2T_1 -T_0 +2T_\bath \log\( \frac{T_1+T_\bath}{2T_\bath} \) \) - \frac{1}{4\pi T_\bath} \log\( \frac{6E_s}{c T_1} \sqrt{\frac{T_1}{\Tb}} \)  + \cdots \,,
\end{equation}
which is irrelevant to the choice of $\tilde{u}_3$. And note that the RHS is positive when $T_\bath \gtrsim T_p$ as we illustrated around \eqref{eq:tp}.

%% file: sections/discuss.tex

In this paper, we continued our investigation \cite{Chen:2019uhq} of the \aims\ model \cite{Almheiri:2019hni,Almheiri:2019psf} describing a joining quench in a
doubly-holographic framework. The most interesting questions concern the
two-dimensional dynamics describing black hole evaporation (or growth). Invoking
holographic duality (twice), the generalized entropy becomes purely geometric
and its evaluation is tractable in this dynamical setup. In the
three-dimensional holographic dual, the black hole geometry contains a Planck
brane where Jackiw-Teitelboim gravity is localized. At finite temperature, there
is a new ingredient: a horizon in the three-dimensional bulk, beyond which the
second asymptotic boundary purifies the two-dimensional thermal state in the
bath. Despite this difference, we have shown that the Page curve still exhibits
three distinct phases (quench, scrambling, and late-time equilibration), as in
the zero-temperature case. However, there are several new qualitative features
in both the scrambling and late-time phase.

As in the zero temperature case, the quantum extremal surface remains at the
bifurcation point in the initial quench phase and then jumps out from the original
horizon in the scrambling phase where the generalized entropy shows a(n almost)
linear increase with the physical time. From the first holographic level, the
increase in entropy is due to the two-way exchange of quanta between the bath
and black hole, which is why the linear increase is proportional to $T_1+T_b$.
From the perspective of the doubly holographic model, the increase in
generalized entropy is related to the end-of-the-world brane falling deeper
into the bulk towards the horizon of the three-dimensional black hole.

After the Page time, the system enters the late-time phase in which the black
hole approaches an equilibrium state with the black hole. However, the evolution
of the black hole is determined by the temperature of the thermal bath. For a
bath with a temperature which matches that of the post-quench black hole
$T_\bath=T_1$, this equilibration is immediate and the generalized entropy is
constant throughout this phase. For a lower temperature bath with $T_\bath <
T_1$, the black hole evaporates and loses some of its mass, similar to the zero
temperature case in \cite{Almheiri:2019psf,Almheiri:2019hni,Chen:2019uhq}. Since
the black hole not only emits Hawking radiation but also receives the thermal
radiation from the bath, the black hole can also grow when the bath temperature
satisfies $T_\bath > T_1$. At the extremely late time, the system will finally
equilibrate with the bath temperature, and the entanglement entropy approaches
its equilibrium value. Figure \ref{fig:Page_curve} illustrates these three
possible scenarios.

We also found that the position of the late-time extremal surface relative to
the event horizon of the black hole depends on the temperature of the bath. In
the evaporating black hole models (with a bath at zero temperature)
of~\cite{Almheiri:2019psf,Almheiri:2019hni,Chen:2019uhq}, the late-time extremal
surface lies {inside} the horizon -- in fact, it lies inside the horizon
throughout the entire evolution of the black hole. Correspondingly, the
information of the region outside of the black hole could not be reconstructed
by \QML + bath. On the other hand, in the equilibrium configuration studied
in~\cite{Almheiri:2019yqk}, the extremal surface is located {outside} of the
event horizon. The equilibrium case studied in section~\ref{sec:equilibrium}
reproduces this behavior with the QES located outside the event horizon -- see
eq.~\reef{eq:soot}. Hence in these cases, the information just outside of the
event horizon could be reconstructed by \QML + bath after the Page time.
Moreover, at {any} temperature, the black hole eventually equilibrates with the
bath, and the system is qualitatively similar to the equilibrium case. Indeed,
for any temperature, after a time of $ku \geq
\log\left(\left|1-\frac{T_1^2}{T_b^2}\right| \sqrt{\frac{\pi T_1}{8k}}\right)$,
the late-time extremal surface crosses the horizon and stays outside as the
system equilibrates. Furthermore, for black hole temperatures $T_1$ very close
to the bath temperature $T_\bath$, \ie $\frac{|T_\bath-T_1|}{T_1} \leq
\sqrt{\frac{2k}{\pi T_1}}\, \exp\!\[\frac{T_1-T_0}{2T_1}\]$, the QES is already outside
of the event horizon at the Page time. One may ask why the behaviour of our black
holes where only one side is in equilibrium with the bath matches that of the
eternal two-sided black holes studied in \cite{Almheiri:2019yqk}, where there is
an equilibrium with a thermal bath on both sides. However, this is relatively
obvious from the holographic perspective since the HRT surfaces are really
probing identical portions of the three-dimensional bulk geometry in the island
phase for both cases.

As noted above, the appearance of QES outside of the horizon was first found in
\cite{Almheiri:2019yqk} for an eternal AdS$_2$ black hole coupled with a thermal
bath. This same behavior was also seen in higher dimensional holographic systems
\cite{Rozali:2019day,QEItwo}. A similar phenomenon is also found at black holes
in asymptotically flat spacetime, \eg \cite{Gautason:2020tmk,Hashimoto:2020cas}.
A dynamical QES crossing the horizon (similar to our present results) was also
found for an evaporating black hole in JT gravity \cite{Hollowood:2020cou}. As
discussed around eq.~\reef{eq:paroxetine}, while the QES may extend outside of
the horizon, it is never very far from the horizon. These results may imply that
we should consider some quantum corrections to the event horizon in order to
extend the boundary of the interior of black hole, \eg taking the stretched
horizon \cite{Susskind:1993if} as a surrogate for the event horizon. Then the
QES can be seen to stay outside the classical event horizon but inside the
stretched horizon \cite{Gautason:2020tmk}. However, let us add that in the
higher dimensional holographic systems studied in \cite{Chen:2020uac,QEItwo},
this effect can be understood in terms of entanglement wedge nesting
\cite{Headrick:2013zda,Wall:2012uf}

After deriving the Page curve with three phases as shown in figure
\ref{fig:Page_curve}, we further focused on investigating the ability of various
subsystems consisting of {\QML} and different parts of the bath interval to
reconstruct the black hole interior -- see figure \ref{fig:desvenlafaxine} for the
competing channels for every case. As we first demonstrated in the equilibrium
case of section~\ref{sec:equilibrium}, the reconstruction of black hole interior
always requires at least part of the purification of the bath. Of course, the
key difference from the scenario with the evaporating black hole coupled to a
zero temperature bath \cite{Almheiri:2019psf,Almheiri:2019hni,Chen:2019uhq} is
that our bath here begins in a mixed state before the quench whereas in the
previous studies the bath begins in a pure state (\ie the CFT vacuum). Hence,
part of the purification of the bath becomes essential for interior
reconstruction when the bath temperature is higher than the critical temperature
$T_p \sim \frac{1}{2}\( T_1 + T_0 \) \lesssim T_1$, as given in
eq.~\eqref{eq:tp}. This requirement arises for two reasons: First, the thermal
bath radiation in the interval containing the Hawking radiation must be purified
to distill information about the black hole interior. Second, after the quench,
thermal radiation from the bath falls into the black hole entangling the black
hole interior with radiation in the purifier. That is, part of the entanglement
initially shared between the bath and its purifier is transferred to the black
hole interior and the purifier. So information about the black hole interior is
spread to the purification although, of course, none of the Hawking radiation
enters this region.

A simple example where the importance of the purifier was seen was the case
where the black hole and the bath were in equilibrium, \ie with
$T_\bath=T_1>T_p$. In this case, the reconstruction of the black hole interior
with $\QM_{\mt{L}}$, a finite bath interval $[\sigma_1,\sigma_2]$ at some time
$u$, and a restricted portion $[0,\sigmapure_3]$ of the purifier at another time
$\upure_3$ was considered in section \ref{sec:smoke}. There, the bound
\eqref{merry4} on the purifier interval size $\sigmapure_3$ necessary for
reconstruction can be given a physical interpretation in figure
\ref{fig:duloxetine} as the requirement that $[0,\sigmapure_3]$ captures
purifier quanta entangled with out-going thermal bath radiation in $0 < y^- <
y_2^-$, shown in red in the left panel of figure \ref{fig:duloxetine}. Given the
thermofield double preparation of the bath and purifier, the relevant purifier
quanta are those marked by dashed wavy lines in the right panel of figure
\ref{fig:duloxetine}. The bound \eqref{merry4} then corresponds to the minimal
interval in the purifier which captures these quanta. Namely, if the bath
interval has a length that is only above-critical by a few thermal lengths, then
the requisite purifier interval must capture essentially all of the quanta
marked in the right panel of figure \ref{fig:duloxetine}, \eg see the blue
interval. If the bath interval exceeds the critical length with a large margin
$\sigma_2-\sigma_1-\Delta_\turn$, then the amount of the marked quanta that must
be captured by the purifier interval is reduced proportionately, \eg see the
green interval. This discussion, however, leaves open the question of why the
$0<y^-<y_2^-$ section of bath thermal radiation is important to begin with. One
might argue that the bath radiation in $y_1^-<y^-<y_2^-$ obfuscates the Hawking
radiation captured by the bath interval $[\sigma_1,\sigma_2]$, so that purifying
this section of bath thermal radiation is beneficial. One may also argue that
$0<y^-< y_\QES^-$ contains thermal bath radiation eaten by the quantum extremal
island, so its purifier would contain information about the island. But, it also
seems that the bath radiation in the in-between range $y_\QES^-<y^-<y_1^-$ is
not pertinent. In particular, if one is free to discard the purifier quanta for
this radiation, then it should be possible to reduce the interval length of
$[0,\sigmapure_3]$ beyond what is allowed by \eqref{merry4} in some cases where
$\sigma_2-\sigma_1$ exceeds $\Delta_\turn$ by many thermal lengths.

One may ask why the previous effects are not always important. That is, why is
there a critical bath temperature $T_p$ below which no portion of the purifier
is needed to recover the black hole interior. Certainly, there are many physical effects that come into play here, \eg the redundancy of the encoding of the black hole interior in the Hawking
radiation \cite{Chen:2019uhq}, 
but remarkably the critical temperature $T_p$ can be derived with the following simple intuitive argument:\footnote{This argument and the following calculations are similar in spirit to the calculations in Appendix B of \cite{Penington:2019npb}. We thank Geoff Penington for discussing this point with us.} Recall that in the usual black hole evaporation (with $T_\bath=0$), the Page phase arises when the naive entropy of the Hawking radiation exceeds the Bekenstein-Hawking entropy of the black hole. Of course, we now understand that this conflict is resolved by the formation of a quantum extremal island, and hence a portion of the black hole interior is reconstructible in this phase. When the black hole is coupled to a finite temperature bath, the appearance of the critical temperature $T_p$ indicates that islands form for lower bath temperatures but not for higher temperatures, when keeping track of modes in the mixed state of the bath (along with \QML). But in turn, we can understand this as indicating that for $T_\bath<T_p$, one reaches an inconsistency where the naive entropy of the bath (including the Hawking radiation and also \QML) exceeds the entropy of the black hole and the bath purifier. But no such inconsistency arises for $T_\bath>T_p$. Examining this latter perspective in more detail below then allows us to derive the critical temperature $T_p$.

That is, we consider the necessity of islands at late times in the evolution of the system. First, we observe that the (coarse-grained) Bekenstein-Hawking entropy provides a bound on the (fine-grained) entropy of \QMR, with\footnote{Violation of the Bekenstein area bound in the island region is a necessary condition for the appearance of QEIs~\cite{Hartman:2020khs}. In the following argument, we begin by assuming that it holds and so no QEI forms, even in the far future.}
\begin{equation}
\begin{aligned}
  S_{\QM_\mt{R}}
  \lesssim &\ S_{\UV} + S_\mt{BH}(\Teff)
            \xrightarrow[u\to\infty]{} S_{\UV} + S_\mt{BH}(\Tb)\,.
            \label{eq:wereNoStrangers}
\end{aligned}
\end{equation}
Here, $S_{\UV}$ denotes a UV-divergent contribution due to the separation between \QMR{} and the
bath, $\Teff$ is the effective temperature \eqref{eq:Teff} of the  JT black hole on the right, and $S_\mt{BH}$ is the Bekenstein-Hawking entropy \eqref{eq:bekenstein}. The argument now
proceeds as follows: Purity of the complete
system, including the bath's purifier, demands
\beq
S_{\QM_{\mt{L}}\cup\,\Bath}=  S_{\QM_{\mt{R}}\cup\,\pure}\le\ S_{\QM_{\mt{R}}} + S_{\pure}\,, \label{ponder}
\eeq
where $\pure$ denotes the bath's purifier. Here, the inequality follows from the subadditivity of entanglement entropy. Now let us begin by assuming the absence of any islands, in which case,
\begin{align}
  S_{\QM_{\mt{L}}\cup\,\Bath}
  \approx&\ \ S_{\Bek}(T_0) + S_{\Bath}\,.
           \label{eq:youKnowTheRules}
\end{align}
Note that here, we are implicitly including the entire bath region and so we must regulate the size of the latter to avoid having an IR divergence in  $S_{\Bath}$. Further, combining the bound
\eqref{eq:wereNoStrangers} with the subadditivity inequality in eq.~\reef{ponder}, we also have
\begin{align}
  S_{\QM_{\mt{R}}\cup\,\pure}
  \le&\ S_{\QM_{\mt{R}}} + S_{\pure}
       \lesssim S_{\UV} + S_{\Bek}(\Tb) + S_{\pure}\,.
       \label{eq:andSoDoI}
\end{align} 
It remains to approximate the difference $S_{\Bath}-S_{\pure}$. Just after the joining quench, $S_{\Bath}$ can be expressed as the sum of three contributions: $S_{\pure}$, the UV contribution $S_{\UV}$, and a shock contribution,\footnote{To obtain the following expression for $S_{\shock}$, we may compare, for example, the $x^\pm\in \rm{\Rtwo}$ and $x^\pm\in \rm{\Rfour}$ cases of eq.~\eqref{eq:absolut}. Further, we have chosen $1/T_1$ to be a `typical'   length scale for the $x^+$ coordinate. Other choices differing by $O(1)$ factors from this will not significantly modify the result of this argument.} \ie
\begin{align}
S_{\Bath}(u=0)= S_{\pure} + S_{\UV} + S_{\shock}\,,
\quad{\rm where}\ \ S_{\shock}
  \approx \frac{c}{6}\log\frac{E_S}{c\, T_1}\,.
  \label{eq:aFullCommitments}
\end{align}
Now while $S_{\pure}$ remains constant, $S_{\Bath}$
changes\footnote{Alternatively, one may
  discard from this argument, all of the `bystander' thermal quanta entangled to each other in the bath and purifier regions, which have not yet fallen into the black hole. In this case, $S_{\Bath}$ and $S_{\pure}$ both increase, respectively due to the absorption of Hawking radiation and being entangled with bath radiation lost to the black hole. What is important is that the difference $\partial_u(S_{\Bath}-S_{\pure})$ evolves according to the RHS of \eqref{eq:toLove}.} due both to the absorption of Hawking radiation at temperature $\Teff$ and the loss of thermal radiation to the black hole (purified by quanta in $S_{\pure}$) at temperature $\Tb$. To be precise, we have
\beq
  \partial_u S_{\Bath}
  \approx \frac{\pi c}{6}\,\(\Teff-\Tb\)\,.
     \label{eq:toLove}
\eeq
Now combining eqs.~\reef{eq:aFullCommitments} and \reef{eq:toLove}, we find at late times
\beqa
  \lim_{u\to\infty}S_{\Bath}-S_{\pure}
  &\approx& S_{\UV} + S_{\shock}
     + \frac{\pi c}{6} \int_0^\infty du\; (\Teff-\Tb) 
    \label{eq:whatImThinkingOf} \\
  &\approx& S_{\UV} + \frac{c}{6}\log\frac{E_S}{c\, T_1}
     + \frac{\pi c}{6k}\left[
     T_1-\Tb + O\left( \frac{(T_1-T_{\bath})^2}{T_1} \right )
     \right]
\nonumber
\eeqa
Finally, combining
eqs.~\eqref{eq:youKnowTheRules} and \eqref{eq:andSoDoI} together with eq.~\eqref{eq:whatImThinkingOf}, we find that our
assumption of no islands leads to
\begin{align}
  \Tb
  \gtrsim& \frac{T_0+T_1}{2} + \frac{k}{2\pi}\log\frac{E_S}{c T_1},
\end{align}
where we recognize the RHS as the expression \eqref{eq:tp} for $T_p$. Thus, $T_p$ corresponds to the bath temperature above which the inequality \reef{ponder} can be satisfied at late times without introducing any island.
Conversely, to satisfy the entropy bound \reef{ponder} for $\Tb<T_p$, an island must be introduced at sufficiently late
times and as a result, the black hole interior may be reconstructed from \QML{} and the bath alone, without the bath's purifier. Alternatively, for $\Tb>T_p$, reconstructing the black hole interior requires additional information from the purification.
This argument  provides further intuition for understanding the critical temperature $T_p$ than perhaps offered by the
initial calculations leading up to \eqref{eq:tp} in section \ref{sec:purif}.

Furthermore, for the lower bath temperatures $T_\bath < T_p$, we found that with the
subsystem comprising only \QML and a finite bath interval, as shown in figure
\ref{fig:desvenlafaxine}c, it is possible to reconstruct the black hole interior
in section \ref{sec:lowerTem}. The length of the minimal bath interval for
reconstruction increases with the physical time and approaches a linear increase
as shown in eqs.~\eqref{linear_sigma2} and \eqref{finaly2plus}, and as
summarized in figure \ref{fig:bath_intervals}. After including the purification
in the subsystem as presented in figures \ref{fig:desvenlafaxine}b and
\ref{fig:desvenlafaxine}d, we considered the reconstruction of the black hole
interior in section \ref{sec:purif2} with a general bath temperature $T_\bath$,
\ie interior reconstruction also becomes possible for $T_\bath > T_p$. We first
found that the black hole interior is reconstructible with any bath interval
above the shock-wave with a length larger than $\Delta_{\rm turn} \sim
\frac{T_1-T_0}{4k T_1}$, given in eq.~\eqref{eq:areYouSure} for the equilibrium
case. For the evaporating and thermalized black hole, the interval length
required for interior reconstruction increases with time as shown in
\eqref{eq:derivativsigma2}. Since late time behavior should be similar to the
equilibrium case, one finds as expected, the minimal interval length for late
times asymptotes to a finite constant which is defined as $\Delta \sigma_{\rm
  turn}$ in eq.~\eqref{sigmaturn}. The two above results are illustrated in
figure~\ref{fig:sigmaturn}.

Recent explorations on QES and Page curves inspire the island formula for the
quantum systems coupled to gravity \cite{Almheiri:2019hni}. Although we do not
explicitly apply the island formula in our analysis, it is clear that the island
region emerges in the recoverable channel, as shown in figure
\ref{fig:desvenlafaxine}. Without knowledge of the island formula, we can also
derive the same results and desired Page curve because we can apply the RT
formula in the doubly holographic models. In other words, RT formula knows about
the existence of the island. On the other hand, it is also possible to get the
right answer by noting the entropy of a subsystem in a {\it pure state} equals
the entropy of its complementary part. For example, we can easily find that the
entropy of QM$_\mt{R}$ after Page transition is defined by $S_{\QES-1}$ with
$y_1$ on the AdS boundary. Taking the pure state as the whole system, we simply
know $S_{\mt{QES}-1}$ also defines the generalized entropy of QM$_{\mt{L}}$,
entire bath interval, and its purification (see figure \ref{fig:desvenlafaxine}a
), which implies the Page curve for that subsystem. However, this approach does
not work for mixed states because the entropy of a subsystem in a {\it mixed
  state} generally does not agree with the entropy of its complementary part.
Let's construct a mixed state as an example by tracing out the bath's
purification. Then the complementary subsystem of QM$_{\mt{R}}$ consists of
QM$_{\mt{L}}$ and only the entire bath interval. Correspondingly, the
generalized entropy of this complementary system is defined by the minimal
entropy between the two channels (see figure \ref{fig:desvenlafaxine}c with
$\sigma_2 \to \sigma_{\mt{IR}}$ )
\begin{equation}
  \begin{split}
    S_{\mt{N}}&=S^{\gen}_{\QES''} +S_\halfLine \,, \qquad \text{No Island}\,,\\
    S_{\mt{R}}&=S^{\gen}_{\QES-1} +S_{\mt{IR}} \,, \qquad \text{With Island}\,.\\
  \end{split}
\end{equation} 
It is obvious that neither of the above two terms equals the entropy of \QMR,
\ie $S^{\gen}_{\QES-1}$. More importantly, we have shown $S_{\mt{N}}$ is always
preferred when $T_\bath \gtrsim T_p$, which indicates the entanglement wedge of
the corresponding subsystem with QM$_{\mt{L}}$ and any thermal bath interval
does not contain the island region.

As a final remark, let us comment on an important lesson from our results for
the reconstruction of the black hole interior. It is obvious that the emitted
Hawking radiation carries out information about the black hole. Although all the
Hawking radiation is only stored in the finite interval $\[0 ,
  \sigma_{\rm{shock}}(u)\)$, our studies on the reconstruction for a black hole
  coupled to a finite temperature bath indicate that the information describing
  the black hole interior is not contained solely within this part of the bath
  (along with \QML). Rather we see that in this situation, the black hole and
  Hawking radiation (\ie $\[0 , \sigma_{\rm{shock}}(u)\)$, the bath region) are
    entangled with a complicated environment comprising {\QML}, the remaining
    bath interval and the bath purifier, and hence the information about the
    black hole interior is distributed in a complicated way over the whole
    system. Of course, as identified above, the new physical mechanism
    contributing to the information flow in the present situation is the
    incoming radiation falling from the bath onto the black hole, which
    entangles the black hole interior with the purifier (and possibly distant
    regions in the finite temperature bath). For example, we found that when the
    bath temperature satisfies $T_\bath > T_p$, reconstruction always needs the
    purification even if we already have all of the Hawking radiation and \QML.
    On the other hand, we also found that the QM$_\mt{L}$ plus only a smaller
    bath interval $\[0 , \sigma_2(u)\]$ with $e^{ \pi T_1 (u -u_{\Page})}\gg 1$
    and $\sigma_2<\sigma_{\rm{shock}}(u)$ is also sufficient to recover the
    information of the black hole interior when $T_b<T_p$ in section
    \ref{sec:lowerTem} (see the right panel of figure \ref{fig:bath_intervals}).
    This means that we actually do not require all of the Hawking radiation. The
    information inherited in the ignorable (early-time) Hawking radiation
    located at $\[\sigma_2, \sigma_{\rm{shock}}\]$ is shared by other parts of
    the system. This reflects the redundancy of the encoding of the black hole
    interior in the Hawking radiation discussed in \cite{Chen:2019uhq}.
